\documentclass{report}
\usepackage{graphicx} % Required for inserting images
\usepackage[utf8]{inputenc}
\usepackage{geometry}
\usepackage{authblk}
\usepackage{graphicx}
\usepackage{float}
\usepackage{mathrsfs}
\usepackage{multirow}
\usepackage{amssymb}
\usepackage{amsmath}
\usepackage{mathtools}
\usepackage{pgfplots}
\usepackage{longtable}
\usepackage{hyperref}
\usepackage{url}
\usepackage{graphicx}

\usepackage{caption}
\usepackage{subfigure}

\graphicspath{ {./Images/} }

\newgeometry{
	inner=1.5in, % Inner margin
	outer=0.75in, % Outer margin
	top=0.75in, % Top margin
	bottom=0.75in, % Bottom margin
}

\title{Risk forecasting using LSTM-MDN\\ a Neural Network approach to Value-at-Risk forecasting\\}

\author{Nico Herrig}

\date{August 2023}

\begin{document}
\begin{titlepage}
    \begin{center}
        \vspace*{1cm}

        \Huge
        \textbf{Risk forecasting using \\
        Long Short-Term Memory \\ Mixture Density Networks}

        \vspace{0.5cm}
        \Large
        a Neural Network approach to Value-at-Risk forecasting

        \vspace{1.5cm}

        \textbf{Nico Herrig}

        \vspace{4cm}

        \includegraphics[width = 0.8\textwidth]{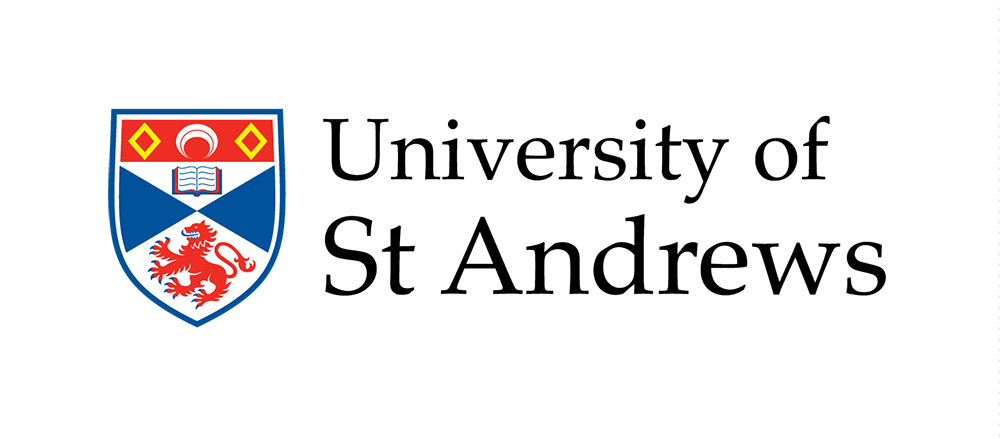}

        \vfill{}

        A thesis presented for the degree of \\
        Master of Science (MSc.)

        \vspace{0.8cm}

        \large
        \textbf{Supervisor:\\
        Dr. Valentin Popov\\
        }

        \vspace{0.2cm}
        
        University of St. Andrews \\
        Department of Mathematics and Statistics \\
        15.08.2023

    \end{center}
\end{titlepage}

%-----------------------PRE - MAIN ---------------------------------------

%Abstract here
\chapter*{Declaration}
I, Nico Herrig, declare that this thesis, including around 13700 words, as well as the work presented in it are my own. I confirm that:
\begin{itemize} 
\item This work was done wholly or mainly while in candidature for a postgraduate degree at this University.
\item Where any part of this thesis has previously been submitted for a degree or any other qualification at this University or any other institution, this has been clearly stated.
\item Where I have consulted the published work of others, this is always clearly attributed.
\item Where I have quoted from the work of others, the source is always given. With the exception of such quotations, this thesis is entirely my own work.
\item I have acknowledged all main sources of help.
\item Where the thesis is based on work done by myself jointly with others, I have made clear exactly what was done by others and what I have contributed myself.\\
\end{itemize}

 \par \vspace{8cm}
\noindent Signed:\\
\rule[0.5em]{25em}{0.5pt} % This prints a line for the signature
 
\noindent Date: \textbf{06.08.2023} \\
\rule[0.5em]{25em}{0.5pt} % This prints a line to write the date
\cleardoublepage

\chapter*{Abstract}
\pagenumbering{roman}
Inspired by recent breakthroughs in the field of machine learning, the objective of this work was to implement Long Short-Term Memory mixture density networks (LSTM-MDNs) for Value-at-Risk forecasting and compare their performance with three established models (historical simulation, CMM and GARCH) based on a defined backtesting procedure. Further emphasis was laid on the neural network's ability to account for volatility clustering and its practicality for a real-world implementation. Three different network architecture were tested: a 2-component mixture density network, a regularized version of the 2-component model as proposed by Arimond et al. (2020) and a 3-component mixture model. To the best knowledge of the author, this work is the first to test the latter approach in the context of Value-at-Risk forecasting.
\par \vspace{0.3cm}
Backtesting was performed using three stock indices (FTSE 100, S\&P 500 and EURO STOXX 50) over two different two-year periods (2017\&2018 as calm period and 2021\& 2022 as turbulent period), resulting in six evaluations. Model performance was assessed by testing for the unconditional coverage and the independence assumption. The neural network's ability to account for volatility clustering is validated through a correlation analysis and a graphical evaluation.
\par \vspace{0.3cm}
The results indicate only a modest potential for the neural network approach as it is used in this thesis. While the implemented LSTM-MDNs did not perform well for the 2017/2018 evaluation, they show better overall results for the 2021/2022 (high-volatility) period compared to the benchmark models. It could be shown that the neural networks are able to account for volatility clustering in a comparable way as the GARCH models due to the recurrent LSTM mechanism.
\par \vspace{0.3cm}
Findings generally indicate several issues of the LSTM-MDN models. First, an appropriate model initialization procedure would be required to ensure stable performance. Second, the neural networks tend to rely on a large amount of training data to learn underlying patterns. Overall, the outcome of this thesis shows that while LSTM-MDNs are able to produce adequate risk forecasts, further research and model adjustments aiming to mitigate the key issues mentioned above are required to ensure stable and reliable performance.
\par \vspace{5cm}
\textbf{Keywords:} \textit{Neural networks, LSTM, Long Short-Term Memory ,Value-at-Risk, VaR, Backtesting, Machine learning, MDN, Mixture Density Network, Quantitative finance 
}

% Dedications
\chapter*{Acknowledgements}
Mein aufrichtiger Dank gilt Prof. Dr. Valentin Popov, welcher diese Thesis betreute und begutachtete. Für die hilfreichen Anregungen und die kritische Begutachtung meiner Fortschritte, sowie jegliche konstruktive Kritik möchte ich mich herzlich bedanken.
\par \vspace{0.3cm}
Ein großes Dankeschön gilt zusätzlich meiner Famile sowie meiner Partnerin, welche mich während meiner Zeit in St. Andrews sowohl mental als auch finanziell zu jeder Zeit bedingungslos unterstützt haben. 
\par \vspace{0.3cm}
Des weiteren möchte ich mich bei all jenen Kommilitonen und Freunden bedanken, welche meine Thesis vor Abgabe korrekturgelesen haben sowie für jegliche kritischen Bemerkungen und konstruktiven Verbesserungsvorschläge.

% Table of Contents
\tableofcontents
\listoffigures
\listoftables

\chapter*{List of Abbreviations}

\begin{longtable}[l]{ l p{14cm} } 

Adam & Adaptive Moment Estimation \\
ANN / NN & (artificial) neural network \\
CMM & constant mean model \\
DNN & deep neural network (NN with more than one hidden layer) \\
FNN & feed-forward neural network \\
GARCH & Generalized AutoRegressive Conditional Heteroskedasticity \\
GD & gradient descent \\
HS & historical simulation \\
LSTM & long short-term memory \\
MDN & mixture density network \\ 
ML & machine learning \\
ReLU & Rectified Linear Unit  \\
RNN & recurrent neural network  \\
SGD & stochastic gradient descent \\
VaR & Value-at-Risk \\
\end{longtable}

\chapter*{List of Symbols \& Mathematical Abbreviations}
\subsection*{Latin Symbols}
\begin{longtable}[l]{ l p{14cm} } 
$a(\cdot)$ & activation function (context: section \ref{Activation_Functions_section}) \\
$b$ & bias vector (context: section \ref{chapter_Neural_Networks})\\
$C$ & cell state (context: section \ref{section_LSTM})\\
$d$ & length of a period/ rolling window \\
$e$ & Euler's number $(\approx 2.71828)$ \\
$ELU(\cdot)$ & ELU function \\
$\mathscr{F}$ & information \\
$h$ & context dependent: length of time horizon (section \ref{section_VAR}) or hidden state (section \ref{chapter_Neural_Networks}) \\
$I$ & indicator variable (binary) \\
$J(\theta)$ & a function of parameters $\theta$, equal to the value of a loss function $L(\cdot)$\\
$K$ & number of components (context: section \ref{section_MDN}) \\
$L$ & context dependent: vector of losses, likelihood \\
$L(y, \hat{y},\cdot)$ & loss functions with inputs $y$ (dependent variable) and $\hat{y}$ (predicted value) \\
$l$ & discrete loss (= negative discrete return) \\
$M$ & number of nodes in hidden layer (context: section \ref{chapter_Neural_Networks}) \\
$m$ & rolling average (context: section \ref{chapter_optimization}) \\
$N$ & sample size or vector length \\
$O(\cdot)$ & measure for computational expense \\
$P$ & asset price \\
$R$ & discrete return \\
$\mathcal{R}$ & L2 regularization term \\
$R(\cdot)$ & ReLU function \\
$Smax(\cdot)$ & Softmax function \\
$T$ & length of a time series \\
$u$ & sample from a uniform distribution \\
$v$ & rolling average (context: section \ref{chapter_optimization}) \\
$w$ & weight vector (context: chapter \ref{chapter_Neural_Networks}) \\
$X$ & stochastic process (general notation) \\
$x$ & variable (general notation) \\
\end{longtable}
%-----------------------------------------------------
\subsection*{Greek Symbols}
\begin{longtable}[l]{ l p{14cm} } 
$\alpha$ & context dependent: probability $\in [0,1]$ (section \ref{section_VAR}), model coefficient (section \ref{chapter_GARCH}), scale parameter (section \ref{chapter_optimization}) \\
$\beta$ & context dependent: model coefficient (section \ref{chapter_GARCH}), scale parameter (section \ref{chapter_optimization})\\
$\gamma$ & learning rate (context: chapter \ref{chapter_Neural_Networks}) \\
$\nabla$ & gradient \\
$\epsilon$ & error term \\
$\eta$ & innovation (context: section \ref{chapter_GARCH}) \\
$\theta$ &  neural network parameters (weight and biases)  \\
$\lambda$ & scale parameter for L2 regulatization term $(\mathcal{R})$ \\
$\pi$ & probability $\in [0,1]$ (context: section \ref{chapter_backtesting} \& \ref{section_MDN})\\
$\sigma$ & measure of deviation (standard deviation/ volatility) \\
$\phi(\cdot)$ & sigmoid function \\
\end{longtable}
%-----------------------------------------------------
\subsection*{Distributions}
\begin{longtable}[l]{ l p{14cm} }
$Bern(p)$ & Bernoulli distribution with probability $p\in[0,1]$\\
$D(\psi)$ & probability density function with distribution parameter vector $\psi$\\
$F(\cdot)$ & cumulative distribution function (CDF) \\
$F^{-1}(\cdot)$ & quantile function (inverse CDF) \\
$\mathcal{N}(\mu, \sigma^2)$ &  Normal/ Gaussian distribution with mean 
$\mu\in\mathbb{R}$ and variance $\sigma^2\in\mathbb{R}_{> 0}$\\
$q_{\alpha}$ & $\alpha$-quantile of a probability distribution \\
$\mathcal{U}(a,b)$ & uniform distribution with support at $a\in\mathbb{R}$ and $b\in\mathbb{R}$ \\
$\chi^2(k)$ & chi-square distribution with $k$ degrees-of-freedom \\
$z_{\alpha}$ & $\alpha$-quantile of a Gaussian distribution \\
\end{longtable}

\subsection*{Other}
\begin{longtable}[l]{ l p{14cm} }
$\lceil \cdot \rceil$ & ceiling function, rounds up its input to the next larger integer in case the input is not an integer itself.  \\
\end{longtable}

\chapter*{Software \& Code}
The code this project is built on (written in both \textit{Python} and \textit{R}) as well as results and data tables can be accessed under the respective \href{https://github.com/NicoHerrig95/Quant_Risk_Models}{GitHub Repository}. Detailed explanations on the used technologies can be found in section \ref{section_data_processing} and \ref{section_LSTM_implementation}.

%-------------------------------------------------------------------------
%----------------------- BODY --------------------------------------------

%INTRODUCTION
\chapter{Introduction} \label{chapter_introduction}

\pagenumbering{arabic}
\setcounter{page}{1}
Through recent advancements in the field of machine learning (ML), significant breakthroughs have been achieved in several areas such as natural language processing, robotics and healthcare sciences \cite{mashrur2020machine, verma2022machine}. Recently, these techniques also found their way into the financial industry. Popular models such as support vector machines, clustering algorithms or neural networks are not only used for money laundering prevention \cite{chen2018machine}, but are more and more introduced into the area of financial risk management. With the aim of minimizing losses, financial risk management can be interpreted as a the process of identifying, quantifying and assessing the financial risk of an asset or a portfolio. Risk in this context can be distinguished into four different types. Market risk refers to the potential for financial investments to incur losses due to unfavorable price fluctuations. Credit risk on the other hand is defined as the risk of a financial loss due to a borrower's inability to repay its debt. Operational risk arises from the uncertainties of the day-to-day business of a company, while liquidity risk emerges from the possible incapacity to fulfill payment commitments. Due to recurring market fluctuations and financial crises, risk management has evolved from a simple risk insurance technique to a sub-field of finance which heavily relies on financial mathematics, complex econometric and financial models \cite{10.1093/jjfinec/nbi003}. Since the financial crisis of 2008, several legislative frameworks, such as the adjustment of the Basel 2 accord in 2009 or the Regulation (EU) No 575/2013 form 2013\footnote{The Regulation (EU) No 575/2013 introduced a framework (Basel Traffic Light system) which increases the risk-based capital requirements of banks if their model for risk assessment is inaccurate.}, have been published to create an incentive for financial institutions to increase transparency and develop mathematical models for accurate risk assessment \cite{pepe2013basel}. Moreover, the assessment of uncertainty has been the main focus of risk-related research in the financial industry since decades \cite{segal2015good}.\par
\vspace{0.3cm}
Although being introduced in the 1980s \cite{holton2002history}, Value-at-Risk (VaR) is, besides Expected Shortfall (ES), still the gold standard for measuring market risk. As the Basel Committee requires banks to quantify risk exposure using VaR, it is  a crucial metric in finance used to measure the potential loss within a specific time period with a determined level of statistical confidence \cite{buczynski2023garchnet}. Despite its mathematical simplicity, a variety of different models and approaches emerged from both the industry and research over the years to appropriately quantifying Value-at-Risk. These approaches range from relatively simple yet effective concepts such as the Constant Mean Model (CMM) or the usage of empirical parameters to statistically motivated models like the ones from the ARCH familiy (e.g., GARCH and RiskMetrics) or concepts relying on Extreme Value Theory \\ \cite{manganelli2001value}.\par
\vspace{0.3cm}

Another approach  which is becoming more and more popular is the usage of machine learning (ML) for estimating market risks. ML concepts used for Value-at-Risk forecasting are ranging from the usage of support vector machines \cite{tsyurmasto2014value} to an implementation of generative AI \cite{arian2022encoded}. One of the most promising yet intuitive ML methods in this context are mixture density networks (MDNs). Introduced by Christopher M. Bishop in 1994, the main characteristic of an MDN, combining a conventional neural network and a mixture density model, is its output layer, returning parameters from an arbitrary conditional probability distribution which is conditional on the model input \cite{bishop1994mixture}. Recent research by Arimond et al. combines this type of model with the concept of Long-Short-Term-Memory (LSTM) \cite{arimond2020neural}. The following chapters build up on the idea proposed by Arimond et al. to model Value-at-Risk using a mixture density network in combination with LSTM layers. The aim of this paper is therefore not the development of a complex MDN providing optimised accuracy in its VaR forecasts, but rather to deal with the question on whether such network architecture provides potential to challenge established models in the financial sector from a practical point of view and under different market conditions. 
\par \vspace{0.3cm}
This work will further examine if LSTM-MDNs can appropriately account for volatility clustering in daily market returns. As a recurring pattern of the market, volatility clustering highlights the tendency for large absolute daily returns to be succeeded by returns of comparable amplitude and vice versa \cite{cont2007volatility}, implying non-constant volatility.

\section{Literature Review}
This section is dedicated to presenting research which is comparable to the experimental design of this thesis and thus inspired this work. Papers which are related to theoretical aspects or which are vaguely comparable to the underlying experimental design are presented in the respective chapter. \par \vspace{0.3cm}
Until this day, there is not much research conducted on the usage of LSTM-MDNs for financial risk forecasting. The first work\footnote{To the best knowledge and belief of the author} examining this topic was \cite{arimond2020neural}, who used a LSTM-MDN in combination with a Hidden Markov Model for Value-at-Risk forecasting of equity markets and long-term bonds. Their research was concerned not only with the adequacy of the model in terms of Value-at-Risk forecasts, but further focused on the network's ability to model market regime switches between bull and bear markets. The authors showed promising results of the implemented neural network and further emphasised the importance of large train sets and a balanced incentive structure of the loss function (more on this in section \ref{section_model_architecture}).
\par \vspace{0.3cm}
Building up on these findings, the paper by Karlsson Lille and Saphir (2021) compared the performance of an LSTM-MDN model (using a two-component Gaussian mixture) with the established "mean-variance approach"\footnote{Their approach is in practice comparable to Constant Mean Model presented in section \ref{section_CMM}} and the historical simulation \cite{karlsson2021value}. The authors conclude limited promise of the implemented network for Value-at-Risk forecasting as the model tends to strongly overestimate the risk. Similar to the findings of Arimond et al. (2020), the inclusion of a regularization term in the loss function (balancing the components from the mixture model) led to better model performance and also increased the model's ability to account for market regime switches. 
\par \vspace{0.3cm}
Buczynski and Chlebus proposed "GARCHNet", a combination of a LSTM neural network and a GARCH model \cite{buczynski2023garchnet}. Their findings show that due to the network's ability to model non-linear structures, GARCHNet outperforms the established GARCH models, which are used as benchmark comparison. 
\par \vspace{0.3cm}
Ormaniec et al. (2022) compared LSTM neural networks with different GARCH configurations \cite{ormaniec2022estimating}. The models were tested on both simulated and real-world data. Their work showed promising results for the LSTM model as they outperformed the benchmark GARCH models in the majority of tests. Ormaniec et al. (2022) especially highlighted the neural network's reactivity to shifts in market volatility.
\pagebreak
\section{Research Questions} \label{section_research_questions}
Inspired by and building up on the research presented above, this work shall answer the following research questions: 

\begin{itemize}
    \item[\textbf{1}]: Does the LSTM-MDN architecture provide more accurate VaR forecasts than models typically used in the industry (historical simulation, CMM \& GARCH)?
    \item[\textbf{2}]: If question 1 holds true, does the LSTM-MDN architecture provide enough practicality for a real-world implementation? 
    \item[\textbf{3}]: Do LSTM-MDNs account for volatility clustering? 
\end{itemize}

%Theoretical Framework
\chapter{Theoretical Framework}

\section{Market Theory}
Financial markets enable market participants to buy and sell equity without capital flowing to firms directly, with the stock market being the most prominent of them \cite{bond2012real}. Capital markets are subject to a number of influencing factors such as the economy, politics or socially-cultural factors, leading to price fluctuations of the traded equities and the general market. Financial theory quantifies these price changes as volatility, which describes the fluctuation observed over a specific time period \cite{andersen2006volatility}. High volatility,  expressing strong price fluctuations, is therefore an indicator for a high level of uncertainty in the market. Subsequently, this level of uncertainty needs to be accounted for in financial risk forecasting.\par
\vspace{0.3cm}
In quantitative finance, it is advisable to work with price changes instead of equity prices themselves. Converting prices to respective returns over time provides two important properties. Returns are expected to be centered around zero while we expect prices to be non-stationary as their mean changes over time. Further, returns are scale-free and therefore more comparable and easier to interpret \cite{campbell1998econometrics}. The discrete return $R$ for time point $t$ is defined as \begin{equation} \label{eq_returns} 
R_t = \frac{P_t - P_{t-1}}{P_{t-1}},    
\end{equation} where $P$ denotes the equity price. Although it is advisable for some financial applications to work with the continuous returns $r_t = ln(P_t) - ln(P_{t-1})$ as it provides some preferred properties for time-aggregation, the discrete returns are used in the following chapters. First, discrete returns can be interpreted more intuitively. Second, it can roughly be assumed that $R_t \approx r_t$ for daily returns. As daily stock returns are expected to be centered around 0 and are small in absolute value, it can easily be shown that the difference between continuous and discrete returns is negligibly small\footnote{The difference between continuous and discrete returns is below 0.00096 for 99\% of observation from the data.} as it holds $r_t = ln(1 + R_t)$. In its simplest form, the volatility of a time series of returns can be expressed as its observed standard deviation $\hat{\sigma}$ over a defined time period.\par
\begin{figure}[H]
    \centering
    \includegraphics[width=10cm, height = 8cm]{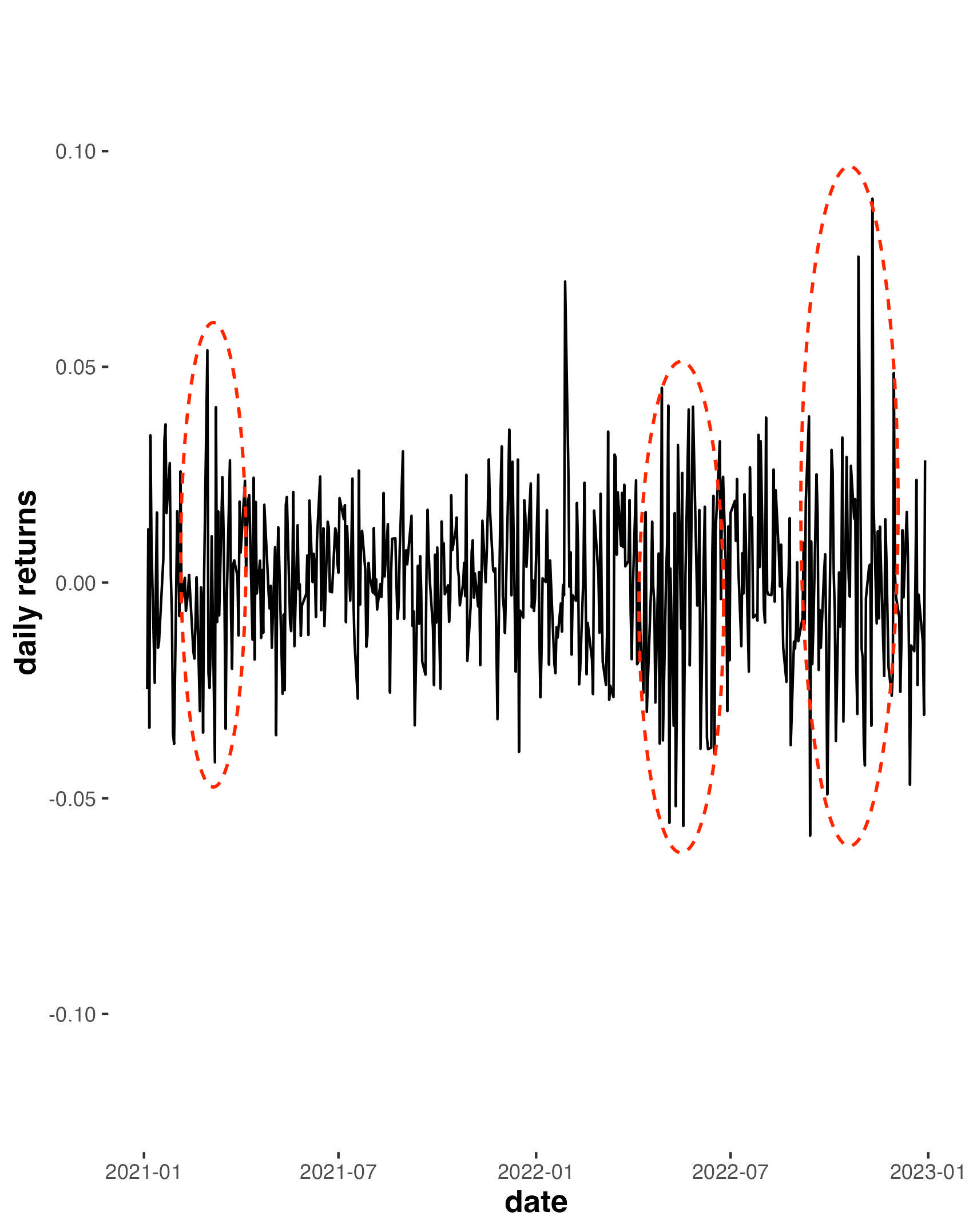}
    \caption{Volatility Clustering}%
{\small Daily returns of AAPL stock - circled areas indicate highly volatile phases (i.e., shifts in price volatility)}
    \label{AAPL_returns}
\end{figure}

\vspace{0.3cm} 
Empirical research showed that financial markets are subject to certain recurring patterns, which are described as stylized facts in the literature. The stylized facts of a certain financial property like daily returns can be understood as \textit{"empirical findings that are so consistent across markets that they are accepted as truth"} \cite{sewell2011characterization}. One of the most important stylized facts for risk management is volatility clustering, which describes the finding that large absolute daily returns are often followed by large absolute daily returns (see figure \ref{AAPL_returns}) and vice versa \cite{cont2007volatility}. Volatility clustering implies that the standard deviation for (daily) returns is not constant over time and therefore conditional on prior returns i.e. their variance. 

Two of the most prominent time series models which account for volatility clustering are the Autoregressive Conditional Heteroskedasticity model (ARCH) \cite{engle1982autoregressive} and  the four years later published Generalized Autoregressive Conditional Heteroskedasticity (GARCH) model \cite{bollerslev1986generalized}, which is further discussed in section \ref{chapter_GARCH}. 
\par \vspace{0.3cm}
Market uncertainty and price fluctuations lead to the necessity to quantify the underlying market risk of an asset for a rational investment decision making. One of the most prominent and important measures is the Value-at-Risk, which is presented in the following section.

%--------------------------- VALUE AT RISK ----------------------
\section{Value-at-Risk} \label{section_VAR}
Introduced in the 1980s, Value-at-Risk is still one of the most popular measures for market risk in the financial industry and is the proposed measure under Basel II and III \cite{sharma2012evaluation}.VaR expresses the maximum amount of monetary value a financial asset may lose in a certain time horizon (in the future), with a given level of confidence \cite{best2000implementing}. Considering a tradable asset at present time. Further, denote the loss for a fixed time horizon in the future $L_{t+h}$, where $h$ is a fixed time frame and $L$ is a continuous variable. Let $F_{L_{t+h}}(l) =  \mathcal{P}(L_{t+h} \le l|\mathscr{F})$ be the conditional cumulative distribution function of $L_{t+h}$, where $\mathscr{F}$ denotes the information until the presence ($t$). Using the definition provided by \cite{mcneil2015quantitative}, the VaR of an asset for a given time horizon $h$ can then be expresses as  
\begin{equation}\label{eq_VaR}
\begin{split}
VaR[\alpha, h] & = \inf\{l\in\mathbb{R}: \mathcal{P}(L_{t+h} \le l|\mathscr{F}_t) \le 1 - \alpha\} \\
& = \inf\{l\in\mathbb{R}:  F_{L_{t+h}}(l) \ge \alpha \} \\
& = F^{-1}_{L_{t+h}}(\alpha).
\end{split}
\end{equation}

It is worth mentioning that the above definition and the following chapters deal with daily losses rather than daily returns, where a daily loss is simply defined as the negative return: $l_t = -R_t$. Further, the Value-at-Risk property is generally reported as a positive value. Equation \ref{eq_VaR} shows that the Value-at-Risk of an asset can generally be defined as the $\alpha$-quantile $(q_{\alpha})$ of its loss distribution. For a symmetric distribution it holds: $-q_{\alpha} = q_{1-\alpha}$. Therefore, the $\alpha$-quantile of the loss distribution is equal to the $1-\alpha$ quantile of the respective  distribution of returns. \par \vspace{0.3cm} 
The current Basel Accords suggest calculating the VaR on a daily basis with a confidence level of 99\% ($\alpha = 0.99$) for an observation period of one year or more \cite{uylangco2016evaluation}. Thus, financial institutions have a need to develop models which determine the cumulative distribution function $F_{L_{t+h}}(l)$ and further estimate $VaR[\alpha = 0.99, h=1]$, where $h$ denotes the number of days in the future. 
 
\section{Benchmark Models}
Both research and the financial industry produced a variety of different approaches to forecast Value-at-Risk. Generally, these approaches are dividable into parametric approaches and non-parametric approaches \cite{karlsson2021value}. In the context of Value-at-Risk, parametric methods assume that losses come from a defined theoretical distribution and make assumptions about the respective distribution parameters, while non-parametric models do not make this assumption and solely rely on empirical data.

\subsection{Historical Simulations} \label{historical_simulations}
The most prominent yet simple non-parametric approach is the historical simulation method. Instead of assuming a theoretical probability distribution, it uses the $\alpha$-quantile from the empirical cumulative distribution function (ECDF) of losses which are sorted in ascending order. \par \vspace{0.3cm} 

Consider a tradable asset and a sample of its ordered daily losses from the last $T$ days: $\{L_t\}$, where $l_{(1)}\le l_{(2)} \le ... \le l_{(T)}$. The empirical Value-at-Risk for $h = 1$ under historical simulation with price $P$ would then be defined\footnote{For a sample of ordered losses $L$ of size $T = 36$, $VaR_{HS}[\alpha = 0.95, h = 1]$ = $l_{(\lceil 0.95*36 \rceil)} = l_{(\lceil 34.2 \rceil)} = l_{(35)}$, i.e. the 35th order statistic of $L$.} as 

\begin{equation}\label{eq_historicalsim}
VaR_{HS}[\alpha, h = 1] = l_{(\lceil \alpha T \rceil)}\times P_t 
\end{equation}
The biggest advantage of this method lies in its simplicity and its freedom from parametric assumptions. A study from 2010 showed that from those instututions which disclosed their method of calculating VaR, 73\% used the historical simulation approach  \cite{perignon2010level}. In many cases, the non-parametric approach performs better than more complex parametric models such as GARCH \cite{sahoo2017comparison}.

\subsection{Constant Mean Model}\label{section_CMM}
In contrast to the historical simulation, the Constant Mean Model (CMM) assumes that returns (and subsequently losses) come from a Gaussian distribution. Generalized on a stochastic process $X_t$ with an expected value $E(X_t) = \mu_X$, a CMM of $X_t$ is defined as
\begin{equation} \label{equ_CMM}
    X_t = \mu_X + \epsilon_t \text{,~~~where } \epsilon_t \overset{\mathrm{iid}}{\sim} \mathcal{N}(0, \sigma^2_{\epsilon}).
\end{equation}
Equation \ref{equ_CMM} implies that $X_t$ is a White Noise process\footnote{A stochastic process is called a White Noise process if a) random variables are uncorrelated ($Cov(X_t, X_{t+h})=0$ for $h \ne 0$) and b) the process has an expected value of 0 ($E(X_t) = 0$).} following a Gaussian distribution (Gaussian White Noise) with a constant mean $\mu_X$ and a variance $\sigma^2_{\epsilon}$. \par \vspace{0.3cm}
In practice, the parameters $\mu$ and $\sigma$ have to be estimated from a time series of past returns. Consider such a series of the past $T$ daily returns until presence. The parameter estimates would then be defined as
\begin{equation}\label{equation_CMM_mean}
    \hat{\mu} = \frac{1}{T} \sum_{t=1}^{T}R_t \text{,~~~for~} t=1,...,T\\
\end{equation}

\begin{equation}\label{equation_CMM_sigma}
    \hat{\sigma}^2 = \frac{1}{T} \sum_{t=1}^{T}(R_t - \mu)^2\\ \text{,~~for~} t=1,...,T
\end{equation}
The returns can then be modelled as 
\begin{equation} \label{CMM_returns_eq}
    R_t = \hat{\mu} + \epsilon_t \text{,~~~where } \epsilon_t \overset{\mathrm{iid}}{\sim} \mathcal{N}(0, \hat{\sigma}^2).
\end{equation}

It follows from equation \ref{CMM_returns_eq}  that $R_t \overset{\mathrm{iid}}{\sim} \mathcal{N}(\hat{\mu}, \hat{\sigma}^2)$. As returns are assumed to be normally distributed under the model, the $\alpha$-quantile of the Gaussian distribution can be used for calculating the respective Value-at-Risk. The VaR for under the Constant Mean Model for a fixed time frame of $h=1$ can be calculated as
\begin{equation} \label{equation_CMM_VAR}
\begin{split}
VaR_{CMM}[\alpha, h = 1] &= -\left(\hat{\mu} + z_{1-\alpha}\sqrt{\hat{\sigma}^2}\right)P_t\\
& = -\left(\hat{\mu} + z_{1-\alpha}\hat{\sigma}\right)P_t \text{~~~,}\\
\end{split}
\end{equation}
where $P_t$ is the asset value at time point $t$.
As for the historical simulation, the CMM has the advantages that it is simple to implement and is computationally inexpensive. However, as it assumes returns to be  normally distributed, it fails to account for the stylized fact that returns tend to follow a leptokurtic probability distribution \cite{sewell2011characterization}. Due to the assumption of a constant variance, the CMM is further unable to account for volatility clustering.

\subsection{GARCH} \label{chapter_GARCH}
The GARCH model shall account for several stylized facts, including volatility clustering and the assumption of returns coming from a leptokurtic distribution \cite{malmsten2010stylized}. As proposed by Robert Engle [1982], a GARCH(p,q) model of returns can be described as the following
\begin{equation}
    R_t = \mu + \epsilon_t
\end{equation}

\begin{equation}
    \epsilon_t = \sigma_t \eta_t
\end{equation}

\begin{equation} \label{GARCH_volaeq}
    \sigma_t^2 = \alpha_0 + \alpha_1\epsilon^2_{t-1}+...+ \alpha_q\epsilon^2_{t-q} + \beta_1\sigma_{t-1}^2 + ... + \beta_p\sigma_{t-p}^2\text{~~~,}
\end{equation}
where the innovation $\eta_t$ is assumed to be i.i.d. distributed, following a defined probability distribution\footnote{e.g., normal distribution, t-distribution, GED, skewed t-distribution} with unit variance and an expectation of 0 \cite{buczynski2023garchnet}. Given normally distributed innovations, it can be shown that a GARCH model reduces to a CMM model if $\alpha_1 = ... \alpha_q = \beta_1 = .... = \beta_p = 0 $ in the volatility equation (Equation \ref{GARCH_volaeq}).

\par \vspace{0.3cm}
Depending on the chosen innovation distribution, the Value-at-Risk can then be calculated by multiplying the product of the negative square root of the volatility equation \ref{GARCH_volaeq} and the respective distribution quantile with the asset price. The inclusion of $\mu$ into the VaR calculation under equation \ref{GARCH_VAR} is negligible as it can be assumed that $\mu \approx 0$.
\begin{equation}
\begin{split} \label{GARCH_VAR}
    VaR_{GARCH}[\alpha, h = 1] &= -\left(\sqrt{\sigma_t^2} \times q_{1-\alpha}\right) P_t\\
    &= -\left(\sigma_t \times q_{1-\alpha}\right)P_t\\
\end{split}   
\end{equation}
\par \vspace{0.3cm}

For practical usage, the literature (\cite{hansen2005forecast, namugaya2014modelling}) shows that a low order GARCH(1,1) process is sufficient for modelling returns and is expected to outperform higher-order models. Based on these findings and for the sake of simplicity, GARCH(1,1) structures are solely used in this paper. The conditional variance from equation \ref{GARCH_volaeq} therefore reduces to 
\begin{equation} \label{GARCH1_1_vola}
    \sigma_t^2 = \alpha_0 + \alpha_1\epsilon^2_{t-1}+ \beta_1\sigma_{t-1}^2 
\end{equation}
A GARCH(1,1) process is assumed to be stationary if $0 < \alpha_1 + \beta_1 < 1$ (the stationarity condition) holds \cite{bollerslev1986generalized}. If this condition is given, the unconditional variance of $\epsilon$, denoted $\sigma^2$, is defined as $\frac{\alpha_0}{1-\alpha_1-\beta_1}$. Thus, it can be shown that 
\begin{equation} \label{GARCH_vola_reduced}
\begin{split}
    \alpha_0 &= \sigma^2 \left(1-\alpha_1-\beta_1\right) \\
    \sigma_t^2 &= \sigma^2 \left(1-\alpha_1-\beta_1\right) + \alpha_1\epsilon^2_{t-1}+ \beta_1\sigma_{t-1}^2 \\
    &= \sigma^2 \left(1-\alpha_1-\beta_1\right) + \alpha_1(R_{t-1} - \mu)^2+ \beta_1\sigma_{t-1}^2 \text{~~~,}\\
\end{split}  
\end{equation}
where $\mu$ is assumed to be zero in this paper. Equation \ref{GARCH_vola_reduced} shows that the the conditional variance is composed of three parts: The unconditional uncertainty expressed as a time-independent constant $\sigma^2$ and scaled by $(1-\alpha_1 - \beta_1)$, the recent shock scaled by $\alpha_1$ and the prior conditional volatility forecast multiplied by $\beta_1$.  A $k$ step-ahead forecast for the conditional variance ($\sigma^2_{t+k}$) can be defined as
\begin{equation} \label{equation_long_term_forecast_GARCH}
    \sigma^2_{t+k} |\mathscr{F}_{t-1} = \sigma^2 + (\alpha_1 + \beta_1)^{k-1}((\sigma^2_t|\mathscr{F}_{t-1}) - \sigma^2)\text{~~~,}
\end{equation}
where $\mathscr{F}_{t-1}$ denotes the information available at time point $t-1$. Considering a long-term forecast for $k \gg 1$, the right side of equation \ref{equation_long_term_forecast_GARCH} would approach $\sigma^2$ with increasing $k$ while $\sigma^2_t \approx \sigma^2$ for $\lim\limits_{k\to \infty}$. The unconditional variance can therefore be interpreted as the long-term variance, while the conditional volatility will increase or decrease proportional to the absolute return at the prior day. This allows the GARCH model to account for the phenomenon of volatility clustering.

\par \vspace{0.3cm}
Besides the order of the process, the distribution of the innovations needs to be defined. For this paper, innovations following a normal distribution are compared to innovations which are GED distributed. Akaike information criterion (AIC) is used to compared  both variations and choose the appropriate option per model. AIC is defined as 
\begin{equation} \label{AIC}
    AIC = -2\ln{L} + 2k \text{~~~,}
\end{equation}
where $L$ is the maximised likelihood of the fitted model and $k$ is the number of model parameters. Equation \ref{AIC} shows that the AIC score is based on a trade-off between the likelihood of the model and its complexity (expressed as number of parameters), where a lower AIC score implies a better trade-off.
\par \vspace{0.3cm}
Clearly, the biggest advantage of the GARCH model over the CMM is its capability to account for a variety of stylized facts, including volatility clustering and fat-tailed return distributions. However, besides being computationally expensive due to the parameters estimation procedure, a weakness of a (plain) GARCH model is that it solely depends on the magnitude of the recent shock rather than its sign (i.e., a negative or positive return). Further, the GARCH does not account for skewness or asymmetry in the return distribution \cite{olowe2009modelling}.

\section{Backtesting} \label{chapter_backtesting}
In general, backtesting describes the process of evaluating how well a method or a strategy would have performed in the past by retroactively applying it to past data. In the context of this work, backtesting can be defined as \emph{"the evaluation of financial risk models using historical data on risk forecasts and profit and loss realizations"} \cite{christoffersen2008backtesting}. In risk management,  backtesting  evaluates the quality of a model by comparing its forecast for time point $t$ with the respective ex-post loss. This procedure further enables the comparison between different models and methods.\par \vspace{0.3cm}
Consider $T$ sequential observations from a time series, where each observation contains the realized loss $l_t$ and the VaR forecast $VaR[\alpha]_{t}$ for a given confidence $\alpha$, which was calculated at time point $t-1$. As proposed by Christoffersen (2008), a binary series $I(\alpha)$ can be computed, where 
\begin{equation}
    I_t(\alpha) = 
    \begin{cases}
        1 & \text{if }l_t > VaR[\alpha]_{t} \\
        0 & \text{if } l_t \le VaR[\alpha]_{t}
    \end{cases}
\end{equation}
for $t = 1,...,T$.
\par \vspace{0.3cm}
When backtesting models for VaR forecasting, two properties are generally of interest. The unconditional coverage property assumes that the probability of a VaR breach (i.e., $l_t > VaR[\alpha]_{t}$) during the backtesting period is $(1-\alpha) \times 100\%$ \cite{campbell2005review}. The property can be expressed as 
\begin{equation} \label{uncon_coverage_assumption}
    P(I_t(\alpha) = 1) =  1-\alpha
\end{equation}
The second property of interest is the independence property of $I(\alpha)$, which claims that the probability of a VaR breach does not depend on a potential breach in the past \cite{campbell2005review}. Under the assumption of independence, $I_t(\alpha)$ is i.i.d. Bernoulli distributed with  $p = 1 - \alpha$.
\begin{equation} \label{binary_series}
   I_{t}(\alpha) \overset{\mathrm{iid}}{\sim} Bern\left(1-\alpha\right)
\end{equation}
A model can be deemed appropriate if it fulfills both properties. A test for the unconditional coverage property is presented in section \ref{section_unconditional_coverage}, while section \ref{section_independence} presents a testing procedure for the independence assumption. A joint test which combines both tests is presented in section \ref{section_joint_test}.

\subsection{Unconditional Coverage Test} \label{section_unconditional_coverage}
A prominent method to test for the unconditional coverage hypotheses from equation \ref{uncon_coverage_assumption}  is the Proportion of Failure test (POF)\cite{kupiec1995techniques}. The POF in its core is a likelihood ratio test which accounts for the difference between the expected proportion of VaR breaches during the backtesting period ($1-\alpha$) and the observed proportion of ex-post breaches ($\hat{\alpha}$). The POF statistic is defined as 
\begin{equation} \label{POF}
\begin{split}
 LR_{POF} &= 2\ln{\left[\left(\frac{1-\hat{\alpha}}{1-(1-\alpha)}\right)^{T-I(\alpha)}\left(\frac{\hat{\alpha}}{1} \right)^{I(\alpha)}\right]}\\
  & = 2\ln{\left[\left(\frac{1-\hat{\alpha}}{\alpha}\right)^{T-I(\alpha)}\left(\frac{\hat{\alpha}}{1} \right)^{I(\alpha)}\right]}\\
 I(\alpha) &= \sum_{t=1}^T I_t(\alpha) \\
 \hat{\alpha} &= \frac{1}{T}I(\alpha) \\
\end{split}
\end{equation}
Equation \ref{POF} indicates that the test statistic grows if the number of observed breaches is more or less than the expected number of breaches, while $LR_{POF}$ takes the value zero if $\hat{\alpha} = 1-\alpha$ . Kupiec also showed that the POF test statistic is asymptotically $\chi^2$ distributed with 1 degree of freedom $(LR_{POF} \overset{\mathrm{asy}}{\sim}\chi^2(1))$. Under the null it is assumed that $E(I_t(\alpha)) = 1-\alpha$, while $E(I_t(\alpha)) \ne 1-\alpha$ holds true under the alternative hypotheses.

\subsection{Independence Test} \label{section_independence}
The independence hypothesis can be tested with the Interval Forecast test \cite{christoffersen1998evaluating}. The test, which was proposed by Christoffersen in 1998, measures the dependency between consecutive days. The test statistic\footnote{often denoted as $LR_{CCI}$ in the literature} $LR_I$  is defined as 
\begin{equation} \label{LR_I}
    LR_I = -2\ln{\left[\frac{\left(1 - \pi \right)^{n00 + n10}\pi_0^{n01+n11}}{\left(1-\pi_0\right)^{n00}\pi_0^{n01}\left(1-\pi_1\right)^{n10}\pi_1^{n11}}  
    \right]}\text{~~~,}
\end{equation}
where $\pi$ in equation \ref{LR_I} defines the probability of having a breach at time point $t$, while $\pi_0$ is the probability of a breach at time point $t$ given no breach at $t-1$ and $\pi_1$ is the probability of two successive breaches.

\begin{gather*}
    \pi = \frac{n_{01} + n_{11}}{n_{00} + n_{01} + n_{10} + n_{11}}\\
    \pi_0 = \frac{n_{01}}{n_{00} + n_{01}} \\
    \pi_1 = \frac{n_{11}}{n_{10} + n_{11}}  \\
\end{gather*}

\begin{itemize}
    \item \textbf{n00} is the number of periods with no breaches followed by a period with no breaches,
    \item \textbf{n10} is the number of periods with breaches followed by a period with no breaches,
    \item \textbf{n01} is the number of periods with no breaches followed by a period with breaches and
    \item \textbf{n11} is the number of periods with  breaches followed by a period with breaches.
\end{itemize}

As for the Proportion of Failure test, it can be shown that $LR_I\overset{\mathrm{asy}}{\sim}\chi^2(1))$. If the null is not rejected, it is assumed that breaches are independent of each other.

\subsection{Joint Test} \label{section_joint_test}
\cite{christoffersen1998evaluating} also proposed a test accounting for both  unconditional coverage and the independence hypotheses, which he describes as (mixed) conditional coverage test. In case both values for $LR_{POF}$ and $LR_{I}$ are known, the test statistic for the conditional coverage (CC) test can be calculated as 
\begin{equation}
    LR_{CC} = LR_{POF} + LR_{I} \text{~~~,}
\end{equation}
where $LR_{CC}\overset{\mathrm{asy}}{\sim}\chi^2(2)$. If the null holds for the joint test, both properties $E(I_t(\alpha)) = 1-\alpha$ and $I_{t}(\alpha) \overset{\mathrm{iid}}{\sim} Bern\left(1-\alpha\right)$ are assumed to be valid. Subsequently, a model can be considered as appropriate if the null is not rejected.

%------------------ NEURAL NETWORKS ------------------------------------
\section{Artificial Neural Networks} \label{chapter_Neural_Networks}
Inspired by human biology, artificial neural networks (ANNs or just NNs) can be understood as large parallel computing systems which are based on the architecture of the human brain \cite{gupta2013artificial}. ANNs are firstly introduced in a  paper from 1958 called \textit{The perceptron: a probabilistic model for information storage and organization in the brain}, which got published by the psychologist Frank Rosenblatt \cite{rosenblatt1958perceptron}. The paper was generally concerned with the question on how information from the physical world is sensed, remembered and ultimately processed into behavioral adjustments by the human brain. In short, Rosenblatts objective was to model the human mechanism of information processing and learning. The framework presented in 1958 shall later on act as a cornerstone for a whole field of research. 

\subsection{Neural Network Architecture} \label{Neural Network Architecture}
The most basic form of an artificial neural network is called a vanilla neural network. The vanilla architecture belongs to the group of feed-forward neural networks (FNNs), where information only travels forward, implying that the connections between nodes do not form a cycle (as in a recurrent neural network). Although the architecture used later on is far more complex than the vanilla architecture, FNNs are a good way to showcase the basic functionality of NNs. 
\par \vspace{0.3cm}
In its core, a vanilla NN is an extension of a linear regression. Considering an input vector $x$ of length $N$ with $x_1,...,x_N$, the equation for a linear regression can be written as
\begin{align*}
    \hat{y} = b + w_1x_1+w_2x_2+...+w_Nx_N \text{~~~,}
\end{align*}
where $w$ is a vector of weights. In the context of machine learning, the well known linear regression intercept term ($b$ in the equation above) is called bias term. The vanilla NN now adds another step or "layer" to this equation. This layer is placed between the input vector and the output and is called a "hidden layer". The hidden layer applies a weight vector $w^{(1)}$ and a bias term $b^{(1)}$ to the inputs and is further defined through the number of neurons or nodes in the layer and the used activation function (see equation \ref{output_hiddenlayer}). The input vector $x$ is now connected to each node of the hidden layer by the weight vector and the respective bias. Let $M$ be the number of nodes in the hidden layer, which can arbitrarily be chosen, and $a(\cdot)$ be the activation function which is the same for every node on a respective layer\footnote{the activation function is often denoted as $\sigma(\cdot)$ in the literature. However, in the context of this paper, such notation could cause unnecessary confusion.} (see section \ref{Activation_Functions_section}). The nodes can be denoted as $h_1, h_2, ..., h_M$. The output of the respective nodes can be defines as
\begin{align*}
    h_1 &= a(b_{1}^{(1)} + w_{1,1}^{(1)}x_1+w_{1,2}^{(1)}x_2+...+w_{1,N}^{(1)}x_N)\\
    h_2 &= a(b_{2}^{(1)} + w_{2,1}^{(1)}x_1+w_{2,2}^{(1)}x_2+...+w_{2,N}^{(1)}x_N)\\
    &...\\
    h_M &= a(b_{M}^{(1)} + w_{M,1}^{(1)}x_1+w_{M,2}^{(1)}x_2+...+w_{M,N}^{(1)}x_N).\\
\end{align*}
The output of each node is simply the activation function $a(\cdot)$ applied to the dot product of the inputs $x$ with the respective weights and bias. The hidden layer output can be re-written as 
\begin{equation} \label{output_hiddenlayer}
h_j = a\left(\sum^N_{i=1} w^{(1)}_{j,i}x_i + b_j^{(1)} \right) \\
\end{equation}
Following the same mechanism, the output of each hidden node is now connected to the output layer. Thus, the output layer applies another activation function $a^{(2)}$, weight vector $w^{(2)}$ and bias $b^{(2)}$. For an output layer with one node (usually the layout for point estimates), equation \ref{output_hiddenlayer} would be extended to result in the predicted value $\hat{y}$:
\begin{equation} \label{vanilla_NN_extended}
\begin{split}
        \hat{y} &= a^{(2)}\left(\sum^M_{j=1}w^{(2)}_{j} a^{(1)} \left( \sum^N_{i=1} w_{j,i}^{(1)}x_i +  b_{j}^{(1)} 
    \right) + b^{(2)}\right) \\
    &= a^{(2)}\left(\sum^M_{j=1}w^{(2)}_{j} h_j + b^{(2)}\right) \\
\end{split}
\end{equation}

\par \vspace{0.3cm} 

Some network architectures have an output layer which produces not one point estimate but a variety of different outputs. One example for this architecture is the mixture density network, which is presented in section \ref{section_MDN}. In such case, the output layer can consist of several different activation functions. Equation \ref{vanilla_NN_extended} can be re-written as 
\begin{equation} 
\begin{split} \label{multi_output_NN}
    \hat{y}_k &= a^{(2)}_k\left(\sum^M_{j=1}w^{(2)}_{j} a^{(1)} \left( \sum^N_{i=1} w^{(1)}_{j,i}x_i + b^{(1)}_{j}
    \right) + b^{(2)}_{k}\right) \\
    &= a^{(2)}_k\left(\sum^M_{j=1}w^{(2)}_{j} h_j + b^{(2)}_{k}\right) 
    \text{~~~,} \\
\end{split}
\end{equation}
where $\hat{y}_k$ denotes to the k-th output and $a_k^{(2)}$ is the corresponding activation function from the output layer.

\par \vspace{0.3cm} 
The general functionality of NNs shown above can be arbitrarily extended to a large number of hidden layers. A neural network consisting of more than one hidden layer is called a deep neural network (DNN). The field of AI developed a variety of different network structures to deal with different types of data, such as feed-forward neural networks,  convolutional neural networks which are used for the analysis of visual imagery, or the recurrent neural networks (RNNs) which are well suited for time series problems \cite{duan2022vanilla}. The latter will be explained later in section \ref{section_LSTM}.

\subsection{Parameter Optimization} \label{chapter_optimization}
As with any machine learning model, fitting a neural networks requires optimizing an objective function based on the training data. The typical procedure in machine learning is the minimization of a defined loss function. Generally, loss functions (denoted as $L$ in the following) take at least two arguments: the predicted target value ($\hat{y}$) and the observed target value from the training set ($y$). In this context, $\hat{y}$ can be thought of as the predicted values under the current model parameter combination $\theta$ (consisting of all weights and biases in the network), as equation \ref{vanilla_NN_extended} shows the dependence of $\hat{y}$ on those. One could write 
\begin{equation} \label{loss_func}
    J(\theta) = L(\hat{y}, y)
\end{equation}
with the aim to minimize $J$ by finding the optimal combination of weights and biases. There is no  best-practice when it comes to choosing an appropriate loss function to minimize, as different functions have certain advantages depending on the problem one tries to solve. However, the loss function needs to be convex and differentiable. For a convex and differentiable function to minimize, one seeks the combination of $\theta$ such that $\nabla J(\theta) = 0$ \cite{karlsson2021value}.  \par \vspace{0.3cm} 
The default algorithm to find such optimal parameter combination is the \textbf{gradient descent} (GD) algorithm. The starting point of the algorithm is an initial combination of weights and biases, denoted as $\theta_0$. These starting values can either be randomly chosen or set via a specified weight initialisation technique \cite{narkhede2022review}. GD employs an iterative approach to compute the next point by using the gradient at the present position, multiplies it by the learning rate $\gamma$ and then subtracts the obtained value from the current position. In machine learning, each iteration is called a step. This procedure is repeated until the algorithm found the minimum of the function. In general, the algorithm evaluates the next combination of weights and biases as 
\begin{equation} \label{gradient_descent}
    \theta_{t+1} = \theta_{t}  - \gamma \nabla J(\theta_{t}) \text{~~~, for } \gamma > 0 
\end{equation}
 
\begin{figure}[H]
    \centering
    \includegraphics[width=\textwidth, height = 6cm]{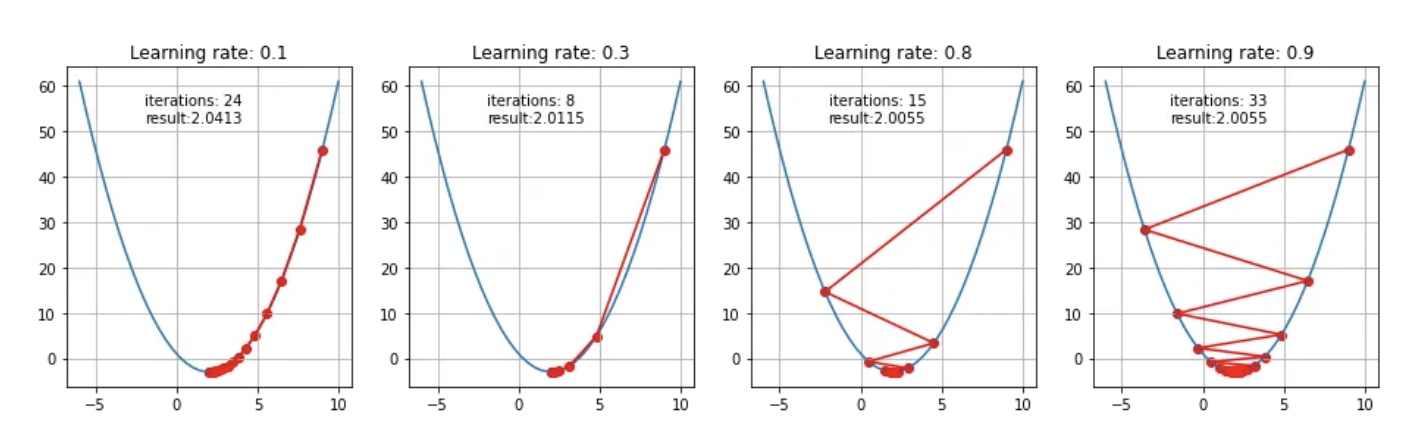}
    \caption{Comparison of learning rates. Source: \cite{kwiatkowskiR}}
    \label{Gradient_descent_picture}
\end{figure}

Equation \ref{gradient_descent} makes it obvious that the learning rate $\gamma$ scales and determines the size of each step. Under a small learning rate, the algorithm is more likely to find the minimum of the function. However, smaller steps lead to an increase in computational time as the algorithm takes longer to converge. In contrast, a large learning rate leads to a lower computational time however the algorithm might not converge at all as it misses to find the optimum. Figure \ref{Gradient_descent_picture} compares different learning rates and the number of iterations needed to find the minimum of a sample equation $x^2-4x+1$.
\par \vspace{0.3cm} 

The graphic above underlines the importance of the choice of $\gamma$ as a well chosen learning rate significantly reduces the number of iterations needed to find the optimum and therefore the computational time. In machine learning, the batch size determines how many observations are evaluated for each step, i.e., for computing the gradient. When all $N$  observations got passed through the NN for training, the model has performed one training epoch. The plain GD algorithm uses all observations per iteration, resulting in a batch size of $N$. The computational cost of calculating the gradient per iteration can therefore be expressed as $O(kN)$, where $N$ is the number of observations in the data and $k$ indicates the number of dimensions \cite{zadeh2015cme}. Subsequently, the algorithm is highly computational expensive for high-dimensional data sets with a large number of observations. The \textbf{stochastic gradient descent} (SGD) algorithm is a prominent alternative to the plain gradient descent, which randomly samples one observations from the data and calculates the gradient solely based on the sample rather than taking the whole $N$ observations for the computation. This reduces the batch size to 1 and computational expense per iteration to $O(k)$. However, as the algorithm uses an approximation of the gradient, the convergence path of the SGD is generally noisier than that of the plain GD. The \textbf{mini-batch gradient descent} is a hybrid of both algorithms presented above, where the batch size can be any value between 1 and $N$.
\par \vspace{0.3cm} 
The adaptive moment estimation optimizer, short \textbf{Adam Optimizer}, approaches the computational complexity problem with the usage of adjustable learning rates based on moving averages of the first and second moments of the gradient. The paper on the Adam algorithm by Diederik P. Kingma and Jimmy Lei Ba, which was presented  at the ICLR\footnote{The International Conference on Learning Representations (ICLR) is one of the most prominent and well regarded yearly conferences on artificial intelligence.} 2015, promised a better performance in terms of computational cost compared to the currently most advanced optimization algorithms such as RMSProp or the SGD algorithm \cite{kingma2014adam}. In a first step, the algorithm computes an estimation of the gradient under the current model parameters ($\nabla J(\theta_{t})$) based on a small sample of observations from the data. Given its computation using just a small batch of the data, $\nabla J(\theta_{t})$ can be regarded as a random variable. The rolling averages of the first and second moment functions of the gradient  are then calculated as 
\begin{equation} \label{ADAM_momentfunctions}
    \begin{split}
        m_{t+1} &= \beta_1 m_{t} + (1-\beta_1) \nabla J(\theta_{t}) \\
        v_{t+1} &= \beta_2 v_{t} + (1-\beta_2) \nabla J(\theta_{t})^2 \text{~~~,}
    \end{split}
\end{equation}
where the scaling parameters $\beta_1$ and $\beta_2$ are typically set to 0.9 and 0.999, respectively, as proposed in the original paper. However, these estimates are biased towards zero as the initial values $m_0$ and $v_0$ are both zero. The algorithm therefore calculates the adjusted estimators as
\begin{equation}
    \begin{split}
        \hat{m}_{t+1} &= \frac{m_{t+1}}{1-\beta_1} \\
        \hat{v}_{t+1} &= \frac{v_{t+1}}{1-\beta_2} 
    \end{split}
\end{equation}
The optimization step can then be calculated:
\begin{equation}
\theta_{t+1}  = \theta_{t} - \gamma \frac{\hat{m}_{t+1}}{\sqrt{\hat{v}_{t+1}}+\epsilon} \text{~~~,}\\
\end{equation}
where $\epsilon$ is a small number to prevent division by zero\footnote{Kingma and Ba [2014] proposed $\epsilon = 10^{-8}$, which is also the common practice for the algorithm in most applications.}. The learning rate $\gamma$, which can be altered manually, is proposed to be set to 0.001 by the original paper \cite{kingma2014adam}. From equation \ref{ADAM_momentfunctions} it gets obvious that the step size decays as the average of the estimated gradient (i.e., its first and second moment) gets smaller. Intuitively, one would expect that the step size is larger for the first iterations and then declines over time. 
\par \vspace{0.3cm}
These optimization algorithms can also be applied to quasi-convex loss functions, which is the case for neural networks with more than one layer \cite{choromanska2015loss}. The non-convexity of a loss function $L(\hat{y}, y)$ introduces the problem of local minima, which can potentially prevent the algorithm from finding the function's true minimum. 
\subsection{Activation Functions} \label{Activation_Functions_section}
Activation functions are specifically employed in artificial neural networks to convert an incoming signal into an outgoing signal. Subsequently, this output signal is utilized as input for the subsequent layer in the sequence \cite{sharma2017activation}. One of the main reasons for their usage is the introduction of non-linearity into the network, which enables it to map non-linear patterns. Another important feature is the input-output mapping between layers, limiting the output of a node to a certain numerical range. Functions which are performing output limitation are also called "squashing functions" \cite{brownleeJ}. The following presents the relevant activation functions which are referenced on in later sections. 
\par \vspace{0.3cm} 
The \textbf{logistic or sigmoid} activation function is defined as 
\begin{equation} \label{sigmoid_function}
    \phi(x) = \frac{1}{1+e^{-x}}\\
\end{equation}

The sigmoid function maps the outputs between 0 and 1 and is therefore  particularly used for classification problems where the output generally represents a probability.
\begin{figure}[H]
\begin{center}
\begin{tikzpicture}
\begin{axis}[xmin = -5, xmax = 5, ymin = -0.1, ymax = 1.1, axis lines = middle, xlabel = $x$, ylabel = $y$, title = {\textbf{Sigmoid function}}]
\addplot[mark = none, red, domain = -5:5]{1/(1+exp(-\x))};
\end{axis}
\end{tikzpicture}
\end{center}
\caption{Sigmoid function}
\end{figure}
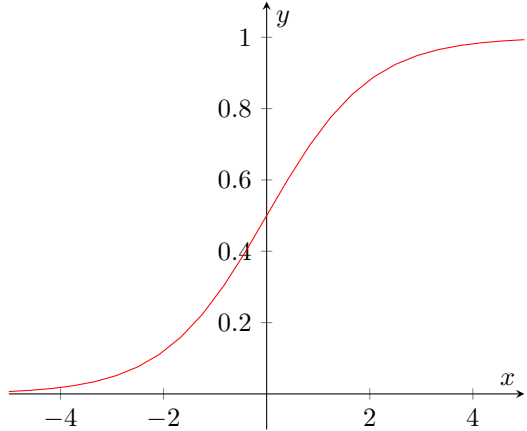
\par \vspace{0.3cm} 
Instead of a limitation between 0 and 1, the hyperbolic tangent activation function, short \textbf{tanh},  maps the function output between [-1, 1] and can subsequently map a negative input to a negative output. 
\begin{equation} \label{tanh}
tanh(x) = \frac{e^x - e^{-x}}{e^x + e^{-x}} \\  
\end{equation}

While both being sigmoidal functions, the hyperbolic tangent function is expected to perform better than the sigmoid function as it possesses  properties which are appealing for its use while model training \cite{kalman1992tanh}.
\begin{figure}[H]
\begin{center}
\begin{tikzpicture}
\begin{axis}[xmin = -5, xmax = 5, ymin = -1.2, ymax = 1.2, axis lines = middle, xlabel = $x$, ylabel = $y$, title = {\textbf{hyperbolic tangent function}}]
\addplot[mark = none, red, domain = -5:5]{(exp(\x) - exp(-\x))/(exp(\x) + exp(-\x))};
\end{axis}
\end{tikzpicture}
\end{center}
\caption{tanh function}
\end{figure}
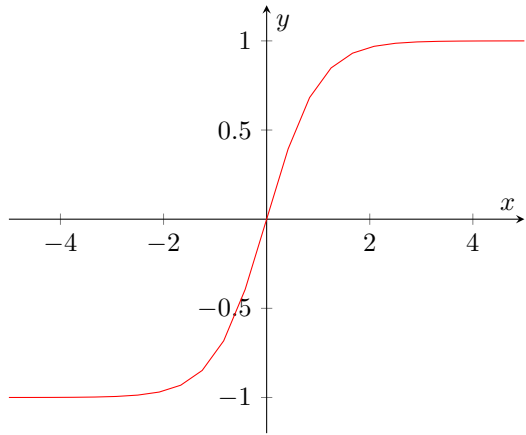
\par \vspace{0.3cm}

\par \vspace{0.3cm}
The \textbf{Softmax} activation function can be understood as a combination of several sigmoid functions, which returns a vector of probabilities instead of a single value \cite{sharma2017activation}. It can therefore be used for multi-class classification problems.  
\begin{equation} \label{Softmax}
    Smax(x)_i = \frac{e^{x_i}}{\sum^K_{j=1}e^{x_j}} \text{~~~, for } i = 1,...,K
\end{equation}

Equation \ref{Softmax} shows that the function produces $K$ outputs which add up to 1, representing the probabilities of each outcome \cite{kagalkar2020cordic}. 
\par \vspace{0.3cm}

The Rectified Linear Unit (\textbf{ReLU}) activation function is a piecewise linear function which takes either 0 or $x$ as value. The function enjoys high popularity within ML research and  is the default for many types of neural networks \cite{brownleeJ_ReLU}. The function is defined as 
\begin{equation}
    R(x) = max(0, x)
\end{equation}

\begin{figure}[H]
\begin{center}
\begin{tikzpicture}
    \begin{axis}[xmin = -5, xmax = 5, ymin = -1, ymax = 5, axis lines = middle, xlabel = $x$, ylabel = $y$, title = {\textbf{ReLU function}}]
        \addplot+[mark=none,red,domain=-5:0] {0};
        \addplot+[mark=none,red,domain=0:5] {x};
    \end{axis}
\end{tikzpicture}
\end{center}
\caption{ReLU function}
\end{figure}
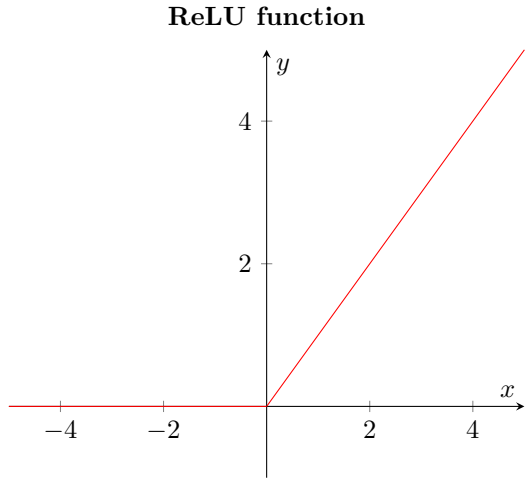

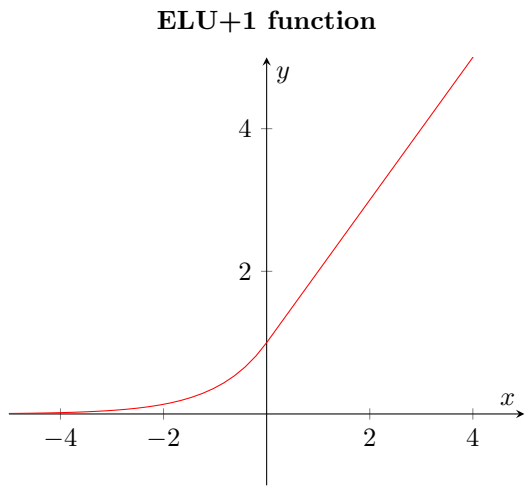
\begin{figure}[H] 
\begin{center} 
\begin{tikzpicture}
    \begin{axis}[xmin = -5, xmax = 5, ymin = -1, ymax = 5, axis lines = middle, xlabel = $x$, ylabel = $y$, title = {\textbf{ELU+1 function}}]
        \addplot+[mark=none,red,domain=-5:0] {exp(\x)};
        \addplot+[mark=none,red,domain=0:5] {x+1};
    \end{axis}
\end{tikzpicture}
\end{center}
\caption{ELU+1 function}
\label{ELU_plus1_graph}
\end{figure}

The Exponential Linear Unit (\textbf{ELU}) activation function was proposed as an improved version of ReLU \cite{clevert2015fast}.
Several studies showed superior accuracy in classification problems compared to ReLU \cite{trottier2017parametric}. The function is defined as 
\begin{equation} \label{ELU}
\begin{split} 
    ELU(x) &= max(0, x) + min(0, e^{x}-1) \\
    &= \begin{cases}
          x &\text{ if } x > 0 \\
          e^{x}-1 &\text{ if } x \le 0 \\
    \end{cases}
\end{split}
\end{equation}
As $e^x$ approaches 0 for very small inputs, the output of ELU is in the range $[-1, \infty]$. Thus, a shifted ELU function, \textbf{ELU + 1}, gives the useful property of mapping outputs to solely non-negative numbers. Figure \ref{ELU_plus1_graph} shows such function.

%vanishing gradient problem
The ReLU or ELU functions are often used to counter the problem of vanishing gradients. As seen in the previous chapter, most optimization algorithms are gradient-based and therefore require to compute first-order derivatives. This vanishing gradient problem is especially common when using sigmoidal activation functions , as they limit inputs of the range $[-\infty, \infty]$ to a small output range. Thus, their first-order derivatives of sigmoidal functions are likely to result in small absolute values, which makes it difficult for the optimization algorithm to update the weight parameters. 
\begin{equation}
    \phi'(x) = \phi(x) \left(1-\phi(x)\right)
\end{equation}
Alternative functions such as ReLU can be used to prevent the vanishing gradient problem as they do not necessarily result in small derivatives. 
\begin{equation}
    R'(x) = \begin{cases}
        0 & \text{ for } x < 0\\
        1 & \text{ for } x \ge 0
    \end{cases}
\end{equation}
However, the usage of ReLU introduces the dying ReLU problem, which refers to the issue of ReLU-activated neurons solely outputting zeros and thus becoming inactive \cite{lu2019dying}. The choice of an activation function therefore depends on the problem one tries to solve or the properties one requires for the output of interest. The properties of Softmax and ELU+1 are especially relevant when using MDNs as they can be used for producing the $\pi$ and $\sigma$ parameters, respectively.

\subsection{Mixture Density Networks} \label{section_MDN}
The idea behind mixture density networks was firstly presented in an eponymous paper by Christopher M. Bishop in 1994 \cite{bishop1994mixture}. The concept of the paper was to combine conventional neural networks with an arbitrary chosen continuous conditional probability distribution. Instead of a conditional mean as output (as it is the case for conventional NNs), the MDN returns parameters for modeling a probability distribution which is conditional on the input vector. Their strength should lie in modeling events coming from different stochastic processes
as well as modeling scenarios governed by different rules \cite{ellefsen2019mixture}. Thus, MDNs could be a promising method to model the non-constant volatility of daily returns.
\par \vspace{0.3cm}
This paper focuses on the implementation of MDNs which are based on a mixture Gaussian distribution, which is composed of several Gaussian distributions. It can be understood as a weighted sum of Gaussian densities \cite{reynolds2009gaussian}. To fully understand the proposed MDN, some notation for Gaussian distributions and mixture models in general is required. A mixture density of $K$ components can be defined as 
\begin{equation} \label{mixture_model}
    P(y|x) = \sum^K_{k=1} \pi_{k}(x)D\left(y|\psi_{1,k}(x), \psi_{2,k}(x),..., \psi_{N,k}(x) \right) \text{~~~,}
\end{equation}
where $\pi_k$ is the mixing parameter of the k-th component, $D$ is the defined continuous probability density function and $\psi_{1,k}(x),...\psi_{N,k}(x)$ is a vector of length $N$  containing distribution parameters from the k-th component conditional on the input. For a Gaussian mixture, each component has two parameters, $\mu(x)$ and $\sigma(x)$. Recall the Gaussian density function
\begin{equation} \label{gaussian_density}
    \mathcal{N}(y|\mu, \sigma^2) = \frac{1}{\sigma\sqrt{2\pi}}\exp{{\frac{-(y-\mu)^2}{2\sigma^2}}}
\end{equation}
The k-th component of a Gaussian mixture can be expresses as $\mathcal{N}\left(\mu_k(x), \sigma^2_k(x)\right)$. Plugging in equation \ref{gaussian_density} in \ref{mixture_model}, the Gaussian mixture can be written as 
\begin{equation} \label{Gaussian_mixture}
    P(y|x) = \sum^K_{k=1} \pi_{k}(x) \mathcal{N}(\mu_k(x), \sigma^2_k(x))
\end{equation}
The  network as proposed by Bishop shall now produce all component parameters for each input, which allows for modeling from the mixture distribution in equation \ref{Gaussian_mixture}. As outputs are generated in the output layer, the concept of mixture density network is to implement a layer into a deep neural network which produces those distribution parameters. As shown in equation \ref{multi_output_NN}, an output layer can include nodes which use different activation functions for more than one output. In case of a Gaussian mixture-based MDN, the output layer needs to include three different nodes for the computation of the $\pi$, $\mu$ and $\sigma$ component, respectively. For a K-component model, the MDN computes a parameter vector of length $K$ for each prediction.
\par \vspace{0.3cm}
Recall equation \ref{multi_output_NN}, which describes an output layer with different activation functions:
\begin{equation*}
   \hat{y}_k = a_k\left(\sum^M_{j=1}w_{j} h_j + b_{k}\right) \text{~~~,} \\ 
\end{equation*}
where $h_j$ is the output from the j-th node in the hidden layer which is placed before the output layer (see equation \ref{output_hiddenlayer}). 
The \textbf{mixture parameter $\pi$} can be interpreted as probabilities representing the contribution of each individual component to the mixture distribution. The parameter is constrained such that 
\begin{equation}
    \sum^K_{k=1}\pi_k(x) = 1 
\end{equation}
As seen in the previous chapter, the Softmax function provides appropriate properties for such constrain as it produces probabilites which sum up to 1. Thus, the Softmax function is used as activation function for the mixture component vector.
\begin{equation}
    \hat{y}_{\pi} = Smax\left(\sum^M_{j=1}w_{j} h_j + b_{\pi}\right)\text{~~~,} \\ 
\end{equation}
where $\hat{y}_{\pi}$ is a vector including all the mixture parameters for the $K$ components ($\hat{y}_{\pi} = [{\pi}_1, {\pi}_2, ..., {\pi}_K]$). Let $z_k^{\pi}$ be the $k$-th element of the input vector of the Softmax function, corresponding to the $k$-th component of the mixture model. The mixture components are then calculated as
\begin{equation}
    {\pi}_k = \frac{\exp({z^{\pi}_k})}{\sum^K_{j=1} \exp{(z^{\pi}_j)}}
\end{equation}\par \vspace{0.3cm}
The \textbf{variance parameter $\sigma$} is constrained to be non-negative. In the original paper, Bishop [1994] proposed the usage of an exponential function: $\sigma_k = \exp{(z^{\sigma}_k)}$. Axel Brando Guillaumes introduced the idea of using the $ELU+1$ activation function instead for the variance parameter, as the exponential function above obviously leads to an exponential growth of $\sigma_k$ for comparably large input values \cite{brando2017mixture}. This method was also successfully implemented in comparable research \cite{Georgios_MDN, karlsson2021value}. The MDN which is used for VaR calculations later on therefore uses $ELU+1$ as its activation function for the variance parameters.
\begin{equation}
    \hat{y}_{\sigma} = ELU\left(\sum^M_{j=1}w_{j} h_j + b_{\sigma}\right)+1 \\
\end{equation}
\begin{equation}
    \sigma_k = \biggl(max(0, z^{\sigma}_k) + min(0, e^{z^{\sigma}_k}-1)\biggr) + 1 \text{~~~,} \\
\end{equation}
where $\hat{y}_{\sigma} = [\sigma_1,..., \sigma_K]$. The \textbf{location parameter $\mu$} does not require a transformation in the output layer and thus can be written as
\begin{equation}
    \hat{y}_{\mu} = \sum^M_{j=1}w_{j} h_j + b_{\mu} \\
\end{equation}
\begin{equation}
    \mu_k = z^{\mu}_k \text{~~~,} \\
\end{equation}
\par \vspace{0.3cm}
where $\hat{y}_{\mu} = [\mu_1,..., \mu_K]$. The parameter optimization procedure for MDNs requires the definition of a loss function which takes the vector of predicted values (or in this case predicted distribution parameters) $\hat{y}$ as an argument. Subsequently, the objective function for model training needs to include the parameter vectors $\hat{y}_{\pi}, \hat{y}_{\mu}$ and $\hat{y}_{\sigma}$. Hence, the likelihood function of the mixture model is used for parameter optimization as it obviously contains the distribution parameters. Equation \ref{neg_log_lik_MDN} shows the loss function for training a MDN. As the gradient-based optimization algorithm seeks to minimize a function, the negative log-likelihood of the underlying mixture model is used. The loss function for the MDN can be written as 
\begin{equation} \label{neg_log_lik_MDN}
     J(\theta) =  - \ln{\sum^K_{k=1} \pi_{k}(x,\theta)\mathcal{N}(\mu_k(x,\theta), \sigma^2_k(x,\theta))} \text{~~~,}
\end{equation}
where $\theta$ represents a vector of model weights and biases which have to be optimized.

\subsection{Recurrent Neural Networks \& Long Short-Term Memory} \label{section_LSTM}
In contrast to the plain feed-forward architecture presented in section \ref{Neural Network Architecture}, a recurrent neural network (RNN) is designed to process a (timely) sequence of input values \cite{kim2016chapter}. They are often used for natural language processing (NLP) problems or modern translators which require the interpretation of a sequence of words. Thus RNNs are suited for problems with time-dependent data. The biggest difference between RNNs and feed-forward NNs lies in the feedback connections of the RNN, which is often called \textbf{feedback loop} \cite{fausett1994fundamentals}. Intuitively, the feedback loop can be thought of as a dependence of the network output on prior outputs. 
\par
\begin{figure}[H]
    \centering
    \includegraphics[width=10cm, height = 8cm]{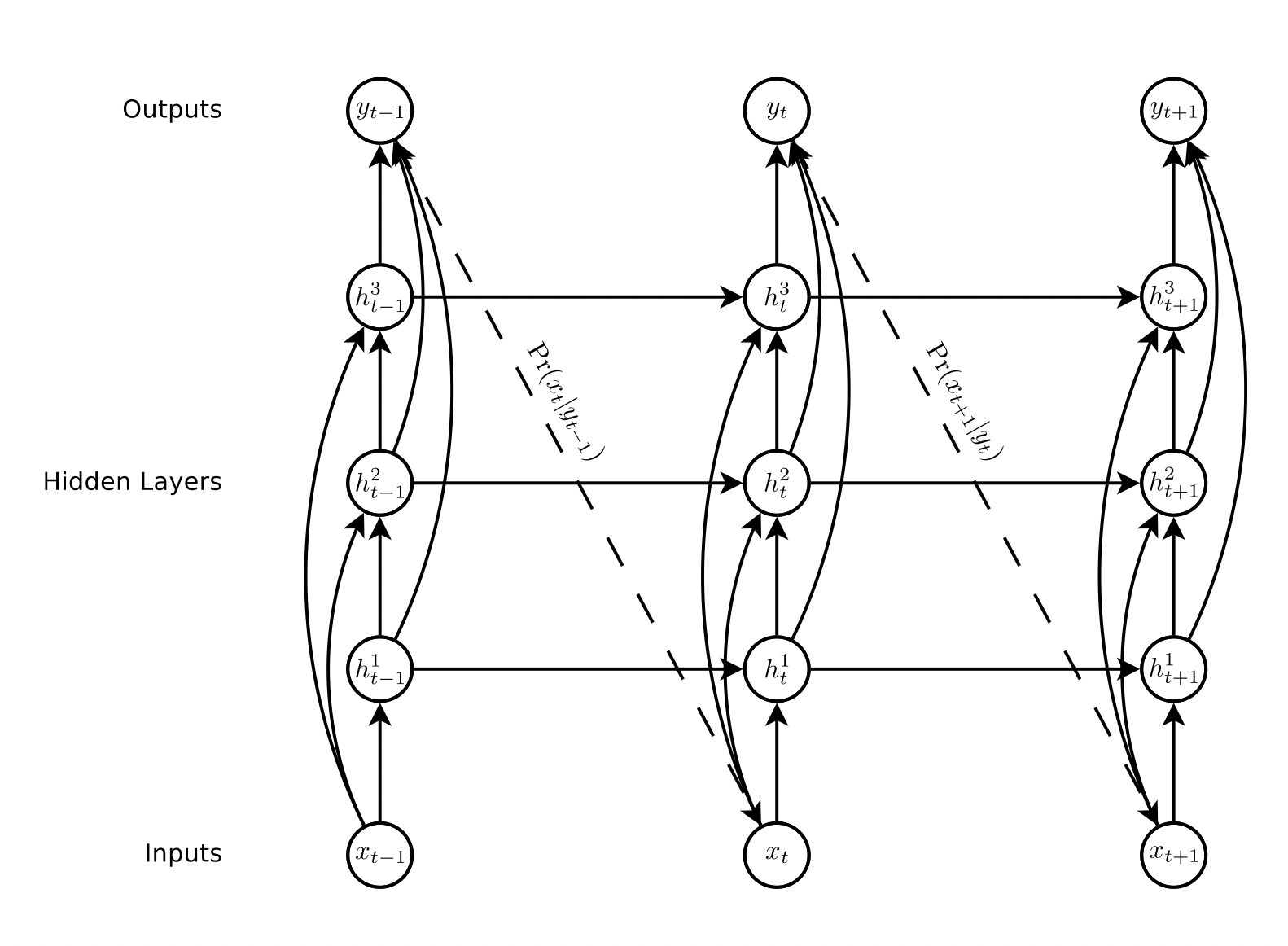}
    \caption{RNN architecture - Source: \cite{graves2013generating}}
     {\small The superscript of $h$ indicates the position in sequence of hidden layers. The dashed lines showcase the feedback loop, i.e., the dependency of the input at $t$ on the output from $t-1$}
    \label{RNN_fig}
\end{figure}

 \vspace{0.3cm}
The feedback loop of an RNN can be shown through simplified notation. The outcome of a hidden node in a FNN at time point $t$ could be expressed as $h_t = a(x_{t}w^{(x)}+b)$, where $w^{(x)}$ is a weight vector applied to input $x_t$ and $b$ represents the corresponding bias vector. The node in a recurrent hidden layer additionally accounts for the state of the hidden node at time point $t-1$. This logic can be expresses in simplified notation as
\begin{equation} \label{RNN_equation}
\begin{split}
    h_t &= a(x_{t}, w^{(x)}, h_{t-1}, w^{(h)}, b) \\
        &= a(x_{t} w^{(x)} + h_{t-1}, w^{(h)}+ b) \text{~~~,}\\
\end{split}   
\end{equation}
where $h_{t-1}$ is the hidden state at time point $t-1$ and $w^{(h)}$ is a weight vector applied to this state.
Equation \ref{RNN_equation} implies that the RNN is capable of detecting timely dependencies as it accounts for previous outcomes. Figure \ref{RNN_fig} graphically shows this logic with an example of a deep neural network with three recurrent hidden layers.

However, due to this feedback loop, the computation of outputs are based on a large number of transformations done by activations functions. This makes RNNs particularly prone to the vanishing gradient problem and generally hard to train, as explained in section \ref{Activation_Functions_section}. 
\par \vspace{0.3cm}
A \textbf{Long Short-Term Memory neural network (LSTM-NN)} is a specific form of recurrent neural network. Proposed in 1997 by Sepp Hochreiter and Jürgen Schmidhuber \cite{schmidhuber1997long}, the LSTM architecture aims to deal with the vanishing gradient problem. The main idea behind the method is the usage of not one but two dedicated paths for the feedback loop. The long-memory path is dedicated to time stamps which lie further in the past, while the short-term path accounts for time stamps which are closer to the current state. Through this architecture, the LSTM-NN has the capability to better model long-term dependencies compared to the plain vanilla RNN architecture presented above. 
\par \vspace{0.3cm}
\begin{figure}[H]
    \centering
    \includegraphics[width=10cm, height = 6cm]{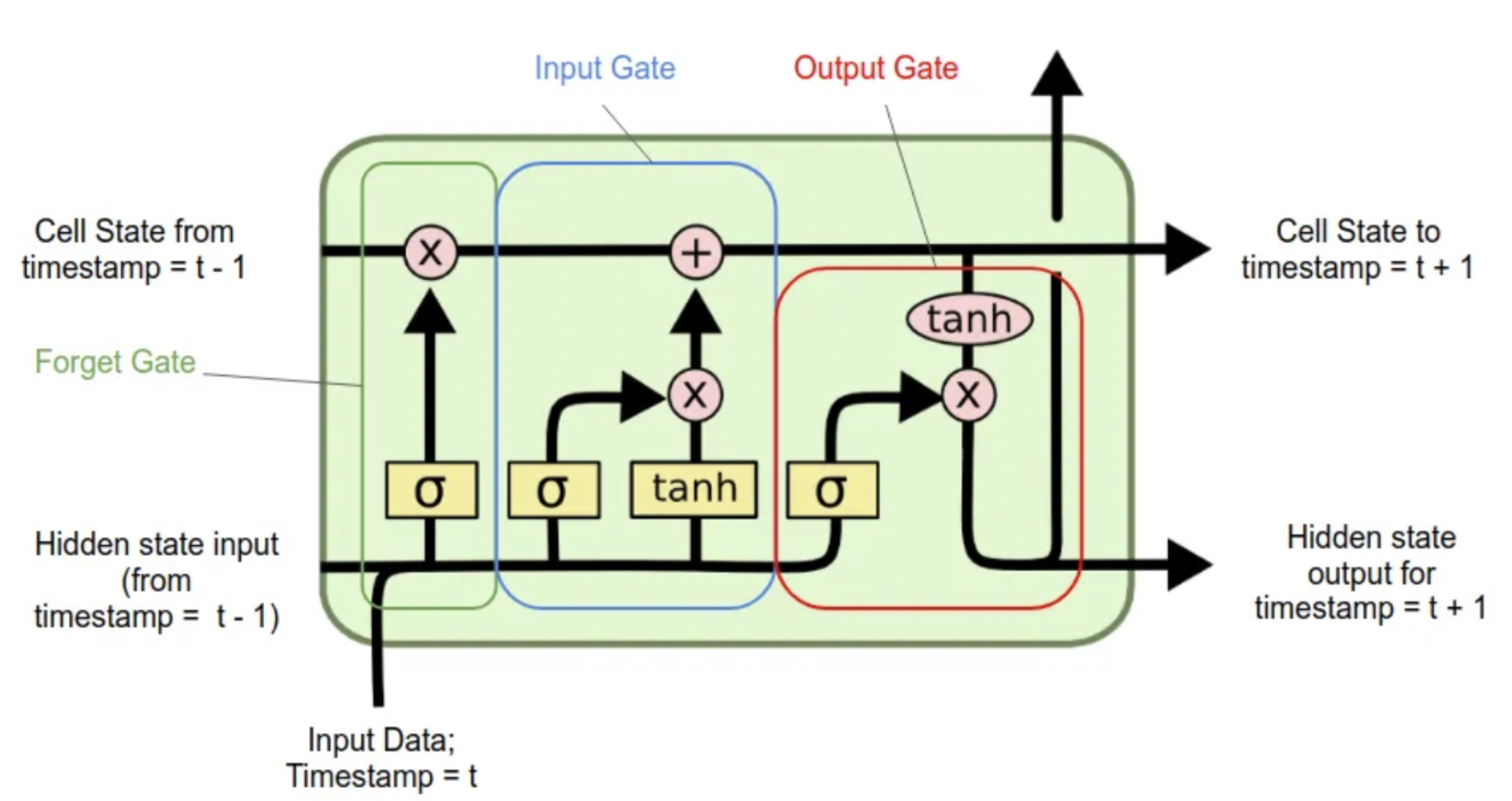}
    \caption{LSTM-cell architecture - source: \cite{Ryan_LSTM}}
    {\small In the context of this graphic, $\sigma$ denotes the sigmoid activation function}
    \label{LSTM_cell_fig}
\end{figure}
The fundamental component of the LSTM architecture is the so-called \textbf{cell state} (denoted $C$ in the following). Christopher Olah\footnote{Christopher Olah is the co-founder of Anthropic, an AI lab focused on the safety of large models.} describes the cell state as \textit{"... kind of like a conveyor belt. It runs straight down the entire chain, with only some minor linear interactions. It’s very easy for information to just flow along it unchanged."} \cite{Olah_2015}. The LSTM cell has the ability to alter the cell state by removing or adding information through the computational process. This regulation is done by the three gate components of the LSTM cell: The \textbf{forget gate},  the \textbf{input gate} and the \textbf{output gate}. 
\par \vspace{0.3cm}
The \textbf{forget gate} consists of a transformation using the sigmoid function. As shown in section \ref{Activation_Functions_section}, the sigmoid function maps its inputs to an output range of [0,1]. Its application in the forget gate can thus be interpreted as the amount of information which shall be removed or "forgotten". The function takes the input vector $x_t$ and the prior hidden state $h_{t-1}$. This operation can be expressed as 
\begin{equation}
    f_t = \phi(w^{(f)}[x_t, h_{t-1}]+b^{(f)}) \text{~~~,}
\end{equation}
where $w^{(f)}$ is a corresponding weight matrix and $b^{(f)}$ a corresponding bias vector. 
\par \vspace{0.3cm}
The \textbf{input gate} now regulates which new information shall be added to the cell state. The gate consists of two separate transformation using a sigmoid function and a hyperbolic tangent function, respectively. The sigmoid transformation controls which values shall be updated, using the same principle as the forget gate by generating values which lie between 0 and 1.
\begin{equation}
    i_t = \phi(w^{(i)}[x_t, h_{t-1}]+b^{(i)})
\end{equation}
The tanh transformation is used to create a vector of candidate values for addition to the cell state. 
\begin{equation}
    \widetilde{C}_t = tanh(w^{(c)}[x_t, h_{t-1}]+b^{(c)})
\end{equation}
The candidate vector $\widetilde{C}_t$ is now point-wise multiplied with $i_t$, controlling the amplitude of the new values added to the cell state. After the computation of the forget gate and the input gate, the updated cell state $C_t$  can then be calculated as
\begin{equation} \label{cell_state_computation}
    C_t = f_t  C_{t-1} +  i_t  \widetilde{C}_t 
\end{equation}
The addition sequence expressed in equation \ref{cell_state_computation} shows that the new cell state is composed of point-wise multiplication of the prior cell state with a vector regulating the proportion of information removal (the left-sided product) and a point-wise multiplication of candidate information to be added to the new cell state and another regulation vector (the right-side product).
\par \vspace{0.3cm}
The new hidden state of the LSTM cell is computed in the \textbf{output gate}, which too includes a sigmoid transformation and a hyperbolic tangent transformation. The sigmoid transformation again determines which values are contained in the output.
\begin{equation}
    o_t = \phi(w^{(o)}[x_t, h_{t-1}]+b^{(o)})
\end{equation}
The new hidden state $h_t$ is now calculated by multiplying $o_t$ with the current cell state $C_t$ which gets squashed to the range [-1, 1] by the hyperbolic tangent transformation.
\begin{equation}
    h_t = tanh(C_t) o_t
\end{equation}
\par \vspace{0.3cm}
As shown above, the LSTM architecture allows a NN to model time dependent processes by accounting for the hidden state at $t-1$. Through this mechanism the model is expected to be able to learn time dependent patterns such as volatility clustering, which motivates its usage for financial time series forecasting.

%Methodology
\chapter{Methodology}
This chapter deals with the practical implementation of the models and methods presented above as well as the process of data collection and data processing. Further, it shall provide an understanding of the workflow used for project implementation. The main software implementation was written in \textit{R} \cite{R_language} and \textit{Python} \cite{10.5555/1593511}. Neural networks are based on the \textit{TensorFlow} \cite{tensorflow2015} framework using the \textit{Keras} API \cite{chollet2015keras} for scripting. Further details can be found in the respective sections.

\section{Data Processing and Observation Periods} \label{section_data_processing}
Daily stock price data is sourced from \textit{Yahoo Finance}. The process of data sourcing was automated by using the \textit{get.hist.quote} function from the \textit{tseries} package in R \cite{Trapletti2020-wm}.  The analysis is concentrated on three different stock indices, which represent different geographical regions to avoid any potential geographical dependencies.
\par \vspace{0.3cm}
The \textbf{FTSE 100} index is composed of the 100 largest  companies listed on the London Stock Exchange and is the most important stock index for the the UK economy. For the analysis, a time series of daily closing prices is sourced, spanning from 02.01.2001 to 09.05.2023, resulting in 5686 daily observations. Representing the US economy, the \textbf{S\&P 500} includes the 500 largest companies by market capitalization which are listed on the US stock market. The time horizon used is again 02.01.2001 to 09.05.2023, which provides 5623 daily prices\footnote{The difference in the amount of daily observations for time series of the same time horizon is due to the varying number of trading days per year on different stock markets}. The \textbf{EURO STOXX 50} index is made up of 50 large companies in the Euro-currency area which are considered as the region’s supersectoral leaders \cite{morea2022circular}. Due to a limitation on data availability, the sourced daily prices range over a time horizon from 30.03.2007 to 09.05.2023, including 4053 sequential observations.

\begin{table}[h!]
\centering
\begin{tabular}{||c c c c||} 
\hline
\textbf{Index} & \textbf{Series Start} & \textbf{Series End} & \textbf{Trading Days} \\ [0.5ex] 
\hline\hline
FTSE 100 & 02.01.2001 & 09.05.2023 & 5686 \\ 
\hline
S\&P 500 & 02.01.2001 & 09.05.2023 & 5623 \\
\hline
EURO STOXX 50 & 30.03.2007 & 09.05.2023 & 4053 \\ [1ex] 
\hline
\end{tabular}\par
\caption{Data Sets}
\label{table_Observation_periods}
\end{table}
\par \vspace{0.3cm}
After downloading the closing prices for each index over the time frames shown in table \ref{table_Observation_periods}, each time series was checked for missing values, where no data sequence contained more than 3\% of NA values. The missing closing prices are imputed by linear interpolation. A missing value at time point $t$ is approximated as
\begin{equation}
    \tilde{x}_t = \frac{x_{t-1} + x_{t+1}}{2}
\end{equation}
Next, the discrete daily returns are calculated based on equation \ref{eq_returns}. As daily returns are expected to be centered around zero, a dedicated step for centering the data was not required.
\par \vspace{0.3cm}

\section{Implementation of Benchmark Models}
The technical implementation of Value-at-Risk forecasting for all benchmark models was done in \textit{R}. VaR forecasts are based on a vector of past daily returns of length $d$. Thus, the one day-ahead forecast for $t+1$ is based on the daily discrete returns $R_t, R_{t-1},...,R_{t-d+1}$. As the Basel Accords require a minimum observation period of one year, the rolling window is set to $d = 250$. The one day-ahead forecasts $VaR[h=1, \alpha=0.99]$ which are evaluated through backtesting are thus based on a vector of 250 ex-post returns [$R_t, R_{t-1},...,R_{t-249}$].Per definition, Value-at-Risk quantifies the potential loss of a financial asset, where the asset value (typically a portfolio) is denoted as $P$. For the purpose of the following analysis, VaR is reported as "scale-free" daily loss (i.e., a negative discrete return), assuming $P = 1$.
\par \vspace{0.3cm}
As desciribed in section \ref{historical_simulations}, the Value-at-Risk forecast under the \textbf{historical simulation} method is based on the empirical $\alpha$-quantile of the loss distribution. Subsequently,  $VaR_{HS}[h=1, \alpha=0.99]$ for time point $t+1$ is the negative empirical $1-\alpha$ quantile of the prior 250 ex-post daily returns. Simultaneously, the parameter estimates for the normal distribution (recall equation \ref{equation_CMM_mean} and \ref{equation_CMM_sigma}) under the \textbf{Constant Mean Model} are calculated from $[R_t, R_{t-1},...,R_{t-249}]$. Using the resulting $\hat{\mu}$ and $\hat{\sigma}$, the VaR forecast is then computed for the following day.
\par \vspace{0.3cm}
\textbf{GARCH(1,1)} models are fitted using the \textit{fGarch} package \cite{fGarch_pkg}. In a first step, two GARCH(1,1) models are fitted per stock index and period for the start date of the respective observation period (assuming the start date is $t+1$), where one model uses normally distributed innovations while the other uses GED distributed innovations. The underlying innovation distribution is then chosen based on a comparison of AIC scores. All 6 comparisons indicated that GED distributed innovations are preferable for modeling the respective daily returns, which underlines the assumption of daily returns following a leptokurtic distribution. GARCH(1,1) models with GED distributed innovations are then fit on a vector of returns $[R_t, R_{t-1},...,R_{t-249}]$. Recalling equation \ref{GARCH_VAR}, the parameters of the resulting GARCH process are then used to produce the VaR forecast under the GARCH(1,1) model for $t+1$.

\section{Implementation of the LSTM-Mixture Density Network} \label{section_LSTM_implementation}
The \textbf{LSTM-MDN} architecture was developed in \textit{Python 3} using the \textit{Keras} package. \textit{Keras} is an open source deep-learning library which works as API for several backend architectures. In case of this project, \textit{Keras} works as API for \textit{TensorFlow}, a framework developed by Google, allowing the implementation of various types of models \cite{tensorflow2015}. Further, the \textit{keras-MDN-layer} package was used for the implementation of the mixture density layer \cite{mdn_layer}. 

\subsection{Data Transformation} \label{Data Transformation}
As explained in chapter \ref{chapter_Neural_Networks}, fitting a NN requires training it on a train set. First, each data set is split into a pre-evaluation set and a test set. The test set contains all observations which fall into the respective evaluation period (i.e., the years 2017/ 2018 and 2021/ 2022). The pre-evaluation set is further divided into a train set and a validation set. The train set contains the first 90\% of observations from the pre-evaluation data, which are used for model training. The last 10\% of the pre-evaluation data is used as validation set. The validation set serves as data the fitted model is evaluated on to determine the optimal hyperparameters and model architecture. \par \vspace{0.3cm}
Another important consideration is the number of prior daily returns which shall be used as input for the one day-ahead forecast. Let $d$ denote the number of ex-post daily returns which are included in the input vector $x_t = [R_{t-d+1}, R_{t-d+2},...,R_{t}]$ to produce the forecast $\hat{y}_{t+1}$. Several values for $d$ have been tried, ranging from $d = 100$ to $d = 3$, where both the model performance and the computational expenses were taken into consideration. On average, a vector consisting of the 10 prior daily returns ($d = 10$) showed the best performance while also providing comparably low computational expenses\footnote{for comparison: Karlsson Lille and Saphir[2021] used a rolling window of $d = 20$ while Arimond et al.[2020] used $d = 10$.}. The rolling window of 10 prior observations can be interpreted as the daily returns from the last two weeks. However, recall that due to the LSTM architecture enabling the model to account for long-term memory by taking into consideration the prior hidden state $h_{t-1}$, the model shall be able to learn patterns which depend on returns exceeding the lookback period of $d$ (see section \ref{section_LSTM}).

\subsection{Model Architecture} \label{section_model_architecture}
Designing a (deep) neural network generally comes with the question of choosing optimal hyperparameters. In case of a LSTM-MDN, the most important considerations are
\begin{itemize}
    \item Number of layers
    \item Number of nodes for each individual layer
    \item Activation functions
    \item Optimization process
    \item Number of components $K$ in the mixture model 
    \item Weight initialization
    \item Loss function
\end{itemize} 
Deciding on hyperparameters for a model is often done either by following a best practice from the literature (i.e., rule of thumbs or similar) or by using automated hyperparameter tuning techniques such as grid search or random search. For many parameters the literature does not provide consent about a best practice approach but rather gives different indications which can be used as rough starting point. An automated hyperparameter tuning procedure relies on comparing the average loss values for different parameter combinations. However, this approach is seen as suboptimal as the aim of this thesis lies in presenting a default architecture rather than the development of individually optimized models for each data set. Thus, the procedure of determining a default architecture is done by experimenting with different hyperparameters and manually observing the effects parameter changes have on model performance. Three different architectures are presented at the end of this section, which are then used for training on each data set. 

\par \vspace{0.3cm}
\textbf{Number of layers:} The literature does not provide a sufficient consent on the optimal number of (hidden) layers in a neural network. Goodfellow et al. propose a larger number of hidden layers for DNNs as their research indicate increasing model accuracy with network depth \cite{goodfellow2016deep}. In contrast, Uzair and Jamil found that a DNN with more than 3 hidden layers tends to produce suboptimal results \cite{uzair2020effects}. Gu et al, which research is concerned with asset pricing using machine learning techniques, suggest that \textit{"...neural network performance peaks at three hidden layers then declines as more layers are added."} \cite{gu2020empirical}. An LSTM-MDN has per definition at least one hidden LSTM layer. Typically, the LSTM layer is complimented by at least one additional Dense layer, which is a hidden layer where each node is connected to every node from the previous layer. As the input data is of low dimensionality nor contains any complex features, it is advisable to keep the network architecture as simple as possible. 
After a trail-and-error process evaluating the performance of different combinations of LSTM layers and Dense layers in terms of model accuracy and computational expenses, the combination of a single LSTM layer followed by one Dense layer showed promising results. 
\par \vspace{0.3cm}
\textbf{Number of nodes:} As for the general depth of a network, there is no specific consent on the optimal number of nodes per layer in a DNN architecture. As for the number of hidden layers, an increasing number of nodes might lead to an overfitted model. The literature provides different rule-of-thump approaches, one of them proposing $\ln{N}$ nodes in a neural network, where $N$ is the number of training samples \cite{wanas1998optimal}.  Intuitively, the number of nodes shall be held comparably small as the underlying data does not contain much inherent complexity. After trying out different combinations, ranging from 64 to 2 neurons in the LSTM layer and 100 to 5 neurons in the Dense layer, results showed that the increase in accuracy was marginal after increasing the total number of neurons in both hidden layers over 16. The implemented LSTM-MDNs therefore include 6 nodes in the LSTM layer and 10 nodes in the Dense layer.  
\par \vspace{0.3cm}
\textbf{Activation functions:} The decision of which activation functions to use heavily depends on the underlying problem and the desired properties of the network output. The input layer itself does not apply activation functions. For the MDN layer, the used functions are already listed in section \ref{section_MDN}. The Dense layer uses a ReLU activation function due to its robustness against exploding or vanishing gradients, which turned out to be a problem during model training. The LSTM layer uses a tanh activation function by default. However, the usage of a ReLU activation function instead of the hyperbolic tangent resulted in a higher accuracy, which is in line with the findings from the literature \cite{karlsson2021value}. Thus, an altered version of the LSTM layer using ReLU as activation function is implemented in the LSTM-MDNs.
\par \vspace{0.3cm}
\textbf{Optimization Process:} Based on the ability to adapt the practical learning rate, the Adam optimizer is chosen as optimization algorithm. The implemented NNs use the default parameters of the respective function in Keras, which are $\beta_1 = 0.9,  \beta_2 = 0.999, \gamma = 0.001$, $\epsilon = 1e^{-7}$ and a batch size of 32.  Model training was set up to run for 100 epochs with an implemented early stoppage mechanism which finalises the training earlier if the loss value does not improve over 5 consecutive epochs.
\par \vspace{0.3cm}
\textbf{Mixture components:} The number of components in a mixture model depends on the number of component distributions assumed in the “ground truth”. A part of the literature assumes daily returns coming from two different components, namely a \textbf{bull market} and a \textbf{bear market} \cite{maheu2012components}. Bull markets are expected to be of low volatility compared to the highly volatile bear market periods. Different literature add a third \textbf{static market}\cite{dias2015clustering}. This thesis therefore compares LSTM-MDNs using a two-component mixture model ($K = 2$) with a model using a three-component approach ($K = 3$).
\par \vspace{0.3cm}
\textbf{Weight initialization:} The implemented LSTM-MDNs are using the \textit{GlorotUniform} initializer, which draws initial weights from a Uniform distribution. Recall that $\theta_0$ is composed of initial weight and bias vectors. Let $w^{(j)}$ denote the initial weight vector for the j-th layer at the start of the optimization procedure described in section \ref{chapter_optimization}.  
\begin{equation}
    w^{(j)} \sim \mathcal{U}\biggl(-\frac{\sqrt{6}}{\sqrt{n_j + n_{j+1}}}, \frac{\sqrt{6}}{\sqrt{n_j + n_{j+1}}}\biggr) \text{~~~,}
\end{equation}
where $n_j$ denotes the number of nodes in the j-th layer \cite{glorot2010understanding}. Model biases are initialized with zeros. As the initialization relies on computationally sampling from a defined uniform distribution, the state of Python's sampling function (i.e., its pseudo-random number generator) influences the values in the initial weight vector. The state of this pseudo-random number generator can be fixed by seed setting, which further enables results to be reproducible. The implemented models were initialized using a "best-of-three" approach: All models per evaluation period are trained on three different random states\footnote{seed-setting is fixed for the Python backend with the function \textit{random.seed()}, for the numpy package via \textit{numpy.random.seed()} and for TensorFlow backend via \textit{tensorflow.rand.set\_seed()}.} with the seed values [911, 6969, 9999]. The seed providing the best results on average is then used for implementation.
\par \vspace{0.3cm}
\textbf{Loss function:} The default loss function for the MDN is defined in equation \ref{neg_log_lik_MDN}. Arimond et al. (2020) added a penalty term to the negative log-likelihood function, which accounts for the tendency of the mixture parameters $\pi$ to converge to extreme values (0 and 1, respectively) under a 2-component mixture model. An alternative loss function, introducing an L2 regularization term\footnote{L2 regularization is used as the regularization term proposed by Arimond et al. (2020) led to convergence issues during model implementation.}, can be formulated as
\begin{equation} \label{regularized_loss_function}
\begin{split}
        J_{reg}(\theta) &=  \biggl(- \ln{\sum^K_{k=1} \pi_{k}(x,\theta)\mathcal{N}(\mu_k(x,\theta), \sigma^2_k(x,\theta))}\biggr) + \lambda \mathcal{R} \\
        \mathcal{R} &= \sum^K_{k=1}\pi_k^2
\end{split}
\end{equation}
Due to its exponential property, $\mathcal{R}$ is minimized if $max(\pi_k)$ for $k = 1,...,K$ is minimized. That is, if  $\pi_1 = \pi_2 = ... = \pi_k = \frac{1}{K}$. Thus, the penalty term accounts for large inequalities between the mixture components and takes the value $\lambda \frac{1}{K}$ if the mixture components are equal. If $\lambda = 0$, the loss function reduces to equation \ref{neg_log_lik_MDN}. Trying different values for $\lambda$, the results indicate that scaling with a value of at least 0.1 is required to produce balanced mixture components. Hence, $\lambda = 0.1$ is used for model implementation.
\par \vspace{0.3cm}
Based on the chosen hyperparameters, three different model architectures are implemented for VaR forecasting and backtesting. \textbf{NNet 1} uses an MDN layer based on a two-component Gaussian mixture. As loss function it uses the unregularized negative log-likelihood of the mixture model (equation \ref{neg_log_lik_MDN}).
\begin{center}
    \underline{\textbf{NNet 1: 2-component LSTM-MDN}}  \\
    Input Layer [$x_t = R_{t-9},...,R_{t}$] \\
    LSTM-Layer [6 nodes - activation: ReLU] \\
    Dense-Layer [12 nodes - activation: ReLU] \\
    Output-Layer(MDN) [\textbf{$K = 2$} - $\hat{y}_{t+1} = [\hat{\pi}_{t+1}, \hat{\mu}_{t+1}, \hat{\sigma}_{t+1}]$ - activation: Softmax, ELU+1] \\
    \textbf{\textit{Loss func.: }}\textit{neg. log-likelihood}
\end{center}
Based on the same architecture and the same number of components in the mixture model, \textbf{NNet 2} uses the regularized loss function proposed in equation \ref{regularized_loss_function} instead.
\begin{center}
    \underline{\textbf{NNet 2: 2-component LSTM-MDN w/ L2-term}}  \\
    Input Layer [$x_t = R_{t-9},...,R_{t}$] \\
    LSTM-Layer [6 nodes - activation: ReLU] \\
    Dense-Layer [12 nodes - activation: ReLU] \\
    Output-Layer(MDN) [\textbf{$K = 2$} - $\hat{y}_{t+1} = [\hat{\pi}_{t+1}, \hat{\mu}_{t+1}, \hat{\sigma}_{t+1}]$ - activation: Softmax, ELU+1] \\
    \textbf{\textit{Loss func.: }}\textit{regularized neg. log-likelihood (L2-term)}
\end{center}
\textbf{NNet 3} is based on the assumption that the true distribution of daily returns is composed of three instead of two Gaussian component distributions. 
\begin{center}
    \underline{\textbf{NNet 3: 3-component LSTM-MDN}}  \\
    Input Layer [$x_t = R_{t-9},...,R_{t}$] \\
    LSTM-Layer [6 nodes - activation: ReLU] \\
    Dense-Layer [12 nodes - activation: ReLU] \\
    Output-Layer(MDN) [\textbf{$K = 3$} - $\hat{y}_{t+1} = [\hat{\pi}_{t+1}, \hat{\mu}_{t+1}, \hat{\sigma}_{t+1}]$ - activation: Softmax, ELU+1] \\
    \textbf{\textit{Loss func.: }}\textit{neg. log-likelihood}
\end{center}

\subsection{VaR forecasting based on Monte Carlo sampling}
A common method to forecast Value-at-Risk is the usage of Monte Carlo (MC) sampling techniques. Let $\hat{y}_{t+1} = [\hat{\pi}_{t+1}, \hat{\mu}_{t+1}, \hat{\sigma}_{t+1}]$ denote the parameter estimation from the LSTM-MDN at time point $t$. These parameters are used to generate a large number of samples from a Gaussian mixture model. These samples can be interpreted as artificial returns which are produced by a Monte Carlo simulation following the estimated parameters from the LSTM-MDN. A basic Monte Carlo estimation \cite{glasserman2000efficient} could be described in pseudo code as the following:
\begin{itemize}
    \item[1]: Produce a uniformly distributed proxy sample $u \sim \mathcal{U}(0,1)$
    \item[2]: If $u \in \biggl[\sum^K_{k=1}\pi_{k-1}, \sum^{K}_{k=1}\pi_{k}\biggr]$ with $\pi_0 = 0$:  generate one sample $\tilde{r}  \sim \mathcal{N}(\hat{\mu}_{k, t+1}, \hat{\sigma}^2_{k, t+1})$ 
    \item[3]: Repeat step 1 \& 2 $N$ times. The results is a vector of simulated daily returns $\widetilde{R} = [\tilde{r_1}, \tilde{r_2}, ..., \tilde{r_N}]$ for time point $t+1$.
    \item[4]: Convert $\tilde{R}$ into a vector of simulated losses: $ \tilde{L} = -\tilde{R}$
\end{itemize}
\par \vspace{0.3cm}
The VaR estimation under the LSTM-MDN is thus defined as the empirical $\alpha$-quantile of the simulated loss distribution:
\begin{equation}
    VaR_{NNet}[\alpha, h = 1] =  \tilde{L}_{(\lceil \alpha N \rceil)}\times P_t \text{~~~,}
\end{equation}
where $\tilde{L}$ is a vector of MC-simulated losses for $t+1$ based on the day-ahead estimates [$\hat{\pi}_{t+1}$, $\hat{\mu}_{t+1}$, $\hat{\sigma}_{t+1}$]. For implementation, $N = 100,000$ is used.

\section{Model Evaluation (Backtesting)}
As this paper is concerned with differences in model performance during calm periods compared to turbulent market periods, model performance is tested on two market environments separately. The \textbf{calm market environment} represents a low-volatility market, spanning from 01.01.2017 to 31.12.2018. Although this period acts as a proxy for a calm market period, the time frame includes global market events such as the after effects of the Brexit referendum\footnote{i.e., the 2016 United Kingdom European Union membership referendum from 23. June 2016} or the US-China trade war under the presidency of Donald Trump. The \textbf{turbulent market environment} is capturing a time frame from 01.01.2021 and ending on 31.12.2022. Table \ref{table_std_dev} lists the sample standard deviation of daily returns for the three stock indices over the respective period, which shows that the market during the turbulent period (spanning over more than half of the Covid-19 crisis\footnote{The WHO declared the Public Health Emergency of International Concern (PHEIC) for COVID-19 on 30. Jan. 2020 and ended the PHEIC period on 05. May 2023 \cite{WHO}.}) is subject to a higher level of uncertainty than the calm period.

\begin{table}[H]
\centering
\begin{tabular}{||c c c||} 
\hline
\textbf{Index} & \textbf{$\hat{\sigma}$ calm period} & \textbf{$\hat{\sigma}$ turbulent period} \\ [0.5ex] 
\hline\hline
FTSE 100 & 0.00682 & 0.00926 \\ 
\hline
S\&P 500 & 0.00816 & 0.01226  \\
\hline
EURO STOXX 50 & 0.00768 & 0.01245  \\ [1ex] 
\hline
\end{tabular}\par
\caption{Standard Deviation of daily returns}
\label{table_std_dev}
\end{table}
The benchmark models are used to produce a VaR forecast for each time point in the respective evaluation period. As described, the benchmark models do not require a dedicated training on past data. The NNs are individually trained on the train set corresponding to the respective testing period and stock index. 18 individual models are trained for model evaluation, where each model is trained and tuned on a pre-evaluation set consisting of all observations until the start of the respective evaluation period. Subsequently, models used for the 2021/2022 evaluation are trained on approximately 1000 additional observations compared to their calm period counterparts.
\par \vspace{0.3cm}
For model evaluation, a backtesting procedure is implemented. One-day-ahead Value-at-Risk forecasts $VaR[\alpha=0.99, h=1]$ from each model are evaluated. First, a binary series $I$ is computed as described in equation \ref{binary_series}. Model performance is assessed through an evaluation of the following properties: 
\begin{itemize}
    \item Proportion of \textbf{overshoots} $\bigl(\frac{1}{T}\sum^T_{t=1}I_t(\alpha)\bigr)$
    \item Unconditional Coverage \textbf{UC} (section \ref{section_unconditional_coverage})
    \item Independence \textbf{Ind} (section \ref{section_independence})
    \item Conditional Coverage \textbf{CC} (section \ref{section_joint_test})
\end{itemize}

% Results
\chapter{Results} \label{chapter_results}
The following presents 6 tables consisting of p-values for the test described in the previous section and the proportion of overshoots for each forecast $VaR[\alpha=0.99, h=1]$ under the respective model. Plots which visually put the forecasts in a context to the ex-post losses are further presented. A decision on rejecting the null hypothesis is done at a significance level of 5\%. Further interpretation is provided in chapter \ref{chapter_discussion}.

\pagebreak
\section{Calm Period (2017 \& 2018)}

\subsection{FTSE 100}
For the FTSE 100 data, only the GARCH(1,1) and the neural network architecture 2 and 3 pass the unconditional coverage test, i.e., the null is not rejected at a 5\% significance level. None of the models tend to violate the independence assumption. The joint test leads to the conclusion that the GARCH model and NNet 2 and 3 can be deemed appropriate. Figure \ref{FTSE_results_calm} confirms the strict overestimating behaviour of NNet 1 (further explanation in section \ref{section_discussion_accuracy}).
\begin{table}[H]
    \centering
    \begin{tabular}{|l|c|c|c|c|c|c|}
    \hline
        & $VaR_{HS}$ & $VaR_{CMM}$& $VaR_{GARCH}$ & $VaR_{NNet_1}$ & $VaR_{NNet_2}$ & $VaR_{NNet_3}$ \\ \hline
        \textbf{overshoots} & 2.178\% & 2.178\% & 1.386\% & 0\% & 1.386\% & 0.396\% \\ \hline
        \textbf{UC (p-val.)}  & 0.021 & 0.021 & 0.41 & 0.001 & 0.41 & 0.12 \\ \hline
        \textbf{Ind (p-val.)}  & 0.229 & 0.229 & 0.078 & 1 & 0.657 & 0.899 \\ \hline
        \textbf{CC (p-val.)} & 0.034 & 0.034 & 0.151 & 0.006 & 0.645 & 0.297 \\ \hline
    \end{tabular}
    \caption{FTSE 100 results (calm period)}
\label{FTSE_results_calm}
\end{table}
\par \vspace{0.3cm}

\begin{figure}[H]
\subfigure[NNet 1]{\includegraphics[width=7.9cm]{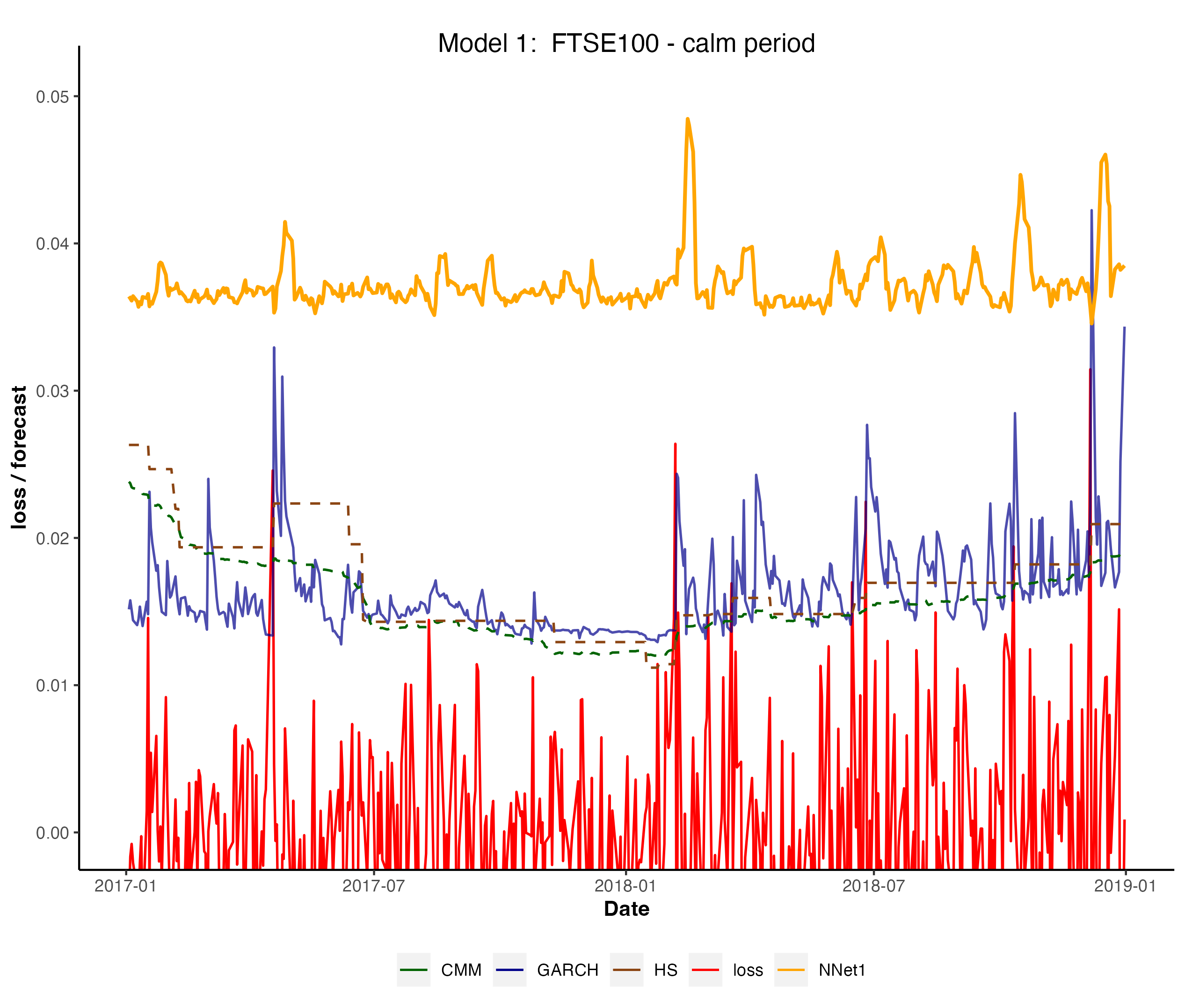}}
\hfill
\subfigure[NNet 2]{\includegraphics[width=7.9cm]{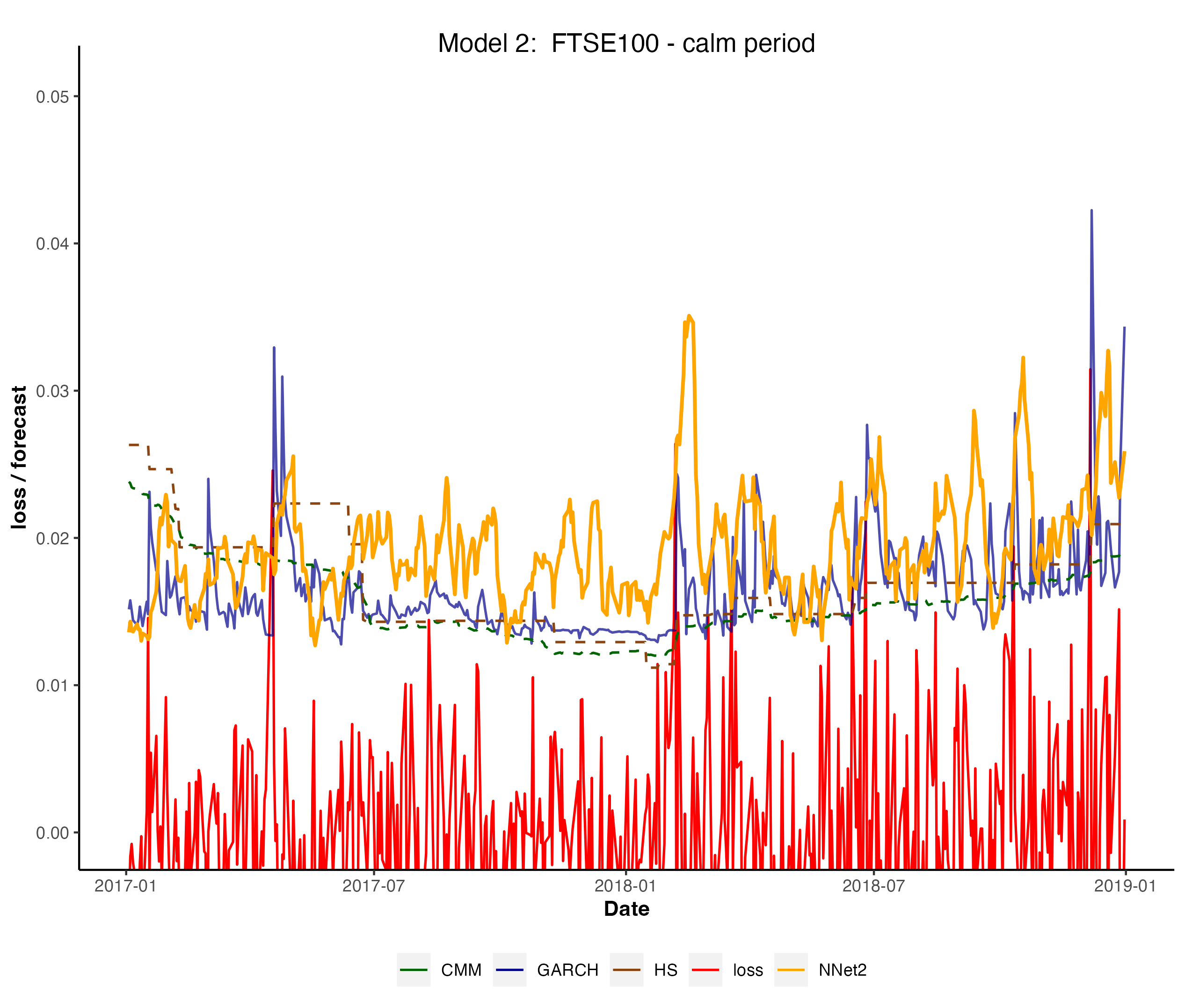}}
\hfill
\subfigure[NNet 3]{\includegraphics[width=7.9cm]{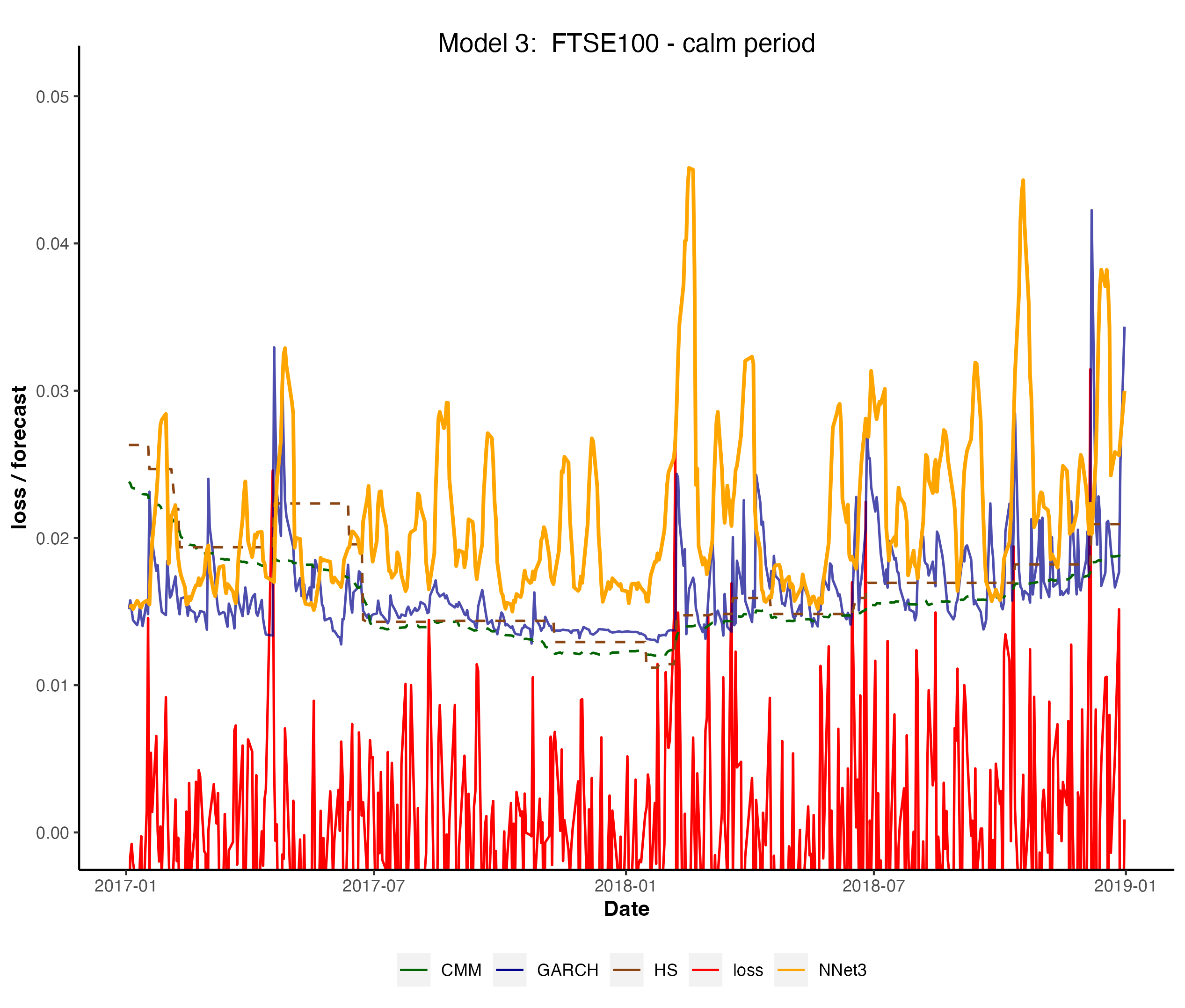}}
\hfill
\caption{VaR forecasts FTSE 100 - calm period}
\label{plot_FTSE_results_calm}
\end{figure}

\subsection{S\&P 500}
The unconditional coverage hypotheses is rejected for all benchmark models and NNet1 as they strongly underestimate risk. The independence test results in rejecting the null for both the CMM and NNet 1. Considering the joint test, the historical simulation as well as NNet 2 and 3 can be seen as appropriate models. Figure \ref{SP_results_calm} (a) confirms the underestimating behaviour of NNet 1 and generally shows large difference in reactivity to volatility clusters of the different neural networks.
\begin{table}[H]
    \centering
    \begin{tabular}{|c|c|c|c|c|c|c|}
    \hline
        & $VaR_{HS}$ & $VaR_{CMM}$& $VaR_{GARCH}$ & $VaR_{NNet_1}$ & $VaR_{NNet_2}$ & $VaR_{NNet_3}$ \\\hline
        \textbf{overshoots} & 1.992\% & 3.586\% & 2.191\% & 2.39\% & 0.797\% & 0.398\% \\ \hline
        \textbf{UC (p-val.)} & 0.049 & 0 & 0.02 & 0.008 & 0.635 & 0.123 \\ \hline
        \textbf{Ind (p-val.)} & 0.185 & 0.023 & 0.231 & 0.002 & 0.8 & 0.899 \\ \hline
        \textbf{CC (p-val.)} & 0.06 & 0 & 0.033 & 0 & 0.865 & 0.302 \\ \hline
    \end{tabular}
    \caption{S\&P 500 results (calm period)}
\label{SP_results_calm}
\end{table}
\par \vspace{0.3cm}

\begin{figure}[H]
\subfigure[NNet 1]{\includegraphics[width=7.9cm]{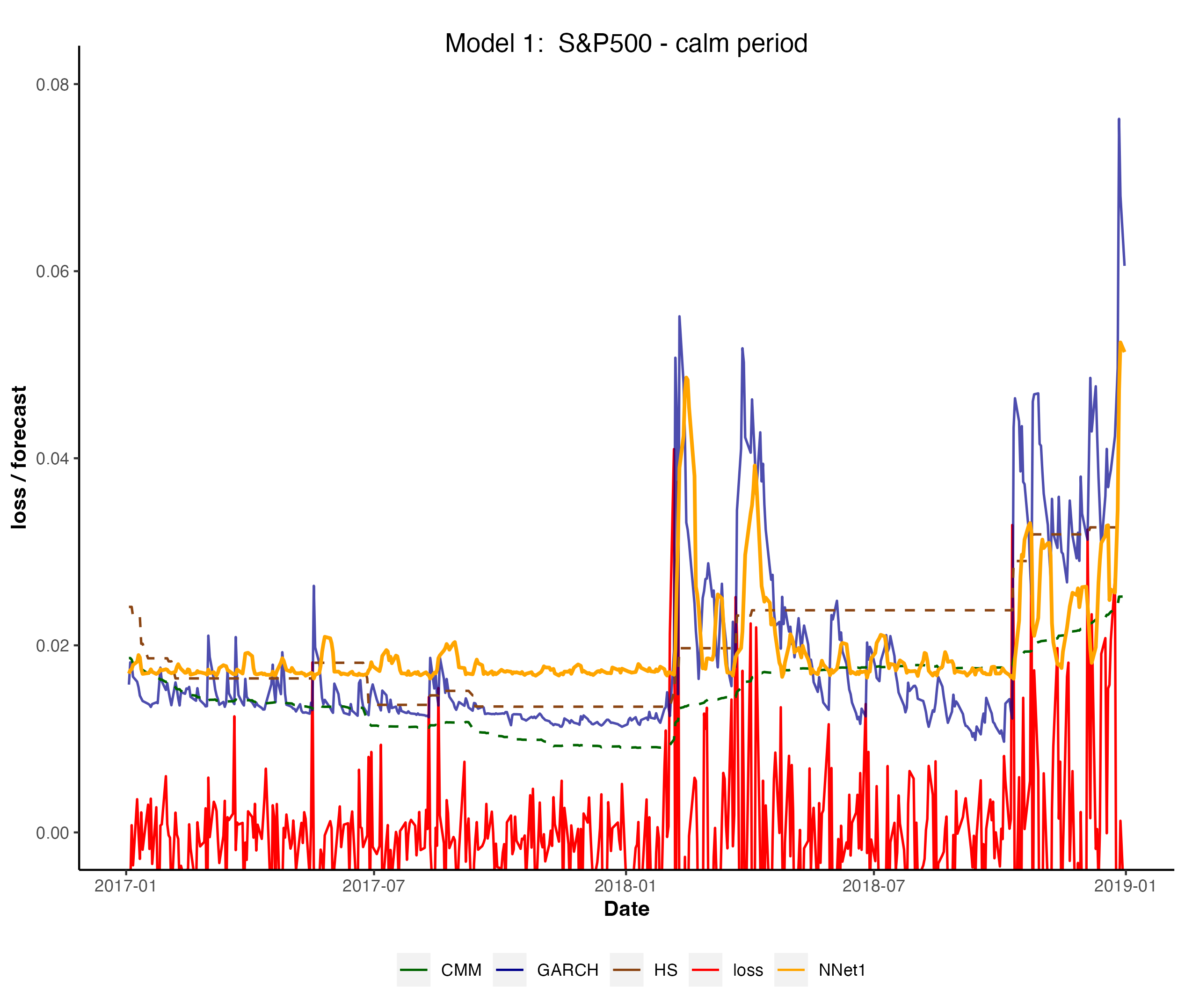}}
\hfill
\subfigure[NNet 2]{\includegraphics[width=7.9cm]{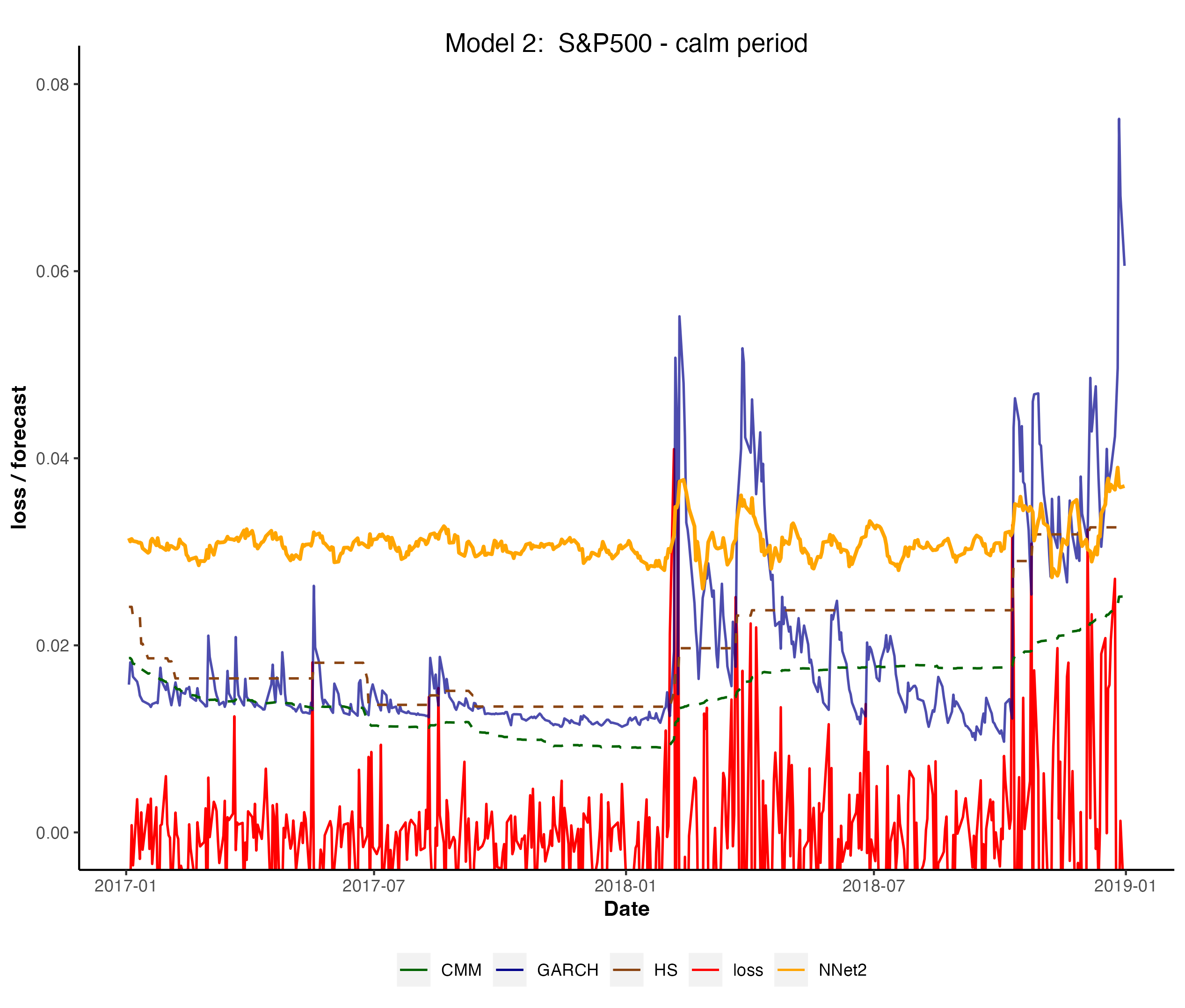}}
\hfill
\subfigure[NNet 3]{\includegraphics[width=7.9cm]{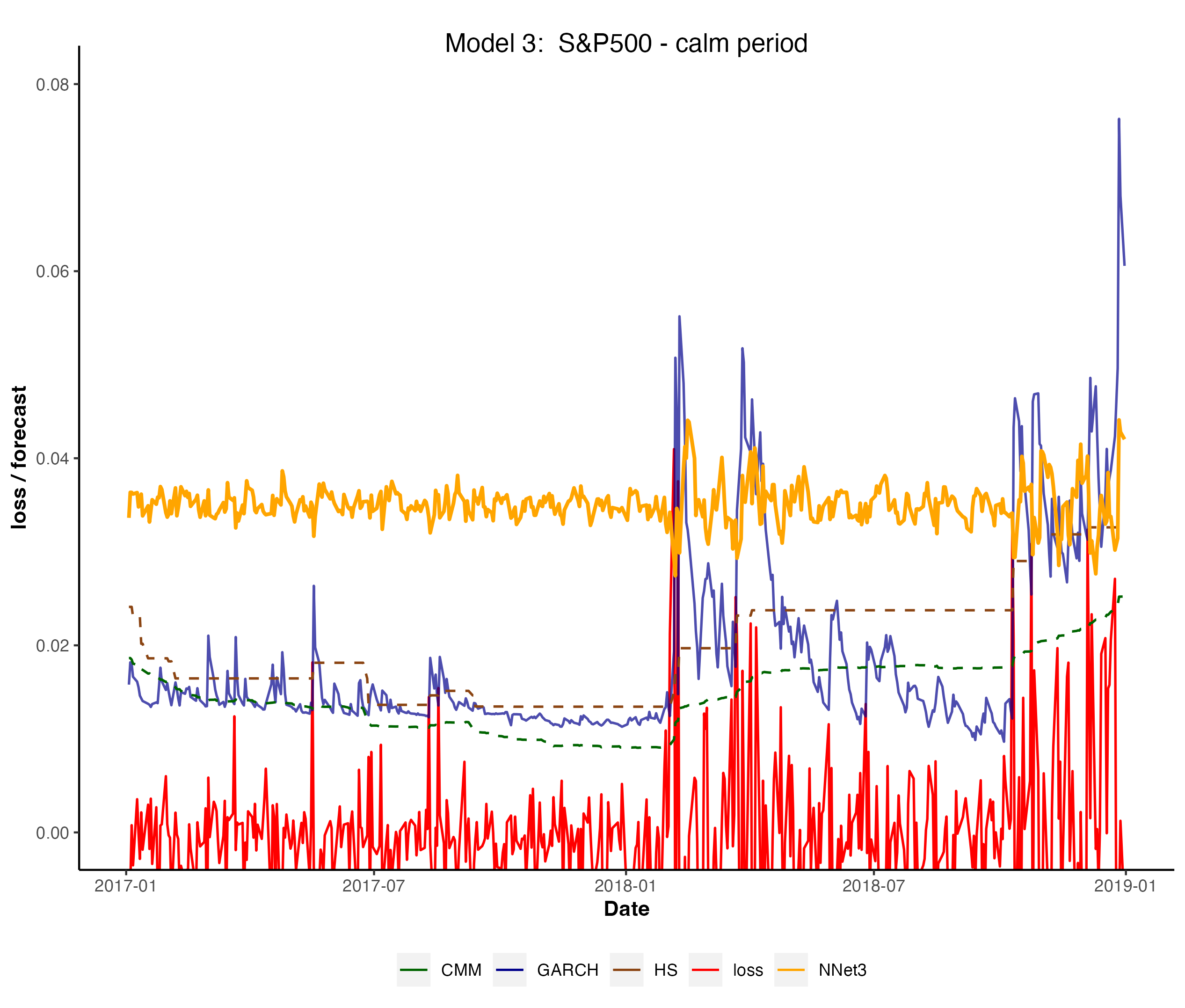}}
\hfill
\caption{VaR forecasts S\&P 500 - calm period}
\label{plot_SP_results_calm}
\end{figure}

\subsection{EURO STOXX 50} \label{EUSTOXX_results_calm_section}
What is striking about backtesting for the calm period of EUROSTOXX50 data is that benchmark models significantly outperform their neural network counterparts. The tests indicate that all benchmark models can be seen as appropriate for the underlying data, while all neural networks strongly overestimate VaR. No NNet model can be seen as appropriately accounting the tail risk. Further, the neural networks do not tend to show any reactivity towards volatility shifts as the VaR forecast stays quite static over the evaluation period (see figure \ref{EUSTOXX_results_calm}). 

\begin{table}[H]
    \centering
    \begin{tabular}{|c|c|c|c|c|c|c|}
    \hline
        & $VaR_{HS}$ & $VaR_{CMM}$& $VaR_{GARCH}$ & $VaR_{NNet_1}$ & $VaR_{NNet_2}$ & $VaR_{NNet_3}$ \\ \hline
        \textbf{overshoots} & 1.4\% & 1.8\% & 1.4\% & 0\% & 0\% & 0\% \\ \hline
        \textbf{UC (p-val.)} & 0.396 & 0.106 & 0.396 & 0.002 & 0.002 & 0.002 \\ \hline
        \textbf{Ind (p-val.)} & 0.656 & 0.565 & 0.656 & 1 & 1 & 1 \\ \hline
        \textbf{CC (p-val.)} & 0.632 & 0.23 & 0.632 & 0.007 & 0.007 & 0.007 \\ \hline
    \end{tabular}
    \caption{EUROSTOXX50 results (calm period)}
\label{EUSTOXX_results_calm}
\end{table}
\par \vspace{0.3cm}
\begin{figure}[H]
\subfigure[NNet 1]{\includegraphics[width=7.9cm]{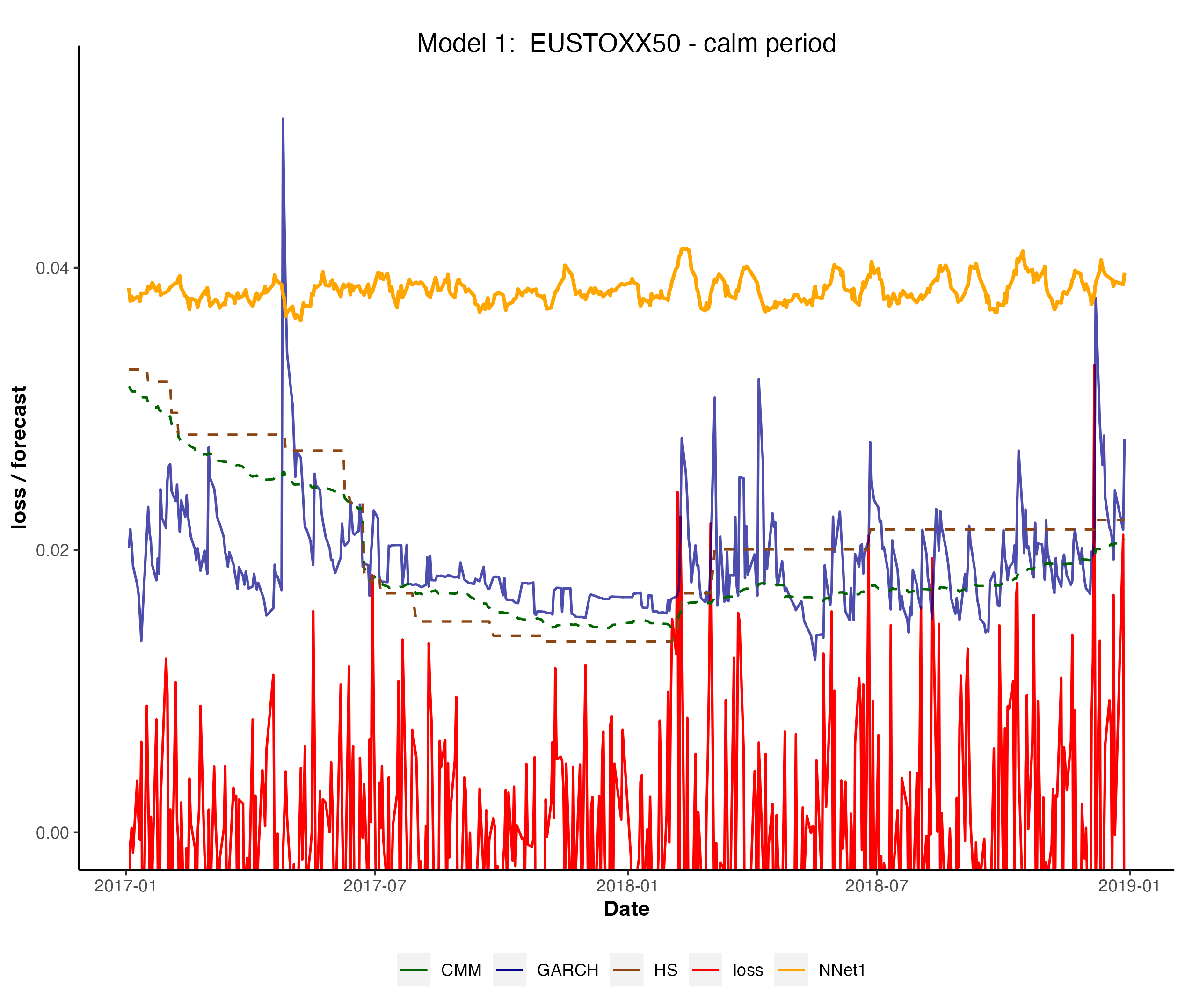}}
\hfill
\subfigure[NNet 2]{\includegraphics[width=7.9cm]{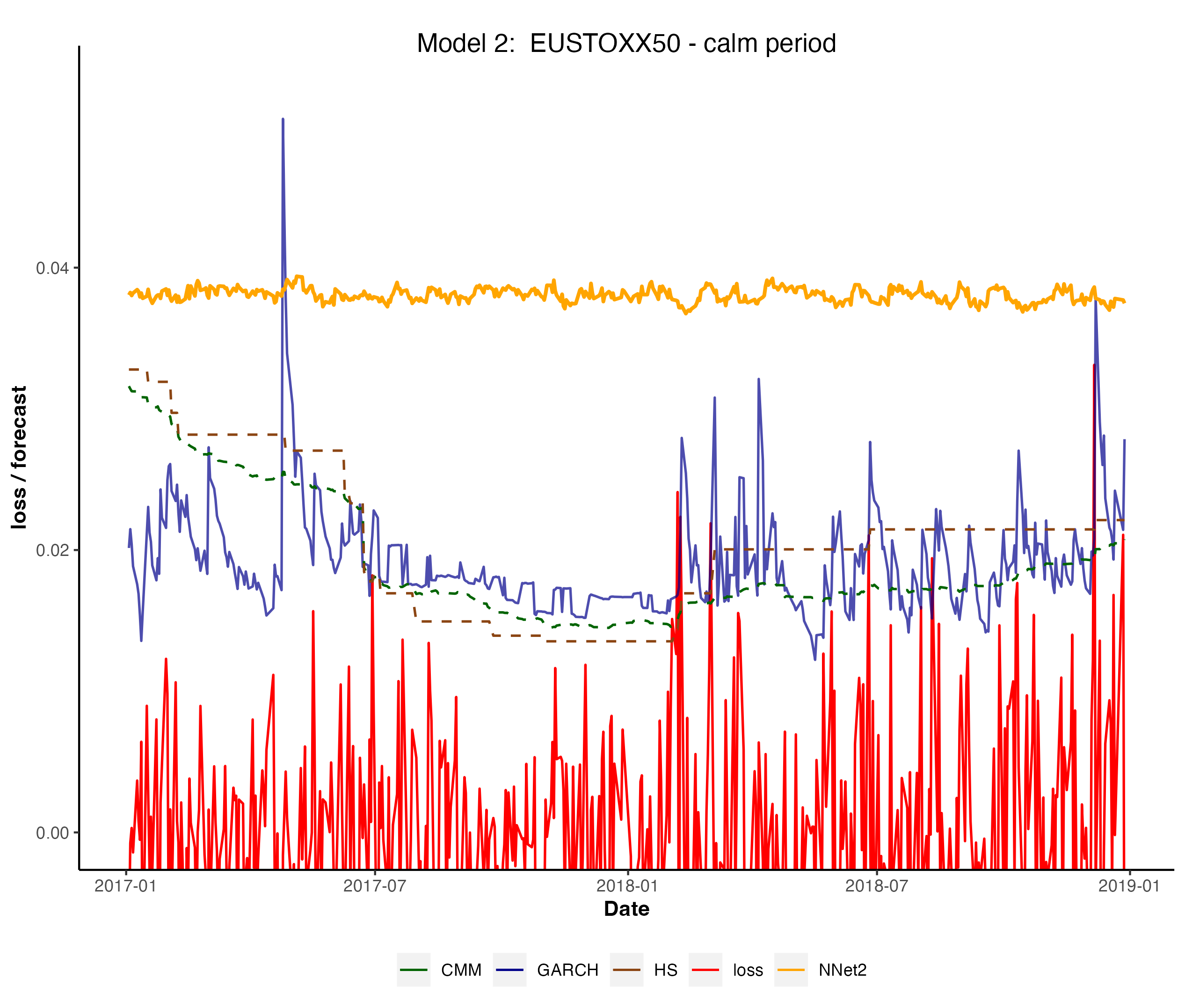}}
\hfill
\subfigure[NNet 3]{\includegraphics[width=7.9cm]{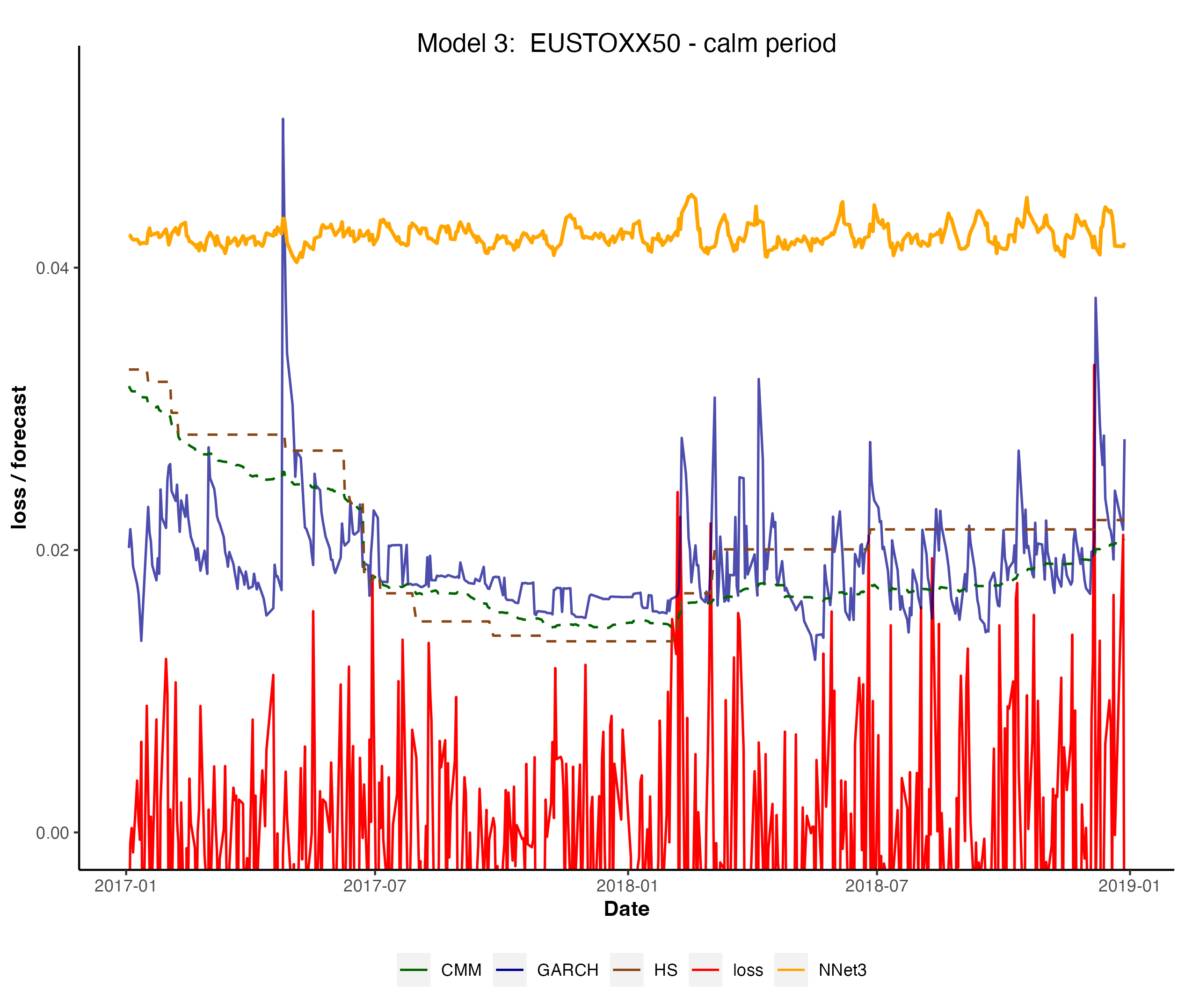}}
\hfill
\caption{VaR forecasts EUROSTOXX 50 - calm period}
\label{plot_EUSTOXX_results_calm}
\end{figure}

\section{Turbulent Period (2021 \& 2022)}

\subsection{FTSE 100}
All evaluated models perform well and produce appropriate VaR forecasts for the FTSE 2021/ 2022 evaluation, while neural networks show a better performance than their counterparts (based on the joint test). NNet 3 (figure \ref{FTSE_results_turbulent}) tends to account for shifts in volatility far stronger than NNet 1 and 2.

\begin{table}[H]
    \centering
    \begin{tabular}{|c|c|c|c|c|c|c|}
    \hline
        & $VaR_{HS}$ & $VaR_{CMM}$& $VaR_{GARCH}$ & $VaR_{NNet_1}$ & $VaR_{NNet_2}$ & $VaR_{NNet_3}$ \\ \hline
        \textbf{overshoots} & 1.193\% & 1.789\% & 1.988\% & 0.994\% & 0.596\% & 1.59\% \\ \hline
        \textbf{UC (p-val.)} & 0.673 & 0.109 & 0.05 & 1 & 0.325 & 0.22 \\ \hline
        \textbf{Ind (p-val.)} & 0.054 & 0.144 & 0.523 & 0.751 & 0.85 & 0.611 \\ \hline
        \textbf{CC (p-val.)} & 0.142 & 0.095 & 0.119 & 0.951 & 0.606 & 0.415 \\ \hline
    \end{tabular}
    \caption{FTSE 100 results (turbulent period)}
\label{FTSE_results_turbulent}
\end{table}
\par \vspace{0.3cm}

\begin{figure}[H]
\subfigure[NNet 1]{\includegraphics[width=7.9cm]{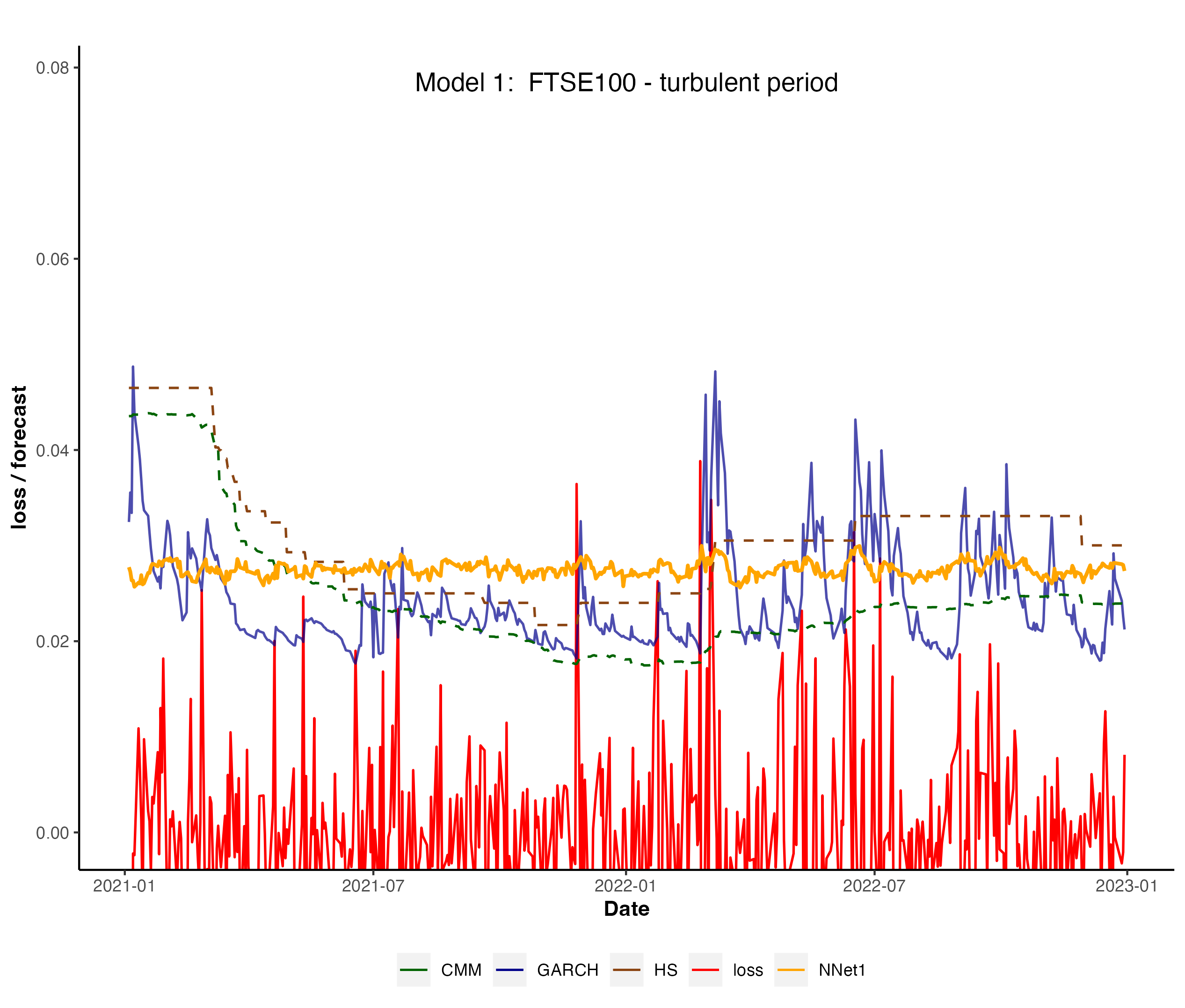}}
\hfill
\subfigure[NNet 2]{\includegraphics[width=7.9cm]{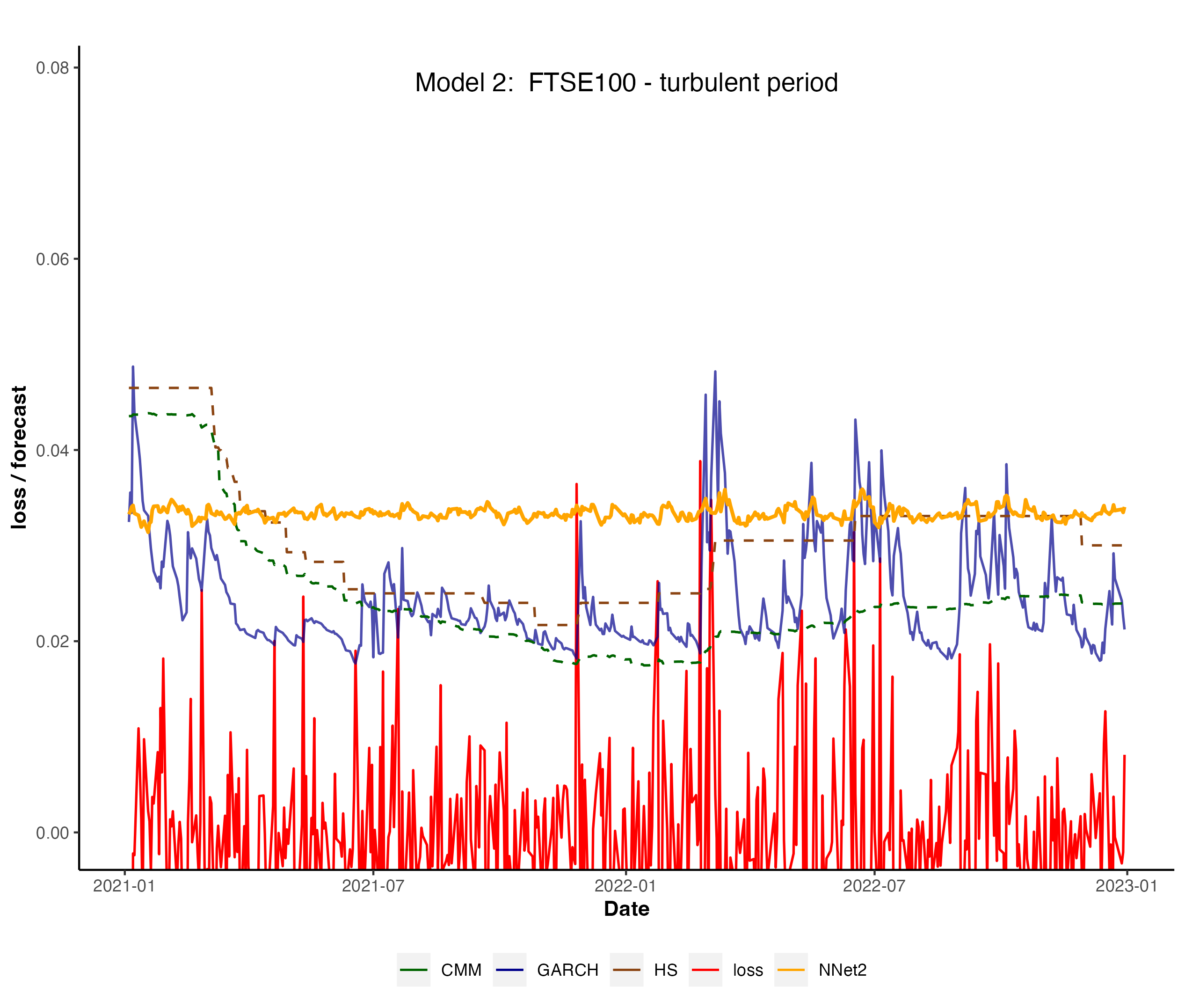}}
\hfill
\subfigure[NNet 3]{\includegraphics[width=7.9cm]{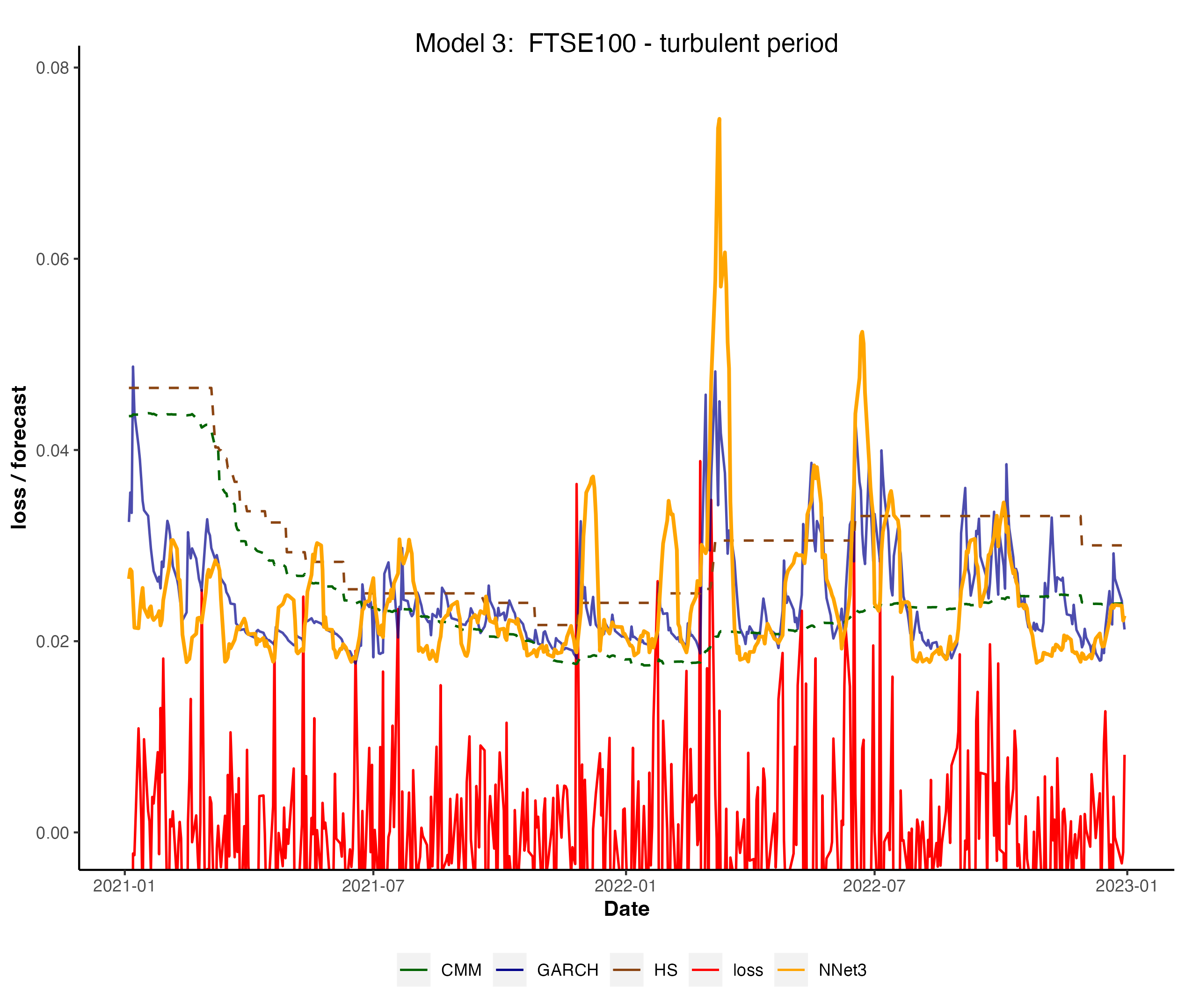}}
\hfill
\caption{VaR forecasts FTSE 100 - turbulent period}
\label{plot_FTSE_results_turbulent}
\end{figure}

\subsection{S\&P 500}
For the UC test, GARCH and NNet 1 show good results while the null barely holds for NNet 2. The NNet 3 architecture leads to a strong overestimation of Value-at-Risk, resulting in no overshootings at all. No model tends to violate the independence assumption. Figure \ref{SP_results_turbulent} confirms the conservative behaviour of NNet 3 while NNet 2 shows the best reactivity volatility clusters.
\begin{table}[H]
    \centering
    \begin{tabular}{|c|c|c|c|c|c|c|}
    \hline
        & $VaR_{HS}$ & $VaR_{CMM}$& $VaR_{GARCH}$ & $VaR_{NNet_1}$ & $VaR_{NNet_2}$ & $VaR_{NNet_3}$ \\ \hline
        \textbf{overshoots} & 2.187\% & 3.777\% & 1.789\% & 1.59\% & 1.988\% & 0\% \\ \hline
        \textbf{UC (p-val.)} & 0.021 & 0 & 0.109 & 0.22 & 0.05 & 0.001 \\ \hline
        \textbf{Ind (p-val.)} & 0.483 & 0.744 & 0.566 & 0.611 & 0.184 & 1 \\ \hline
        \textbf{CC (p-val.)} & 0.054 & 0 & 0.235 & 0.415 & 0.06 & 0.006 \\ \hline
    \end{tabular}
    \caption{S\&P 500 results (turbulent period)}
\label{SP_results_turbulent}
\end{table}
\par \vspace{0.3cm}

\begin{figure}[H]
\subfigure[NNet 1]{\includegraphics[width=7.9cm]{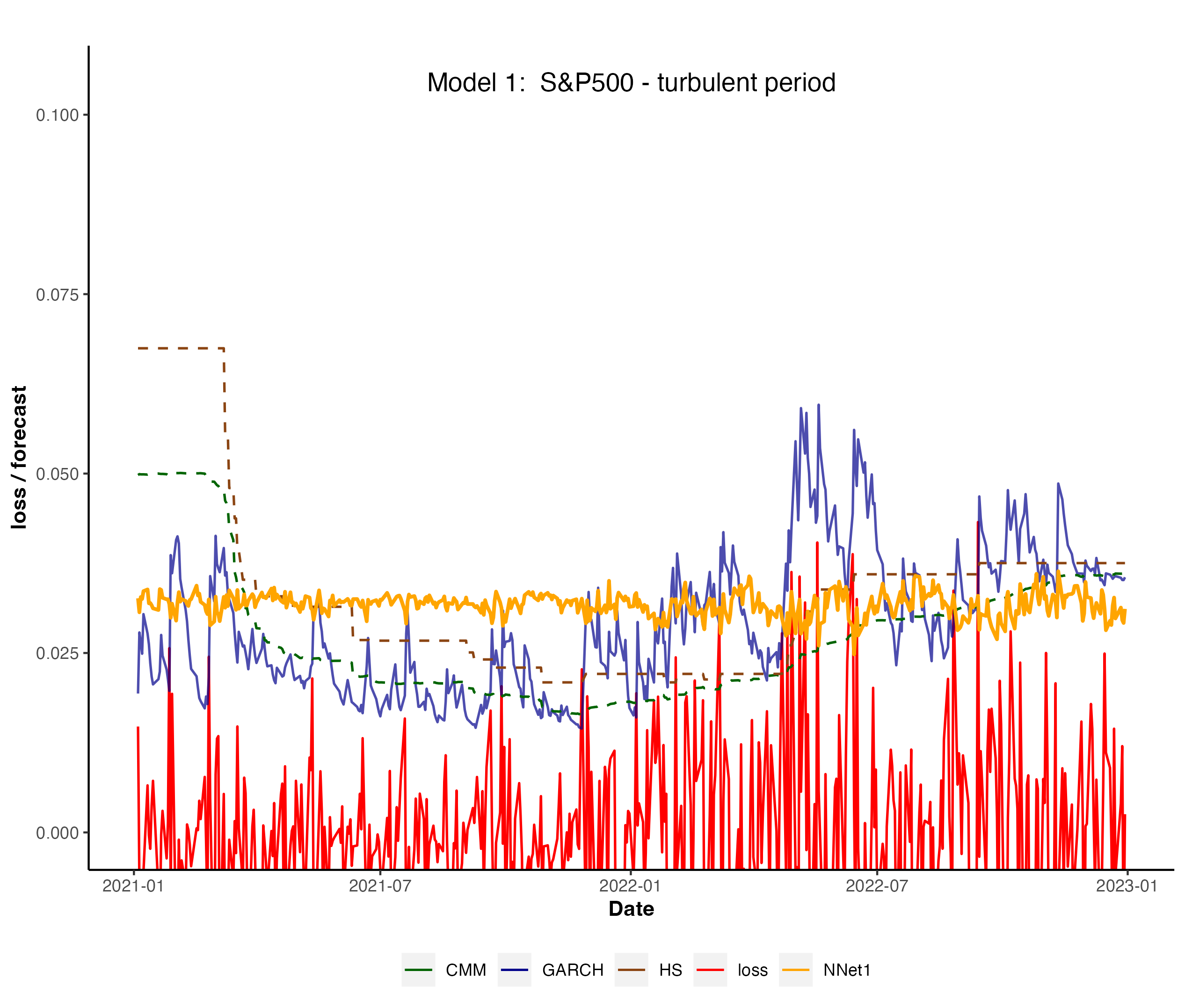}}
\hfill
\subfigure[NNet 2]{\includegraphics[width=7.9cm]{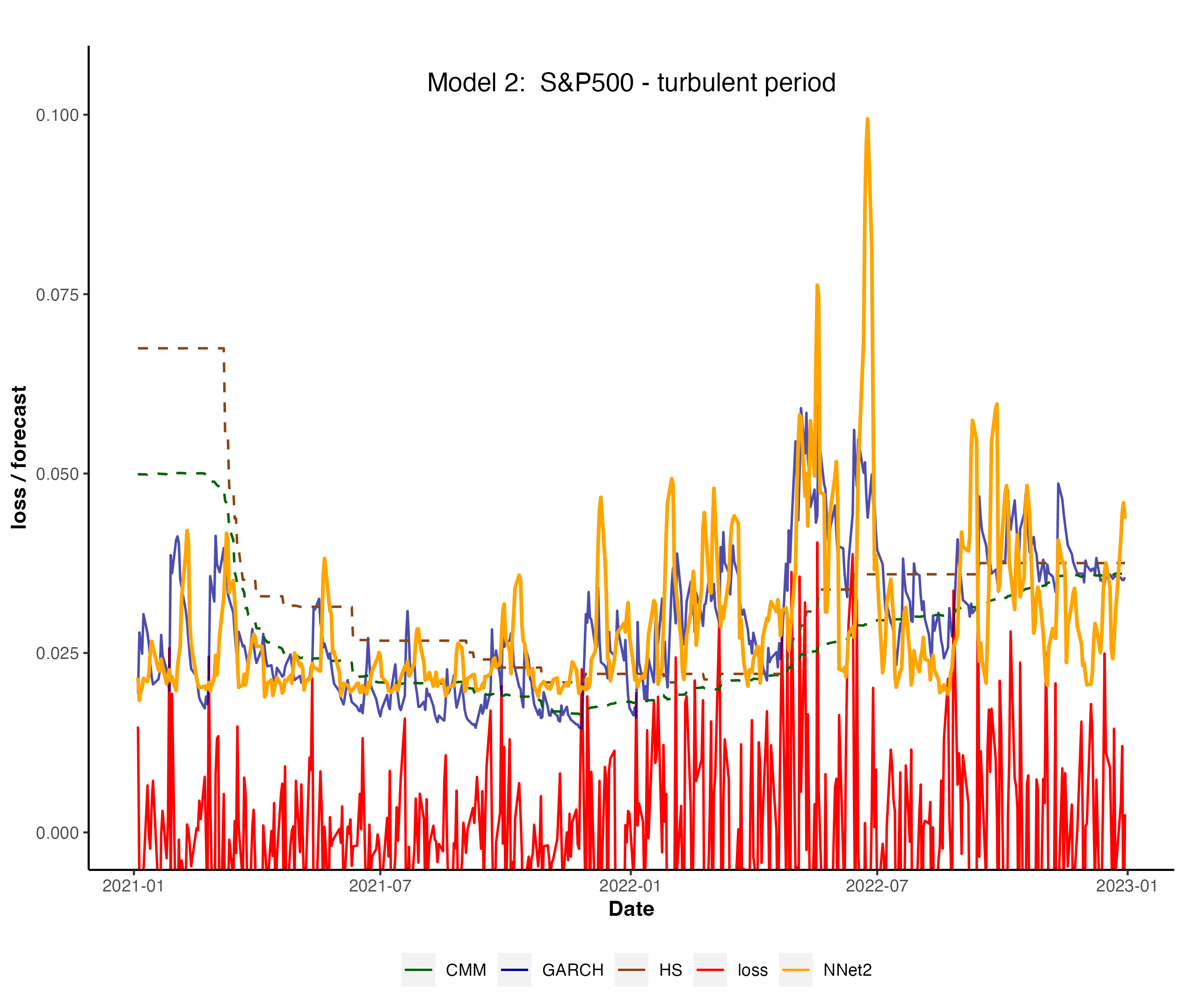}}
\hfill
\subfigure[NNet 3]{\includegraphics[width=7.9cm]{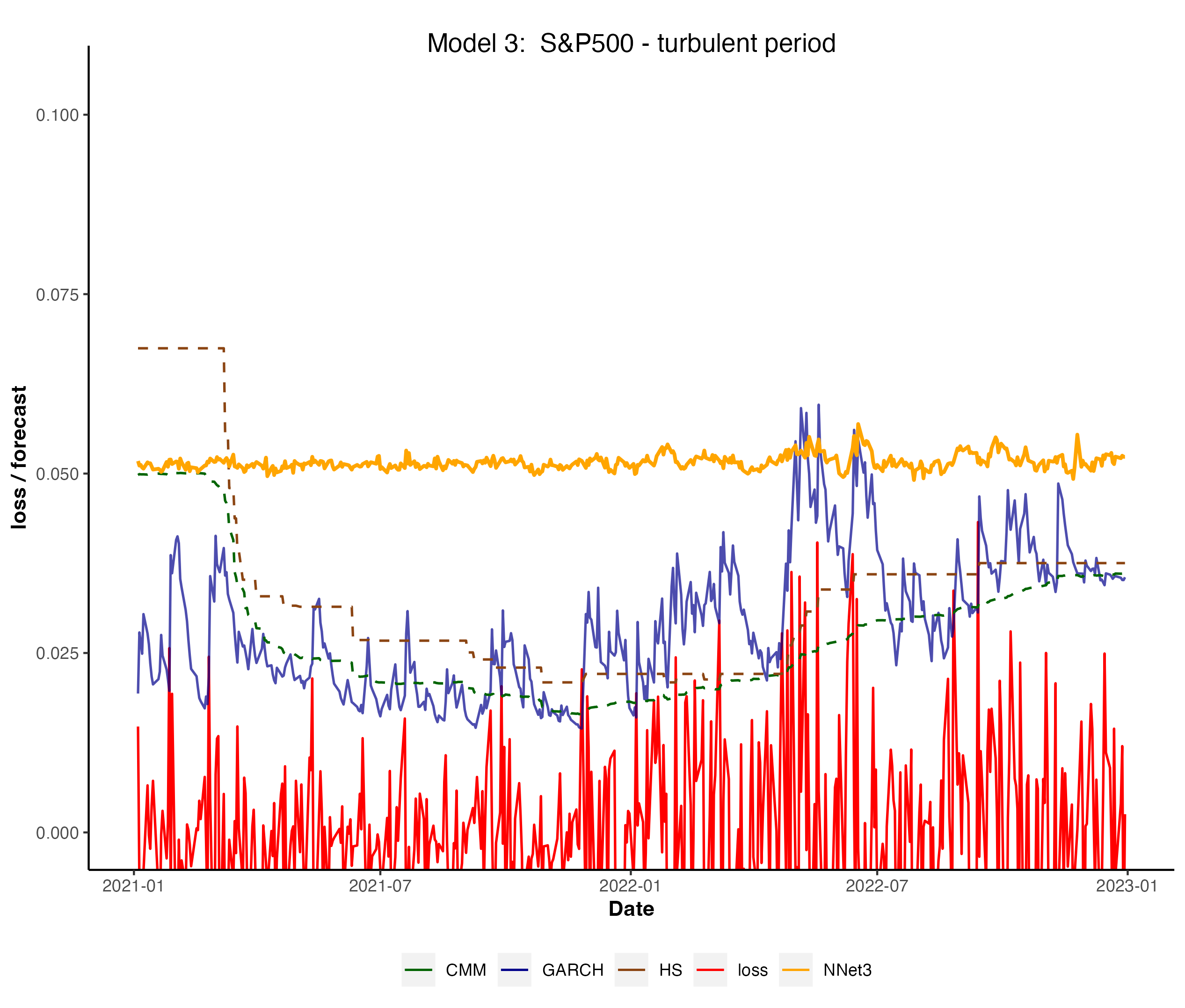}}
\hfill
\caption{VaR forecasts S\&P 500 - turbulent period}
\label{plot_SP_results_turbulent}
\end{figure}

\subsection{EURO STOXX 50}
All models can be seen as appropriate as the null  holds over all  tests for all models, where CMM provides the least accurate results. Compared to the calm testing period for the EUROSTOXX50 data set, the NNs perform well during the turbulent period. Figure \ref{EUSTOXX_results_turbulent} confirms the good results from the backtesting evaluation. All neural networks show good reactivity to changes in volatility.
\begin{table}[H]
    \centering
    \begin{tabular}{|c|c|c|c|c|c|c|}
    \hline
        & $VaR_{HS}$ & $VaR_{CMM}$& $VaR_{GARCH}$ & $VaR_{NNet_1}$ & $VaR_{NNet_2}$ & $VaR_{NNet_3}$ \\ \hline
        \textbf{overshoots} & 0.984\% & 1.969\% & 0.787\% & 0.591\% & 1.378\% & 0.591\% \\ \hline
        \textbf{UC (p-val.)} & 0.975 & 0.053 & 0.617 & 0.315 & 0.418 & 0.315 \\ \hline
        \textbf{Ind (p-val.)} & 0.752 & 0.526 & 0.8 & 0.85 & 0.658 & 0.85 \\ \hline
        \textbf{CC (p-val.)} & 0.951 & 0.125 & 0.855 & 0.593 & 0.653 & 0.593 \\ \hline
    \end{tabular}
    \caption{EUROSTOXX50 results (turbulent period)}
\label{EUSTOXX_results_turbulent}
\end{table}

\par \vspace{0.3cm}
\begin{figure}[H]
\subfigure[NNet 1]{\includegraphics[width=7.9cm]{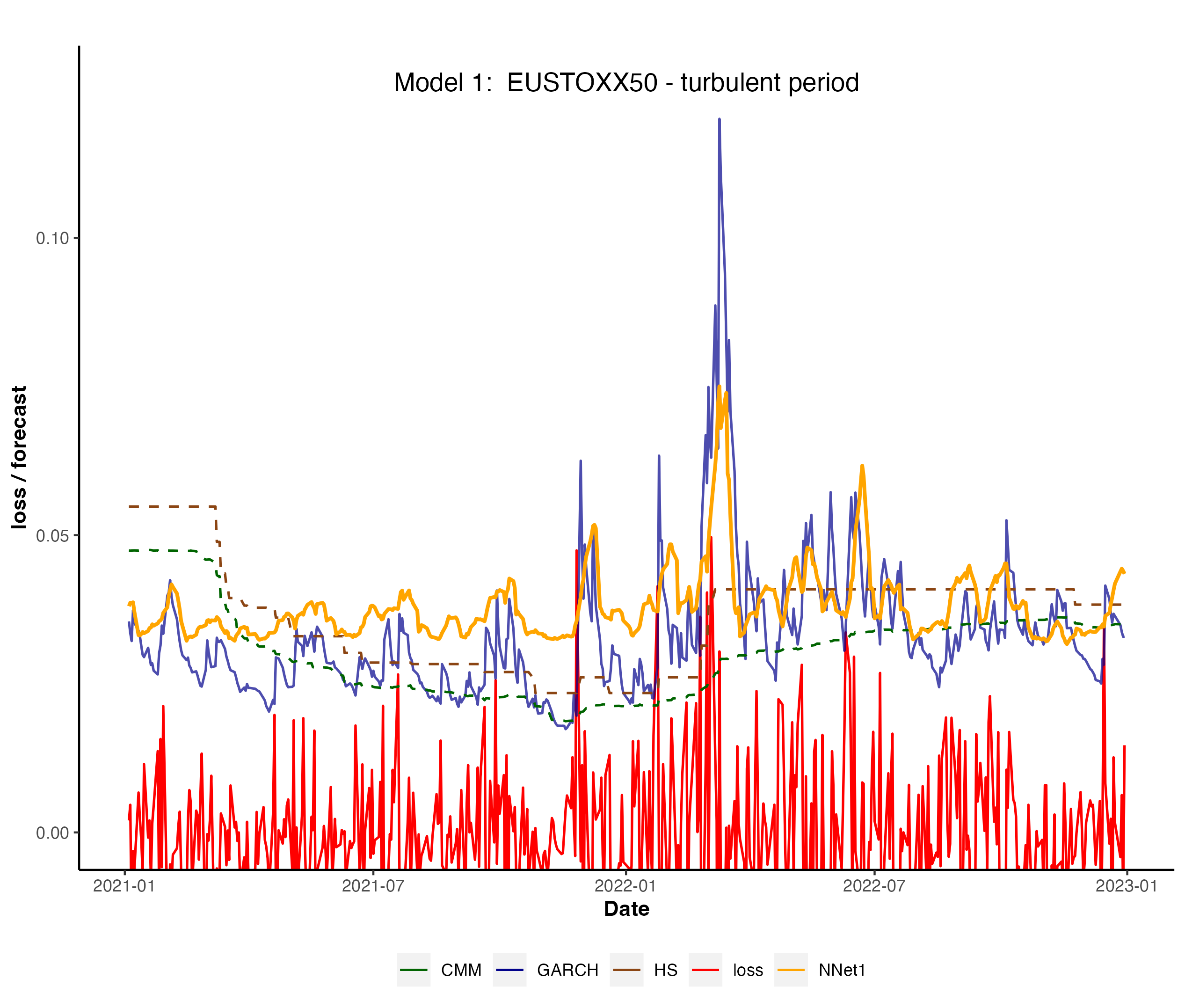}}
\hfill
\subfigure[NNet 2]{\includegraphics[width=7.9cm]{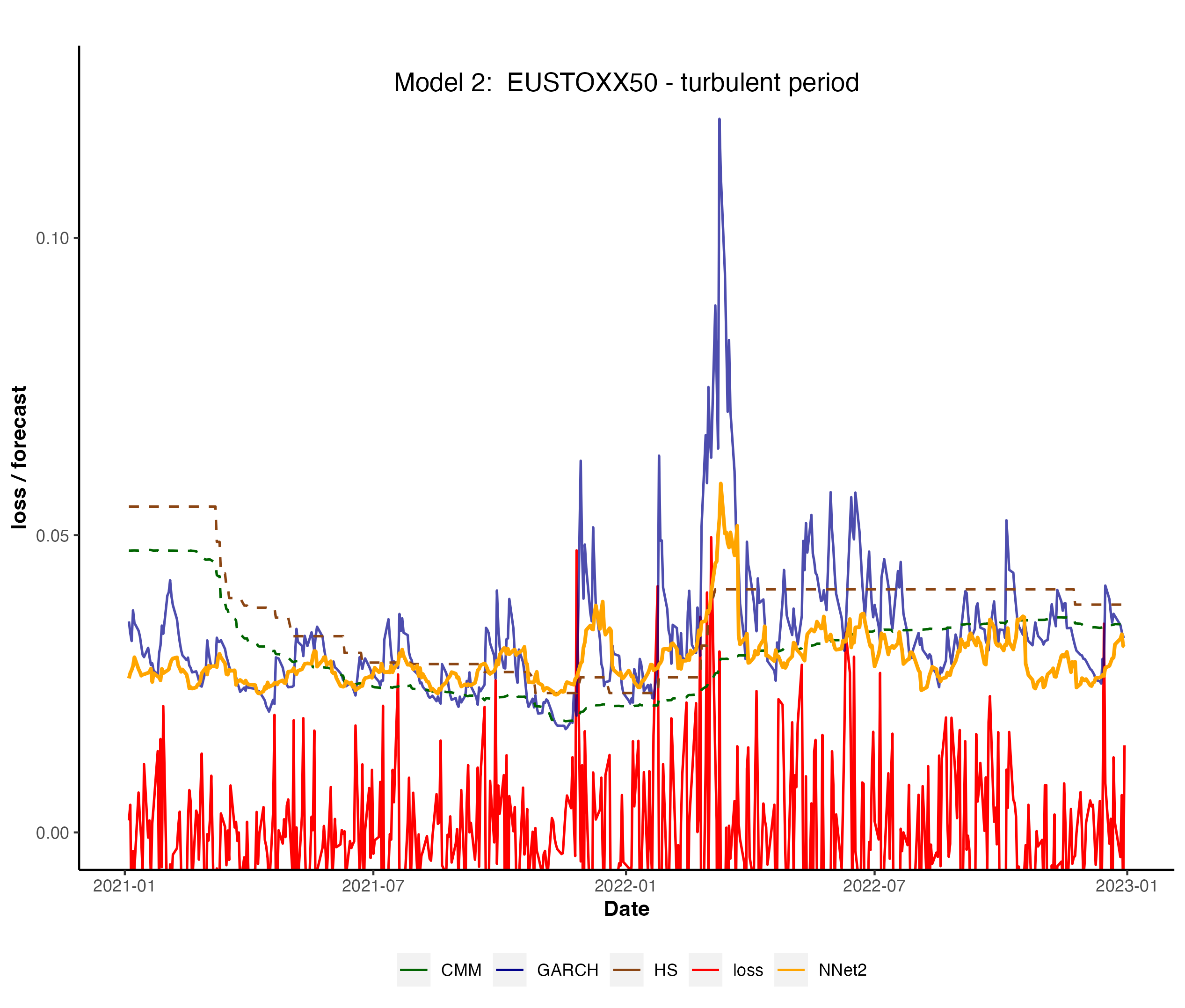}}
\hfill
\subfigure[NNet 3]{\includegraphics[width=7.9cm]{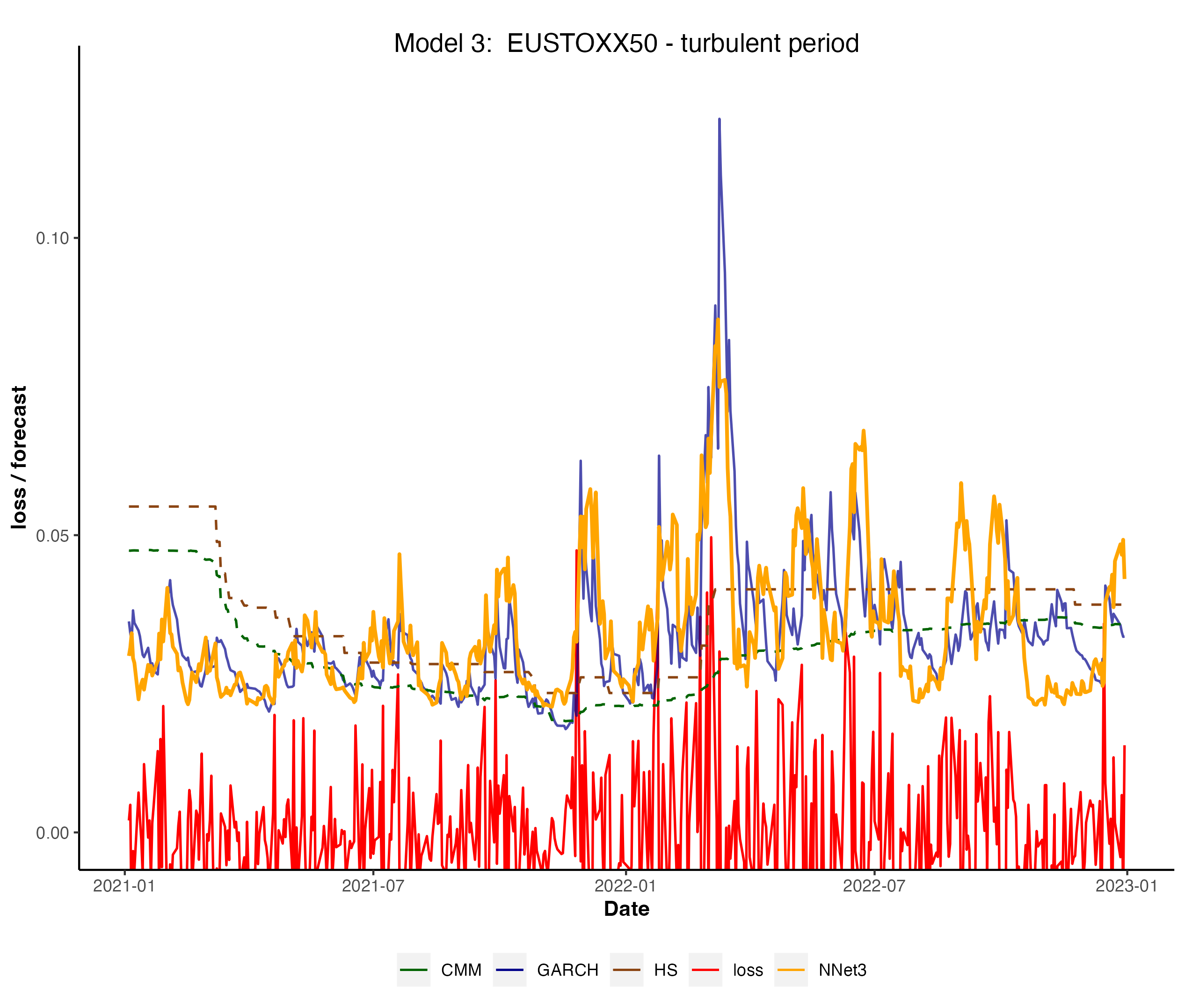}}
\hfill
\caption{VaR forecasts EUROSTOXX 50 - turbulent period}
\label{plot_EUSTOXX_results_turbulent}
\end{figure}

\chapter{Discussion} \label{chapter_discussion}
The following chapter elaborates on results from the previous chapter. First, general findings are discussed with a focus on model implementation and backtesting outcomes. Second, the research questions proposed in section \ref{section_research_questions} are answered. A further subsection proposes possible solutions to the issues of the LSTM-MDN model for VaR forecasting as it is presented in this thesis and suggests further research. 

\section{Findings}
\subsection{Forecast Accuracy \& Model Performance} \label{section_discussion_accuracy}
The backtesting procedure over all six evaluation sets shows mixed results. While some LSTM-MDN models perform well and beat their benchmark counterparts, others significantly underperform or fail to detect any patterns in the data. In contrast to the benchmark models which tend to understimate VaR, the LSTM-MDNs generally behave more conservative. As an example, all benchmark models for the S\&P 500 calm period evaluation strongly underestimate VaR, while NNet 2 and 3 both result in less than 1\% of overshoots. The model assessment also shows that some NNs produce strongly inaccurate forecasts. 
\par \vspace{0.3cm}
One reason could simply lie in a failure to estimate appropriate distribution parameters due to a lack of training data. Although all data sets can be considered large, it is striking that the LSTM-MDN models perform worse during the calm market period (trained on a significantly smaller training set) compared to the turbulent period. The difference in the amount of training data can be seen in table \ref{table_trainset_length}. Considering the joint test as indicator, only 4 out of 9 LSTM-MDNs are deemed appropriate during the calm period compared to 8 out of 9 during the turbulent period. Neural networks trained for the 2017/ 2018 evaluation tend to overestimate the tail risk more often than their counterparts used for the turbulent period and further are more prone to fail to account for volatility patterns in the test data. In comparison, the forecasts during the turbulent period, which are based on larger train sets, tend to be more accurate. A striking example indicating the importance of large training data is the EURO STOXX 50 evaluation for the calm testing period (see section \ref{EUSTOXX_results_calm_section}). All models strictly overestimate Value-at-Risk and result in no breaches over the whole evaluation period. Further, the VaR forecast (figure \ref{plot_EUSTOXX_results_calm}) and the underlying parameter estimates $\hat{\pi}$, $\hat{\sigma}$ and $\hat{\mu}$ (see Appendix A-C) show that the models fail to account for any patterns in the underlying data. The respective models are trained on a data set containing 2189 observations (see table \ref{table_trainset_length}), which is by far the smallest train set used in this thesis. This lack of training data could potentially lead to underfitted models which fail to detect the underlying patterns in the daily returns.
\par \vspace{0.3cm}
Although this finding could be due to an unfortunate coincidence, the results of this work at least indicate that the models tend to perform better when trained on a larger train set. However, a dedicated procedure evaluating model performance for different sizes of train data would be necessary to confirm or falsify this theory.
\begin{table}[H]
\centering
\begin{tabular}{||c c c c||} 
\hline
\textbf{Index} & \textbf{FTSE 100} & \textbf{S\&P 500} & \textbf{EURO STOXX 50}\\ [0.5ex] 
\hline\hline
N train set (calm period) & 3654 & 3601 & 2189 \\ 
\hline
N train set (turbulent period) & 4565 & 4507 & 3090  \\ [1ex] 
\hline
\end{tabular}\par
\caption{Number of observations in each train set}
\label{table_trainset_length}
\end{table}
\par \vspace{0.3cm}
The conservative risk forecasting behaviour of LSTM-MDNs is in line with findings from the literature \cite{buczynski2023garchnet, karlsson2021value, ormaniec2022estimating}. A possible explanation lies in the usage of the ReLU activation function in the LSTM layer \cite{karlsson2021value}. Instead of using the tanh function which maps inputs to an output range [-1, 1], the ReLU output has no upper limitation, which could lead the parameter estimates $\hat{y}$ to be overestimated. Under the usage of ReLU in the LSTM layer, the cell state $C_t$ can thus take any non-negative value, which makes the system prone to outliers as they heavily influence $C_t$ and subsequently the distribution parameter estimates due to the lack of output limitation. To tackle this issue, Karlsson Lille and Saphir (2021) proposed a "shifted model" which accounts for the conservative forecasts by downshifting the estimates.
\par \vspace{0.3cm}
Comparing the three differen LSTM-MDN architectures, the backtesting results show no striking difference in forecast accuracy between the three model architectures nor any other pattern. Compared to the findings of Arimond et al.[2020], the inclusion of a regularization term in the loss function does not result in significantly better forecast accuracy or an increase in the ability to model switches between low-volatile and high-volatile market periods.

\subsection{Reactivity to Volatility Switches} \label{section_reactivity}
Among other question, this thesis is concerned with appropriately accounting for volatility clustering.  GARCH models can account for volatility clustering as shown in section \ref{chapter_GARCH}. The LSTM mechanism is expected to work in a similar fashion as it accounts for time dependence through the inclusion of the prior hidden state $h_{t-1}$ (see section \ref{section_LSTM}). An evaluation of the reactivity to volatility switches is done numerically using a correlation analysis and by visual assessment.  
If the model is able to react to volatility switches, the VaR forecast is expected to be positively correlated with the rolling volatility of losses. The rolling volatility is defined as
\begin{equation}
    \sigma^{(roll)}_{t} = \sqrt{\frac{1}{d-1} \sum_{i=0}^{d-1} (l_{t-i} - \overline{l_t})^2} 
\end{equation}
 for $i = 0,1,...,d-1$, where $d$ is the size of the rolling window and $\overline{l}$ is the mean of ex-post losses $[l_{t-d+1},...,l_{t}]$. The reactivity to short-term volatility changes is tested using a rolling window of $d=5$. Table \ref{correl_table} shows the Pearson correlation coefficient between the rolling volatility of ex-post losses over the respective evaluation period and the VaR forecasts of the neural networks and the GARCH model. 
\par \vspace{0.3cm} 
\begin{table}[H] 
  \begin{tabular}{|c||c|c|c|c||c|c|c|c|}
    \hline
    \multirow{2}{*}{\textbf{Index}} &
      \multicolumn{4}{c}{\textbf{Calm Period}} &
      \multicolumn{4}{c}{\textbf{Turbulent Period}} \\
    & NNet1 & NNet2 & NNet3 & GARCH & NNet1 & NNet2 & NNet3 & GARCH \\
    \hline
    FTSE 100 & 0.220 & 0.384 & 0.400 & 0.658 & 0.388 & 0.319 & 0.608 & 0.661 \\
    \hline
    S\&P 500 & 0.674 & 0.605 & 0.055 & 0.881 & -0.240 & 0.489  & 0.415  & 0.820 \\
    \hline
    EUSTOXX 50 & 0.427 & -0.245 & 0.289 & 0.565 & 0.728 & 0.679 & 0.705 & 0.814 \\
    \hline 
  \end{tabular}\par
\caption{Correlation between rolling volatility ($d=5$) and Value-at-Risk forecasts}
\label{correl_table}
\end{table}
In most cases, the VaR forecasts produced by the LSTM-MDN models show a moderate to strong positive correlation with the rolling volatility, indicating good reactivity. However, the VaR forecasts produced by the GARCH model tend to be stronger correlated. It is assumed that while the LSTM-MDN models clearly show  their ability to account for short-term changes in volatility, they are less reactive than the GARCH model. Further, the test shows a moderate negative correlation with the respective ex-post losses for the   EUSTOXX 2017/2018 NNet 2 forecast as well as the S\&P 500 NNet 1 forecast during 2021/2022. It is expected that this result is caused by a failure of the models to learn the underlying patterns in the data (more on this below).
\par
\begin{figure}[H]
    \centering
    \includegraphics[width=10cm, height = 8cm]{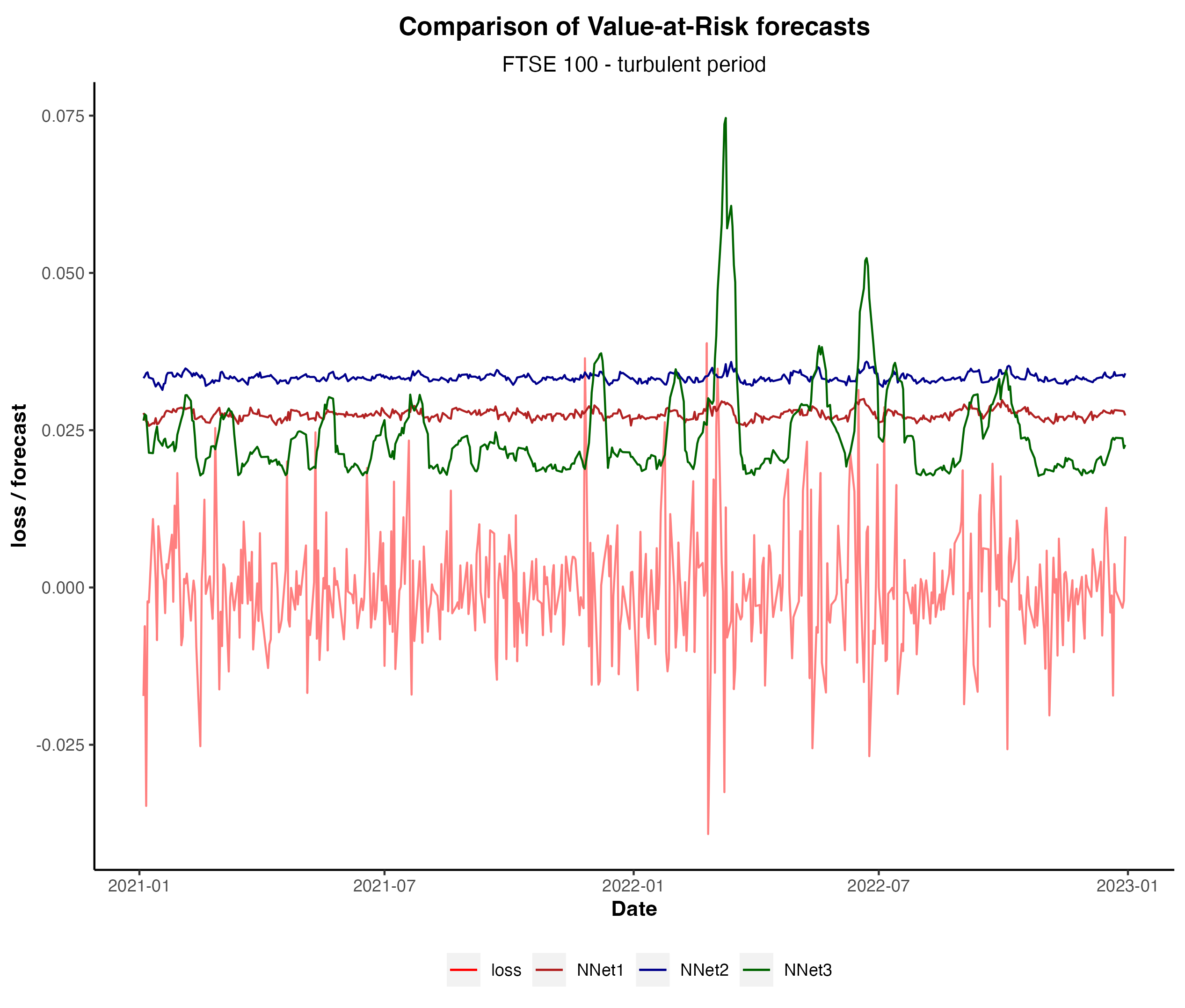}
    \caption{Forecast Comparison  - FTSE 100 (turbulent period)}
    \label{comparison_VaR}
\end{figure}

 \vspace{0.3cm}
A visual assessment is done by graphically analysing the VaR forecasts and the behaviour of their underlying parameter estimates $(\hat{\pi}, \hat{\sigma}, \hat{\mu})$. The forecast (see chapter \ref{chapter_results}) indicate that some models show a good reactivity to changes in the underlying volatility while others fail to adapt to changes. As an example, figure \ref{comparison_VaR} compares the Value-at-Risk forecast from the LSTM-MDN models for the 2021 / 2022 FTSE evaluation. 
Figure \ref{comparison_VaR} shows a large difference in the amplitude of adjustment the models make for fluctuations of the one-day loss. Although NNet 1 and 2 react to shifts in volatility, the amplitude is marginally small. In contrast, NNet 3 shows much better reactivity, which is in line with the results presented in the correlation table \ref{correl_table}.
\par
\begin{figure}[H]
    \centering
    \includegraphics[width=10cm, height = 8cm]{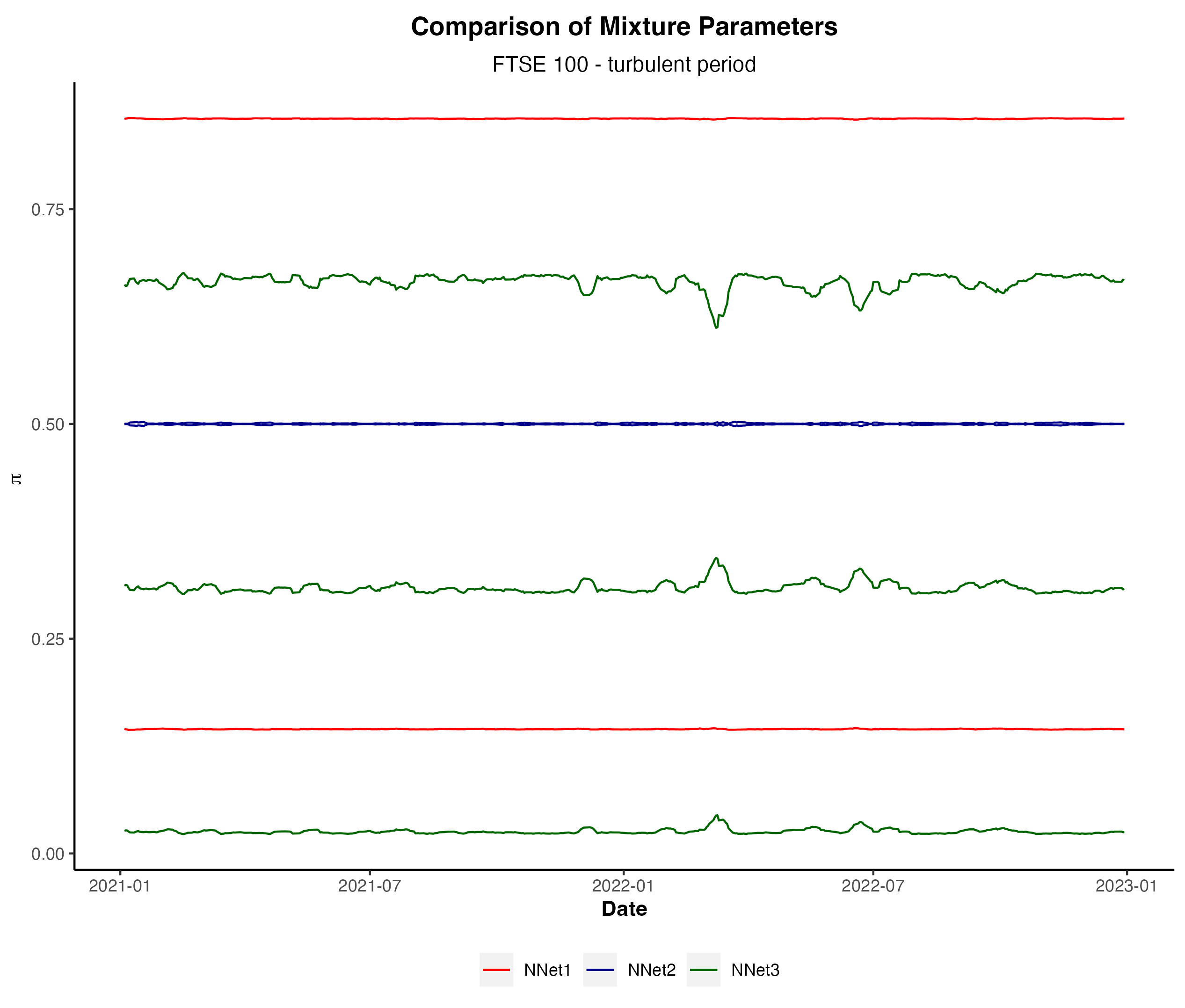}
    \caption{Mixture Parameters- FTSE 100 (turbulent period)}
    \label{comparison_mixture_params_plot}
\end{figure}

 
\pagebreak
Figure \ref{comparison_mixture_params_plot} shows the mixture parameters for all three model architectures. The parameters for NNet 1 approach values of around 0.8 and 0.2, respectively, and stay static over the whole evaluation period. In contrast, the mixture parameters of NNet 2 indeed take balanced values of around 0.5. The estimates for the 3-component model show the strongest variability and tend to behave less static. The plot indicates that the NNet 3 mixture parameters react to volatility switches, with the strongest reaction around March 2022 \footnote{see the increase in volatility at figure \ref{comparison_VaR}}
\par
\begin{figure}[H]
    \centering
    \includegraphics[width=10cm, height = 8cm]{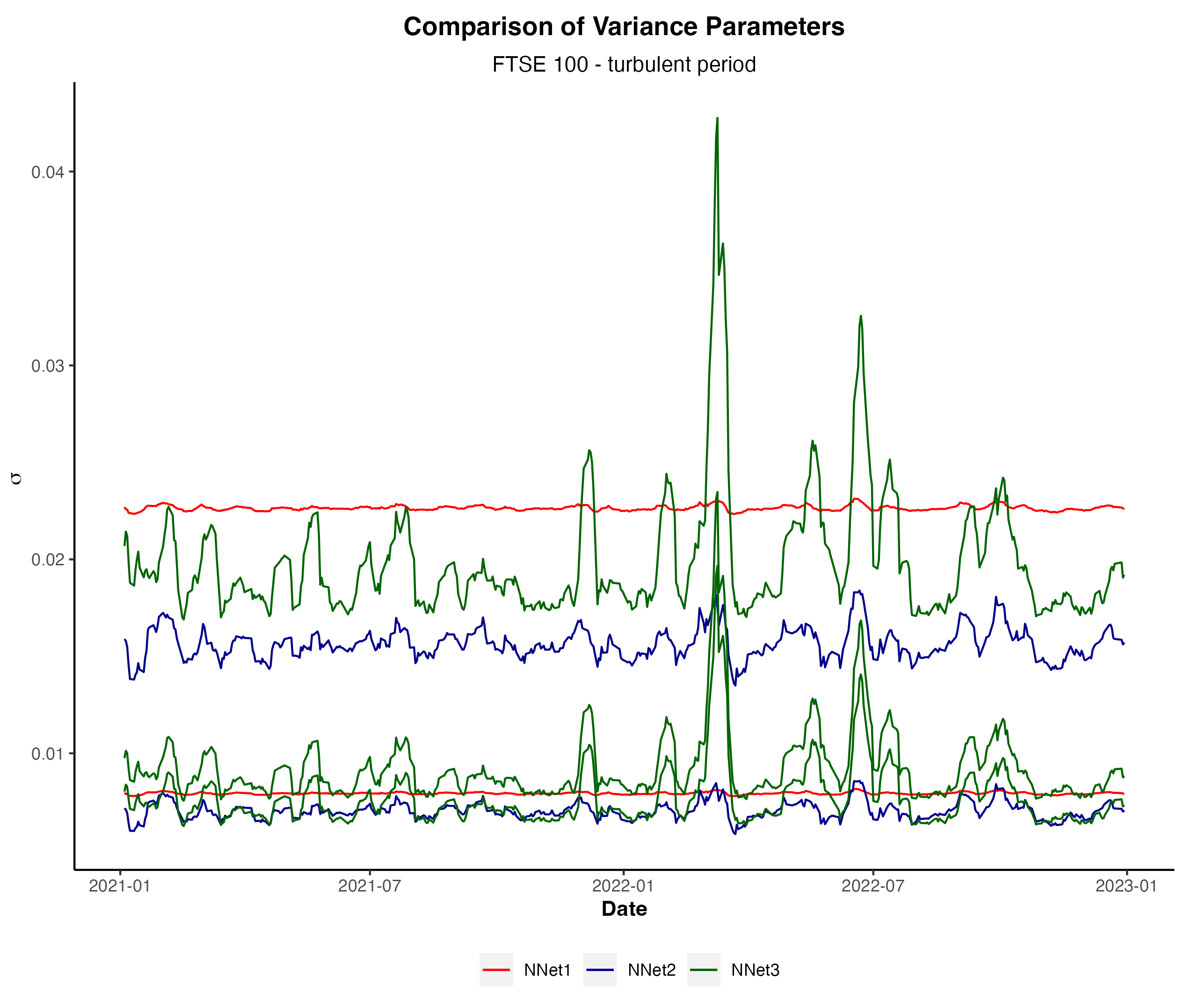}
    \caption{Variance Parameters- FTSE 100 (turbulent period)}
    \label{comparison_sigma_params_plot}
\end{figure}

\begin{figure}[H]
    \centering
    \includegraphics[width=10cm, height = 8cm]{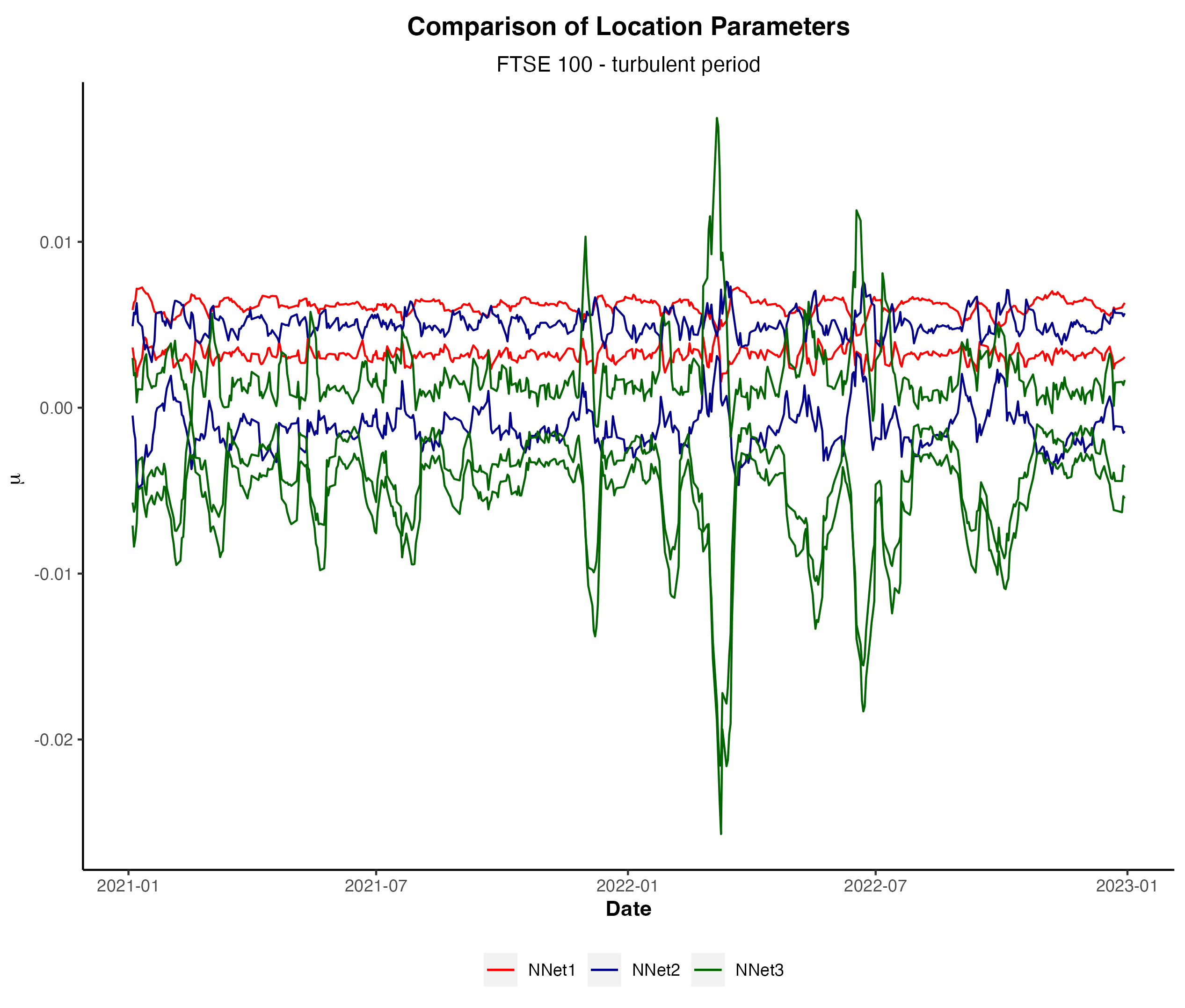}
    \caption{Location Parameters- FTSE 100 (turbulent period)}
    \label{comparison_mu_params_plot}
\end{figure}

 \vspace{0.3cm}
Figures \ref{comparison_sigma_params_plot} and \ref{comparison_mu_params_plot} show the respective variance and location parameters for the three neural networks. While the  $\hat{\sigma}$ estimates from NNet 2 and NNet 3 both show reactivity, the estimate produced by architecture 1 is far more static and barely shows any fluctuation. Again, the response amplitude to volatility changes of NNet 3 is the greatest.
The estimates for $\hat{\mu}$ behave in a similar fashion. While the estimates produced by NNet 3 show the strongest reactivity, the amplitude in changes from NNet 1 and 2 are much smaller, while the location parameter from NNet 1 tends to stay relatively static over the evaluation period.  

\par \vspace{0.3cm}
Figure \ref{comparison_mixture_params_plot} - \ref{comparison_mu_params_plot} shows that the ability to react to volatility switches is mainly achieved by the non-static behaviour of $\hat{\sigma}$ and $\hat{\mu}$ rather than modeling switches through the mixture parameter\footnote{see Appendix A-C, which contain the parameter estimates produced by each neural network.}. Generally, the mixture parameters from NNet 1 tend to converge to strongly imbalanced values, where the dominant component mostly takes a value between 0.9 and 0.999 (see Appendix \ref{Appendix1}). The mixture parameter estimates tend to not largely vary over time and thus stay static. The L2 regularization under NNet 2 balances the two parameters to a value range of around 0.5. However, the mixture estimates using a regularization term also tend to behave static, which contradicts the results produced by comparable models from the literature \cite{arimond2020neural, karlsson2021value}. An evaluation of the mixture components produced by the NNet 3 architecture shows that one component tends to be dominant, taking values between 0.8 and 0.95, while another mixture parameter tend to approach a marginally small value. This behaviour can be interpreted as the model trying to account for strong outliers with one component while the other two components account for the majority of returns. In some cases, the least dominant component approaches 0 which practically reduces the 3-component mixture model to a 2-component model\footnote{see fig. \ref{plot_appendix1_1}(a) and (c) as example}. In comparison with the non-regularized 2-component architecture (i.e., the NNet 1 models), the mixture parameters from the 3-component model tend to be slightly more balanced. 
\par \vspace{0.3cm}
The correlation analysis and the visual example above show that the ability to account for volatility clustering can largely differ even for models which are trained on the same data. As seen in figure \ref{comparison_VaR} - \ref{comparison_mu_params_plot}, NNet 3 accounts far better for volatility switches compared to its peers. A possible explanation for these discrepancies could lie in the model optimization process and a strong dependence on weight initialization.  As described in section \ref{chapter_optimization}, the optimization process of a DNN is a non-convex optimisation problem where the surface of the loss function consists of several local optima. A neural network which converges to a local minimum or saddle point could fail to capture the underlying patterns appropriately. This theory is further elaborated on in the following section \ref{section_weight_init}.

\subsection{Weight Initialisation and Parameter Optimization} \label{section_weight_init}
As described in the Methodology chapter, weight initialization is done by a best-of-three approach (i.e., initializing via three different seeds and picking the best result). For some backtesting procedures, the three different initial parameter vectors $\theta_0$ led to significantly different result. This leads to the assumption that a proper weight initialization procedure is a key factor for ensuring high-quality and reliable forecasts. The urge for a well defined weight initialization process is expected to be linked to the non-convexity of $L$. Findings from \cite{tian2022understanding} show that even single-layer neural networks tend to introduce local optima due to the usage of non-linear activation functions. As a possible explanation it is assumed that some initial weight parameters from $\theta_0$ (depending on the seed value used,  see section \ref{section_model_architecture}) led to a convergence in a local minimum which is far away from the global optimum and thus resulted in a model which produces unreliable and inaccurate forecasts, is unable to recognize underlying patterns in prior returns and further fails to account for volatility clustering. Findings from \cite{arimond2020neural} confirm the importance of weight initialization for optimal model performance in the context of VaR forecasting.
\par \vspace{0.3cm}
Weight initialization is a well known issue in the area of machine learning and thus a variety of research is being done on finding appropriate solutions (e.g., \cite{kumar2017weight} or \cite{mishkin2015all}). The \textit{He-initializer} \cite{he2015delving} could be a promising option as the method is originally developed for the initialization of layers using ReLU activation, which is the case for the networks presented in this thesis. Another widely used approach is the initialization with weights following a Gaussian distribution with zero mean and a given variance ($\theta_0 \sim \mathcal{N}(0, \sigma^2)$), where active research is concerned with the optimal value for $\sigma^2$ \cite{kumar2017weight}. However, the issue of neural network training due to non-convex loss functions is a prevailing issue in the field of ML. A deeper dive into the topic of weight initialization and the optimization of non-convex functions would exceed the limitations of this work.

\section{Research Questions}
\subsection{Question 1: Comparison with established models}
When comparing the benchmark models with the LSTM-MDNs in terms of forecast accuracy, the six backtesting results paint a mixed picture. On average, the neural network approach performs comparably well during the turbulent period. It is worth mentioning that for the S\&P 500 evaluation in 2017/2018, the only models which can be deemed as appropriate assuming a confidence level of 5\% are NNet 2 and NNet 3. On the other hand, the networks are strongly outperformed for the calm period evaluation of the EURO STOXX 50 set. 
\par \vspace{0.3cm}
Overall, the results indicate that LSTM-MDN models tend to perform well and are able to outperform simpler approaches if a) the model is fitted on a large enough train set and b) the underlying evaluation period is highly volatile. While further tests are necessary to confirm claim a), the superiority of LSTM-MDNs over simpler approaches like CMM or historical simulation during periods of high volatility can be explained with its capability to account for recent shocks at $t-1$ due to the LSTM mechanism. However, ARCH-type models are also able to account for recent shocks and thus can account for volatility switches while providing a much simpler structure compared to a neural network. The results do not indicate a superiority of the neural network approach compared to the GARCH(1,1) model.

\subsection{Question 2: Practicality}
Building up on the mixed results from question 1, this section discusses the practicality of an LSTM-MDN implementation for risk management. A first consideration is the complexity of an implementation, which includes computational time and data availability. As described in chapter \ref{chapter_optimization}, computational cost is dependent on several factors. However it is save to say that training LSTM-MDNs is much more time and cost expensive than the presented benchmark models. Further, neural networks generally rely on large sets of training data to adequately produce reliable predictions \cite{raudys1991small}. Assuming this is the case for the MDN-LSTM architecture used for VaR forecasting, model training would require the availability of (daily) asset prices over a long time period. Such a large amount of ex-post returns might not be available for a specific stock or asset. 
\par \vspace{0.3cm}
Another aspects which speaks against the practicality of the neural network approach is the lack of research and validation done on these models for financial forecasting. The first paper\footnote{to the best of the author's knowledge} using an LSTM-MDN for Value-at-Risk forecasting was the published around 3 years ago \cite{arimond2020neural}. More research over a longer time frame is needed to establish best-practices for model implementation and further compare model performance with benchmark models. A large concern for practical usage can be placed on the issues described in section \ref{section_weight_init}. Without an appropriate initialization strategy, a successful implementation of an LSTM-MDN would just perform well "on the off chance". 
\par \vspace{0.3cm}
Although the literature as well as this thesis show given potential in the usage of LSTM-MDNs for risk management under given conditions, their usage by financial institutions cannot be  considered appropriate at this point. Again, comparing the neural network approach with the far simpler GARCH model considering both model performance (accuracy) and aspects of practicality, one would have a hard time to argue in favour of the complex LSTM-MDN approach as presented in this thesis.

\subsection{Question 3: Volatility Clustering}
Findings clearly indicate that the LSTM-MDN architecture is able to account for volatility clustering in a comparable way as models from the ARCH family. Figure \ref{comparison_mixture_params_plot} and \ref{comparison_sigma_params_plot} (as well as Appendix A-C) show that this behaviour is achieved by adapting the conditional variance and location parameter estimates rather than modeling regime switches through the adaption of the mixture component estimate $\hat{\pi}$, which tend to behave more static. On average, the risk forecasts produced by the neural networks tend to show a strong to moderate correlation with the ex-post losses (section \ref{section_reactivity}), underlining their ability to account for volatility clustering. 

\section{Outlook} \label{section_outlook}
Section \ref{chapter_discussion} identifies the size of training data and the dependence on proper weight initialization as main issues of the neural network approach. Mitigating the data requirements could be achieved with regularly refitting the neural networks on all data which is available at time point $t$. A backtesting approach with a refitting strategy would refit the model on a regular intervals (e.g., every 20 time steps) on all available observations until this time point and use the fitted model to forecast for the next interval. The procedure ensures that even during the evaluation period, the train set increases. However, the refitting would come with an increase in computational expenses. Another option is data augmentation, which uses artificially generated data points to enlarge the train set. Research \cite{hellermann2022financial, fons2020evaluating}  show promising results using this technique on financial time series. 
\par \vspace{0.3cm}
Another crucial factor is the dependence on weight initialization, where section \ref{section_weight_init} lists possible options. Arimond et al. (2020) used parameter estimates from a Hidden Markov model in combination with a uniform sampling method as weight initialization, which led to increased accuracy compared to the default method. To the best of the author's knowledge, there is no other method known which appropriately solves this issue. Further research should be concentrated on developing an initialisation strategy for LSTM-MDNs used in the context of financial forecasting, as it is expected to significantly improve model performance.
\par \vspace{0.3cm}
Further, the usage of a mixture model composed of leptokurtic component distributions could be of interest. As explained by \cite{bishop1994mixture}, a mixture density network can model any arbitrary chosen continuous conditional probability distribution. Due to the fact that daily returns follow a leptokurtic distribution \cite{sewell2011characterization}, the usage of GED component distributions or t-distributions could potentially lead to promising results. 

\chapter{Conclusion}
This work investigates the potential of Long Short-Term Memory mixture density networks for the usage of Value-at-Risk forecasting, both from a perspective of a standardized backtesting procedure against established benchmark models and an evaluation of the models' ability to account for volatility clustering. Although the machine learning approach outperforms its benchmark counterparts in some cases and shows promising results for market periods with a higher level of uncertainty (i.e., a higher volatility), the results do not indicate a general superiority of the neural network approach compared to simpler and more established models. Promising results can be reported for the neural network's ability to account for shifts in volatility (\textit{volatility clustering}). As seen in section \ref{section_reactivity}, the neural network architecture is able to account for and react to short-term changes in asset price volatility due to its recurrent LSTM mechanism. However, a correlation analysis (see table \ref{correl_table}) does not indicate a better reactivity than a GARCH(1,1) model.
\par \vspace{0.3cm}
Findings show that model performance is highly dependent on an appropriate initialization procedure, which is a key issue in terms of practicality (see section \ref{section_weight_init}). It is assumed that neural networks without a sufficient initialization process run the risk of converging to a local minimum during the optimization process which is far away from the global optimum, leading to a model which is not able to capture underlying patterns in the data and thus produces unreliable and highly inaccurate forecasts. Further, results indicate that LSTM-MDNs require a large amount of training data to appropriately account and "learn" underlying patterns from the data. A lack of training data makes the neural networks, which generally tend to produce conservative risk forecasts \cite{arimond2020neural}, even more prone to strongly overestimating tail risk.
\par \vspace{0.3cm}
In summary, the findings of this thesis indicate that LSTM-MDN models show limited promise for a practical risk management implementation, at least without any further research or model adjustments as suggested in section \ref{section_outlook}. The results of this thesis underline the general consent from the literature that model complexity and machine learning does not necessarily translate to a better ability to forecast market risk \cite{two_sigma}.
%-------------------------------------------------------------------------
%----------------------- APPENDIX & Bibliography -------------------------

% APPENDIX
\appendix 
\chapter{Appendix 1 - Mixture Parameter Estimates} \label{Appendix1}
\begin{figure}[H]
\subfigure[Mixture Parameters - FTSE (calm period)]{\includegraphics[width=7.9cm]{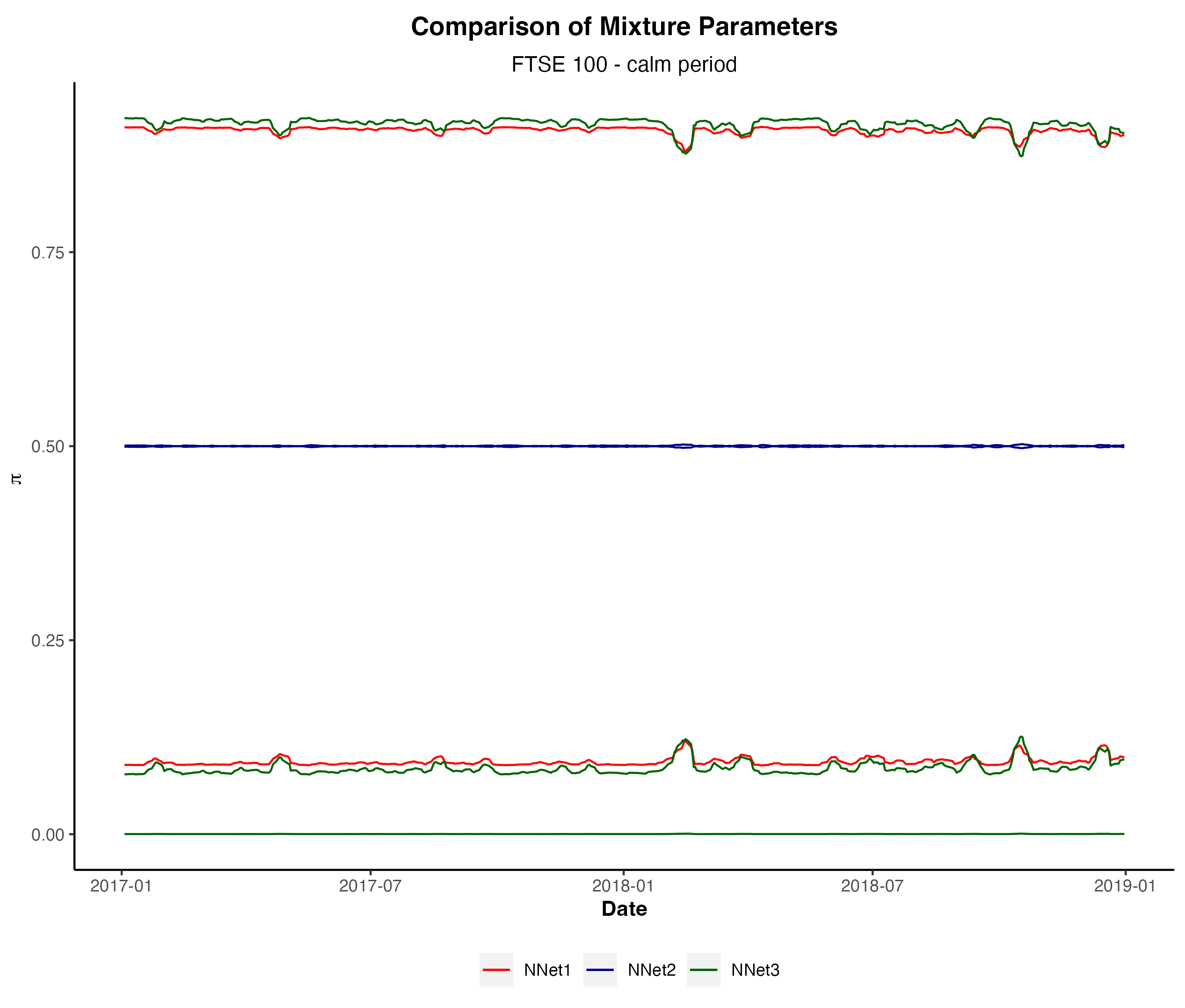}}
\hfill
\subfigure[Mixture Parameters - FTSE (turbulent period)]{\includegraphics[width=7.9cm]{Images/pi_params_FTSE_covid.png}}
\hfill
\subfigure[Mixture Parameters - S\&P (calm period)]{\includegraphics[width=7cm]{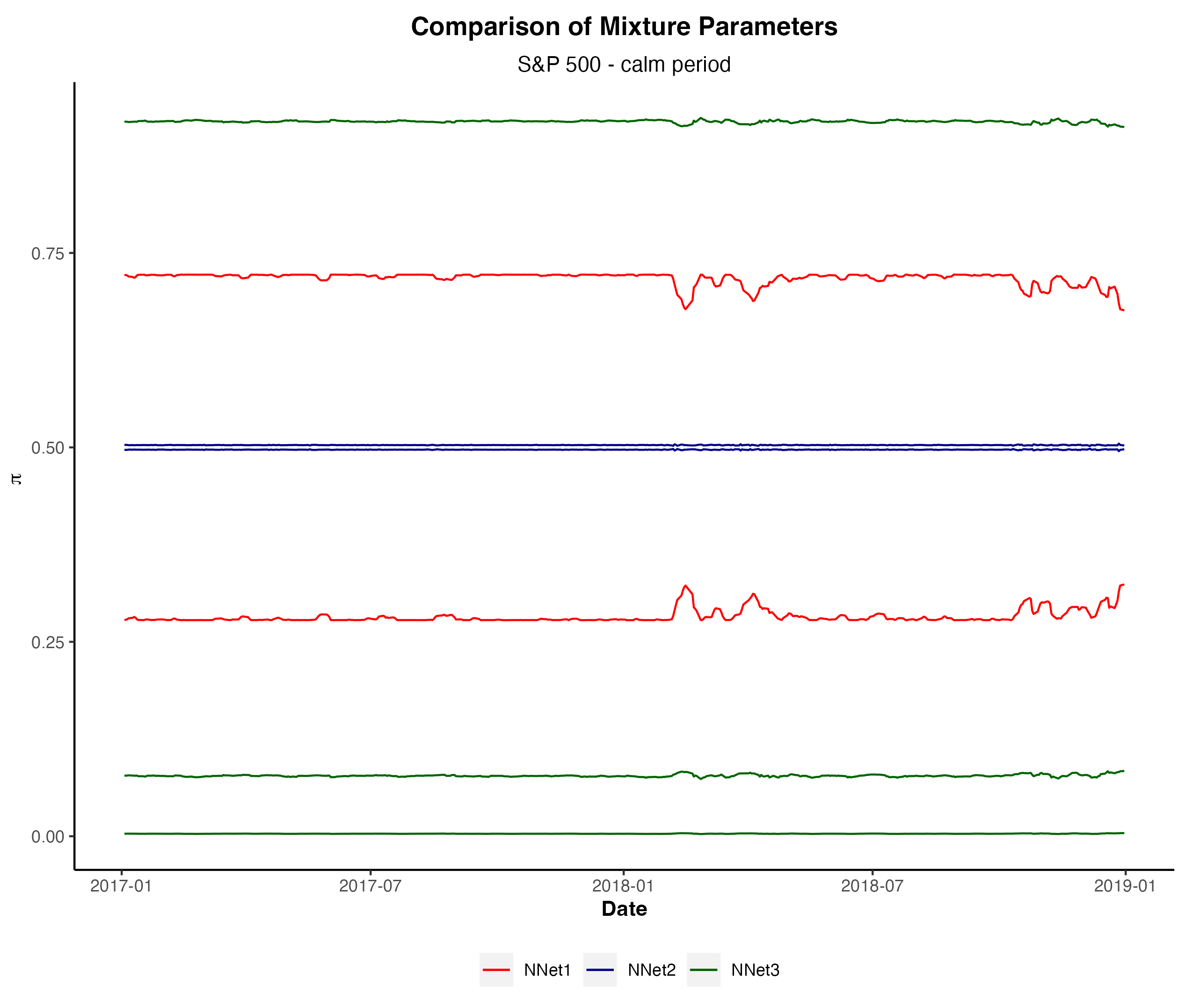}}
\hfill
\subfigure[Mixture Parameters - S\&P (turbulent period)]{\includegraphics[width=7.9cm]{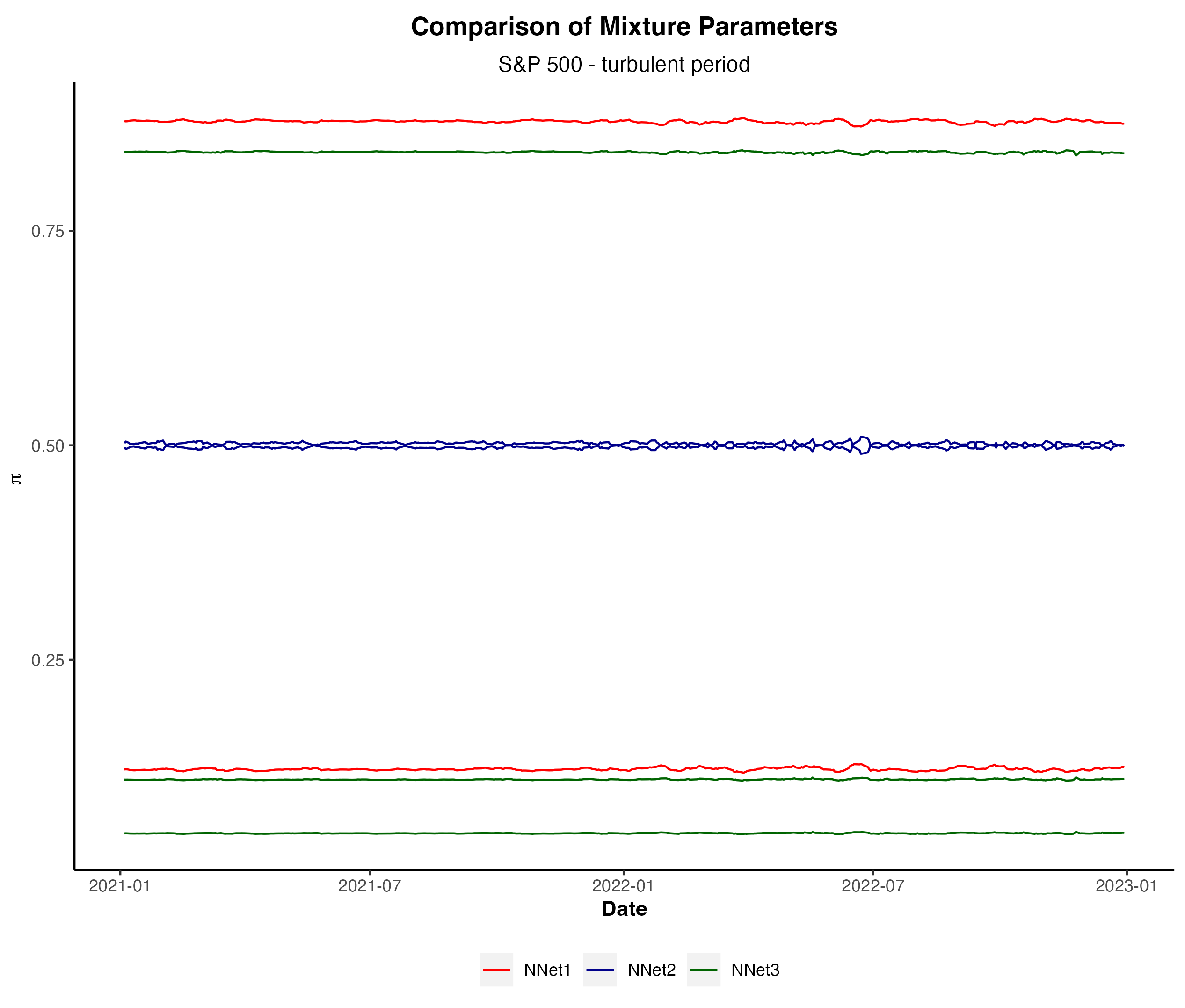}}
\hfill
\caption{Mixture Parameters (1/2)}
\label{plot_appendix1_1}
\end{figure}

\begin{figure}[H]
\subfigure[Mixture Parameters - EUSTOXX (calm period)]{\includegraphics[width=7.9cm]{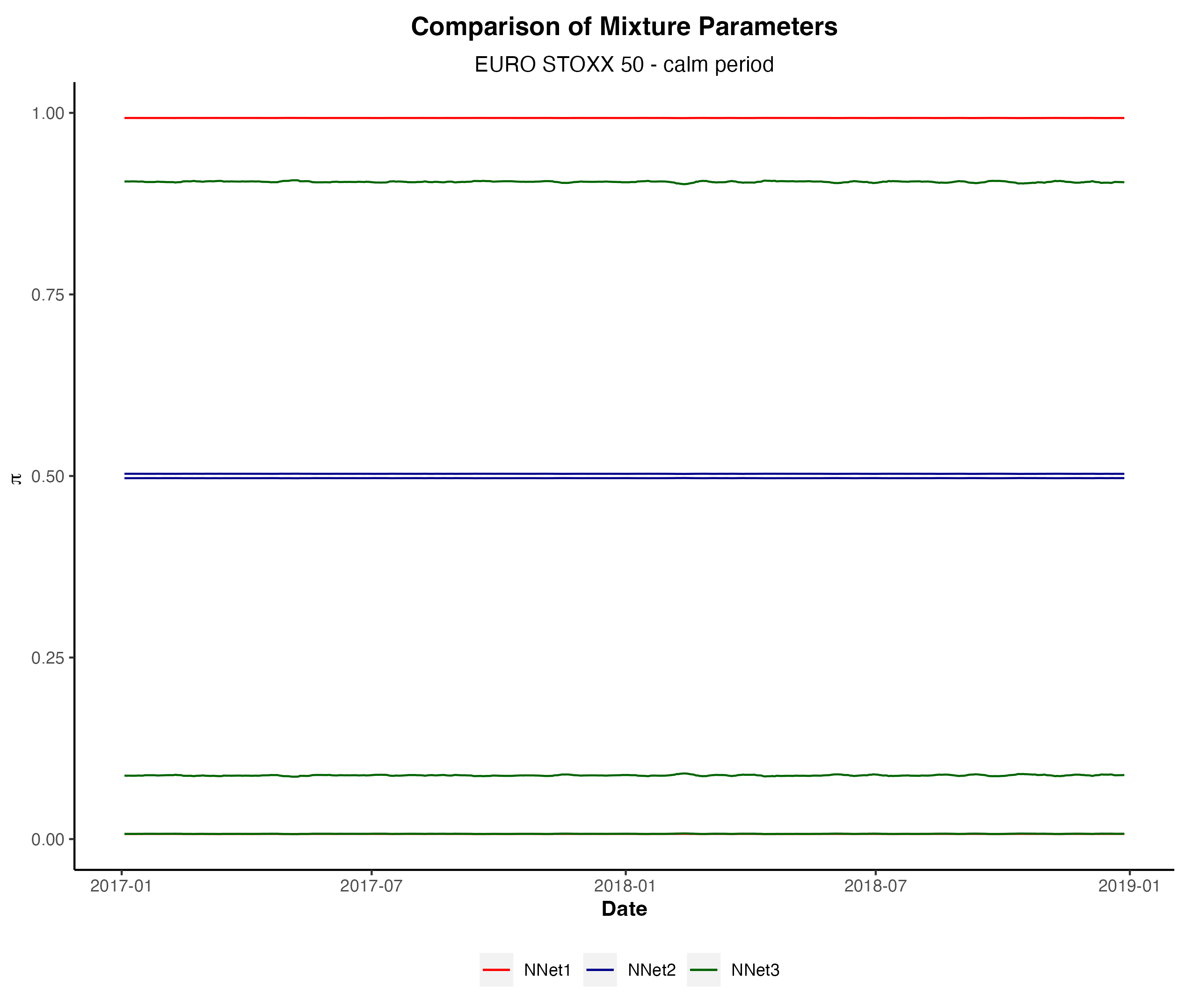}}
\hfill
\subfigure[Mixture Parameters - EUSTOXX (turbulent period)]{\includegraphics[width=7.9cm]{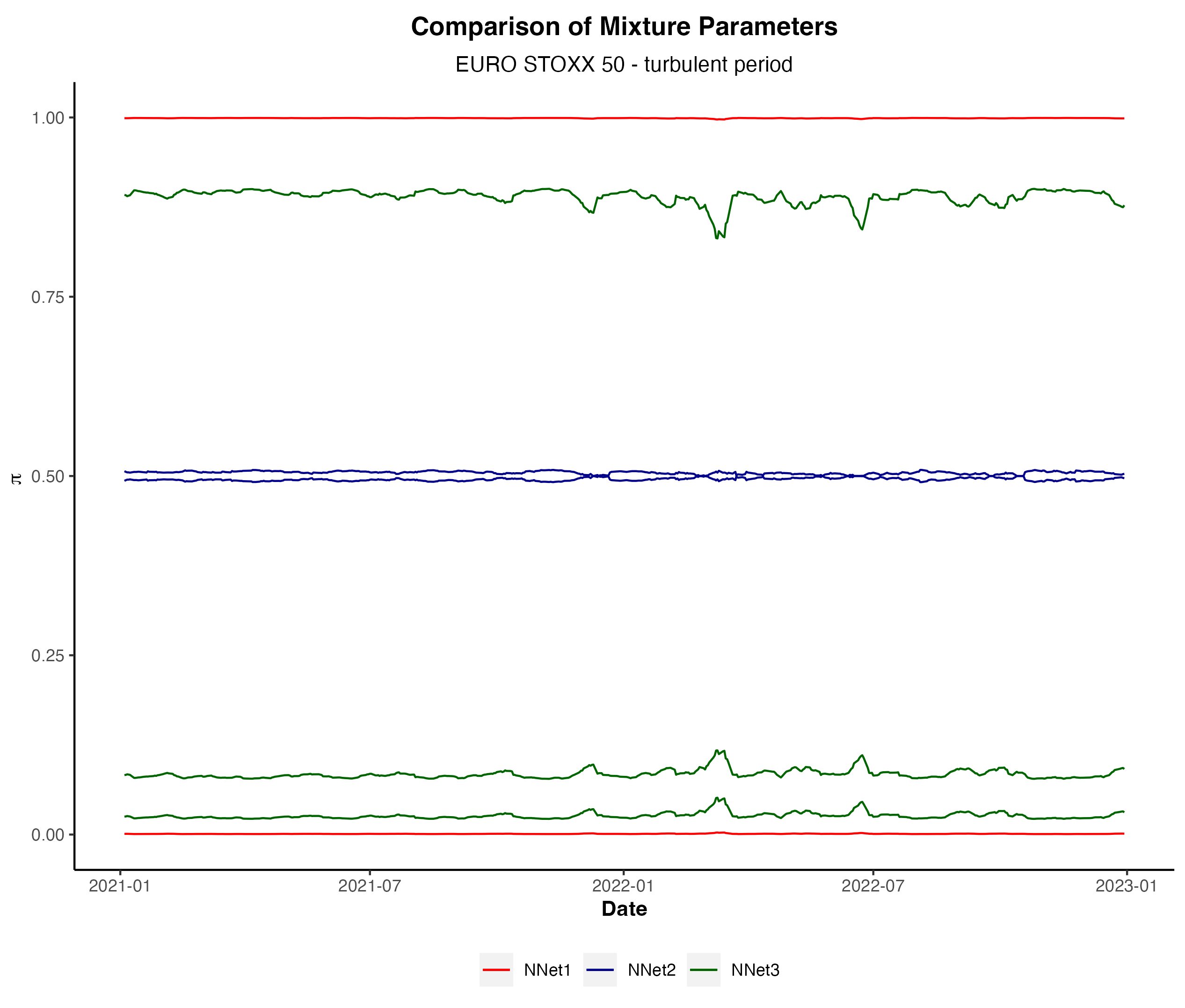}}
\hfill
\caption{Mixture Parameters (2/2)}
\end{figure}
\chapter{Appendix 2 - Variance Parameter Estimates} \label{Appendix2}
\begin{figure}[H]
\subfigure[Variance Parameters - FTSE (calm period)]{\includegraphics[width=7.9cm]{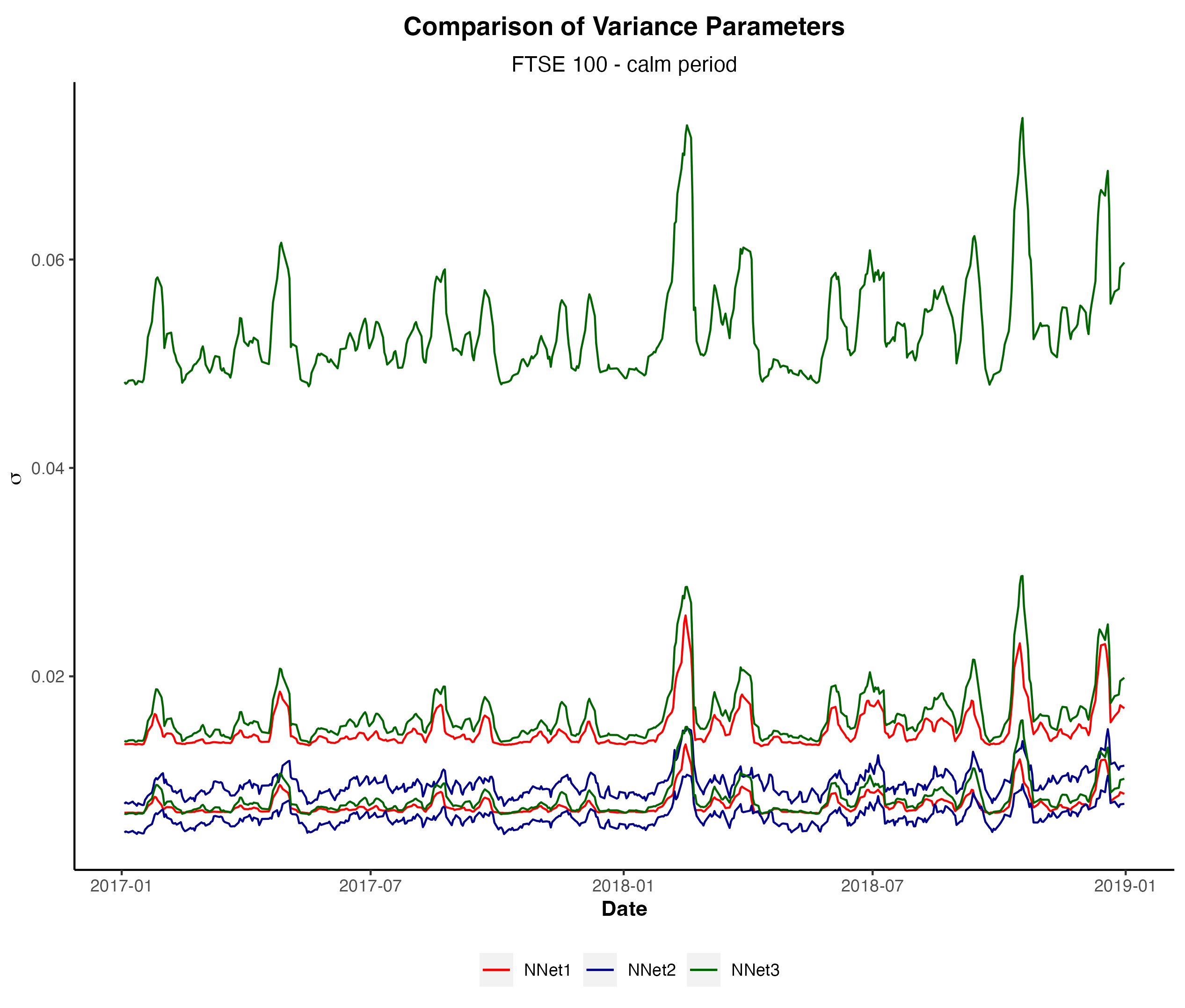}}
\hfill
\subfigure[Variance Parameters - FTSE (turbulent period)]{\includegraphics[width=7.9cm]{Images/sigma_params_FTSE_covid.png}}
\hfill
\subfigure[Variance Parameters - S\&P (calm period)]{\includegraphics[width=7.9cm]{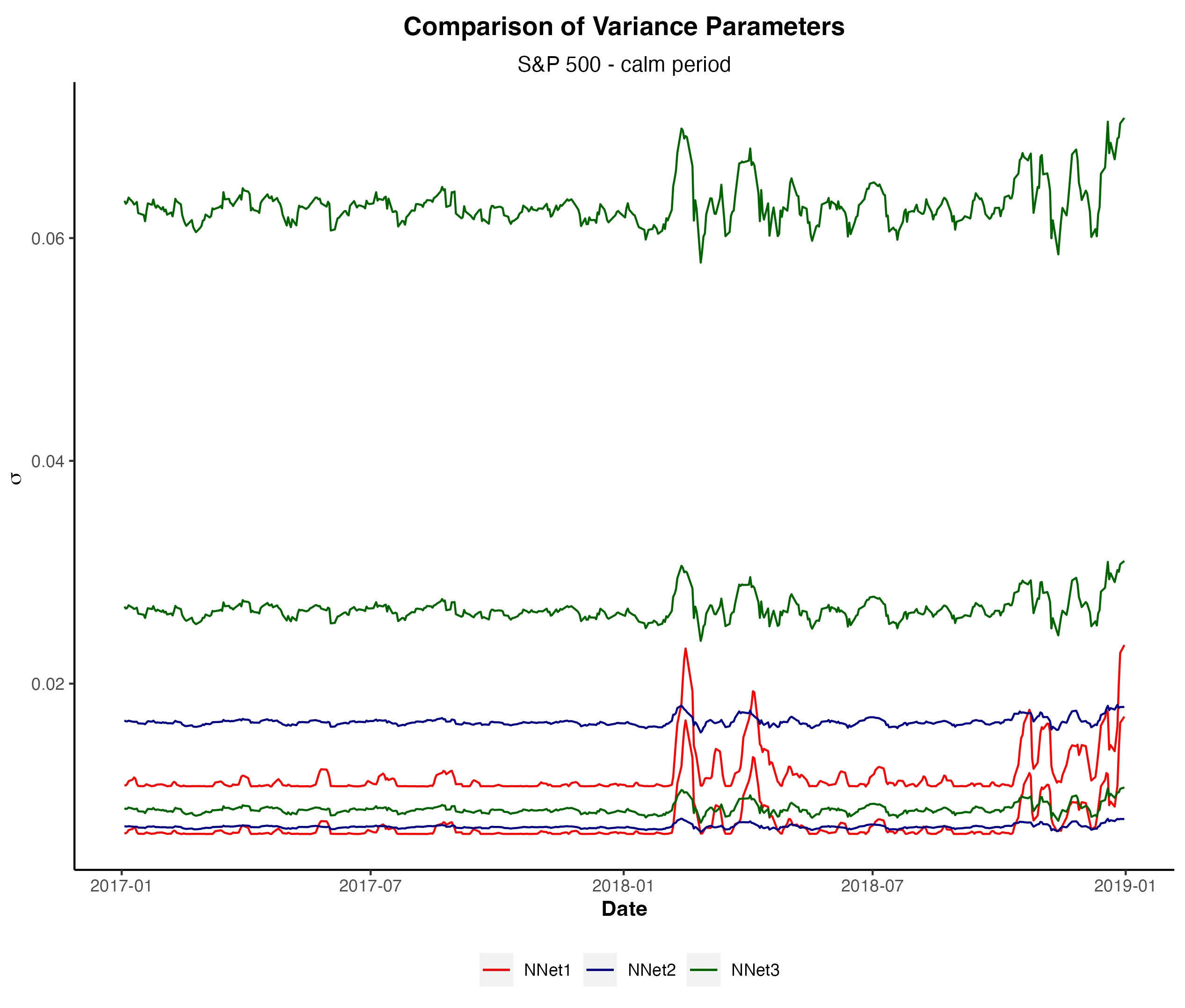}}
\hfill
\subfigure[Variance Parameters - S\&P (turbulent period)]{\includegraphics[width=7.9cm]{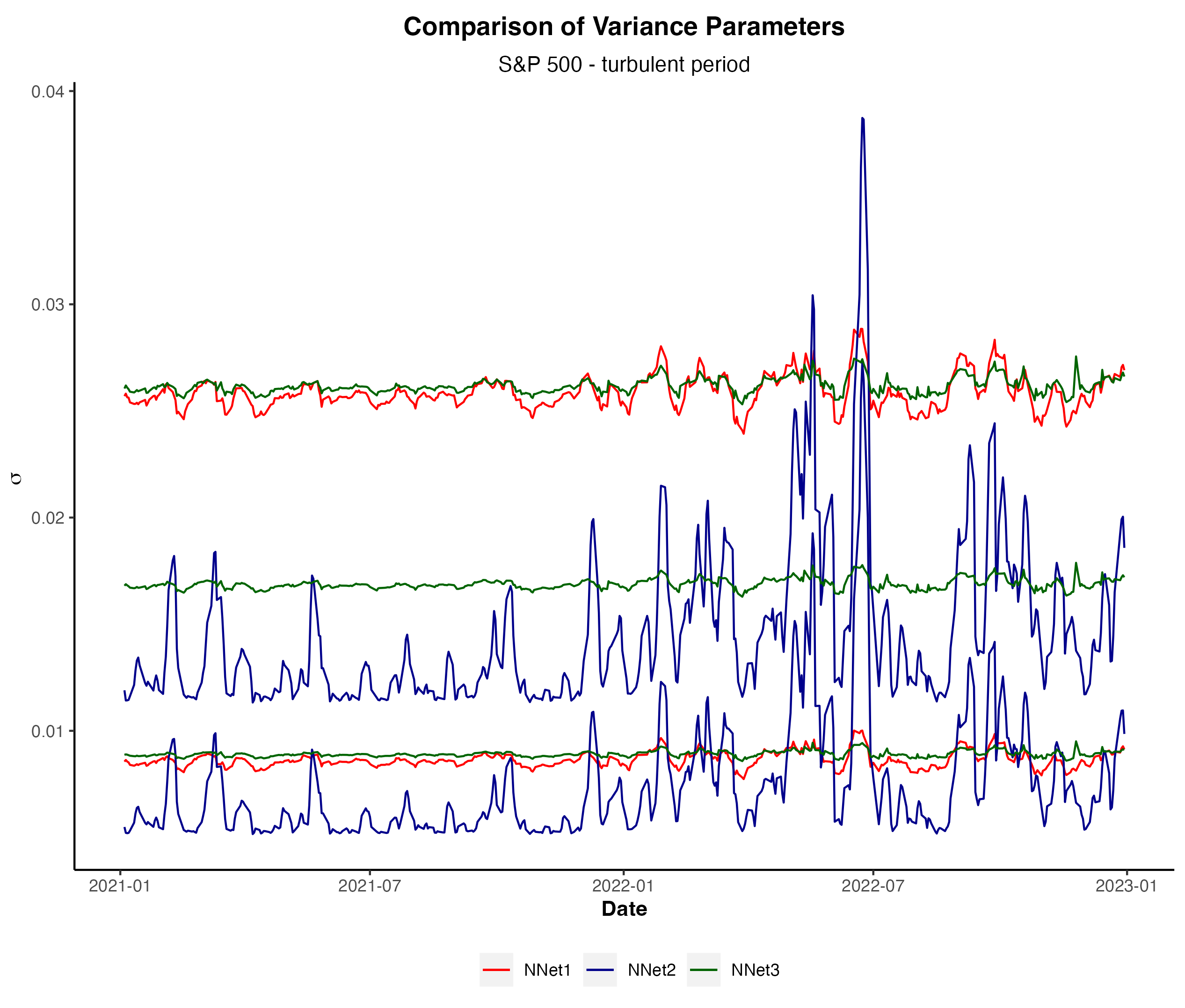}}
\hfill
\caption{Variance Parameters (1/2)}
\end{figure}

\begin{figure}[H]
\subfigure[Variance Parameters - EUSTOXX (calm period)]{\includegraphics[width=7.9cm]{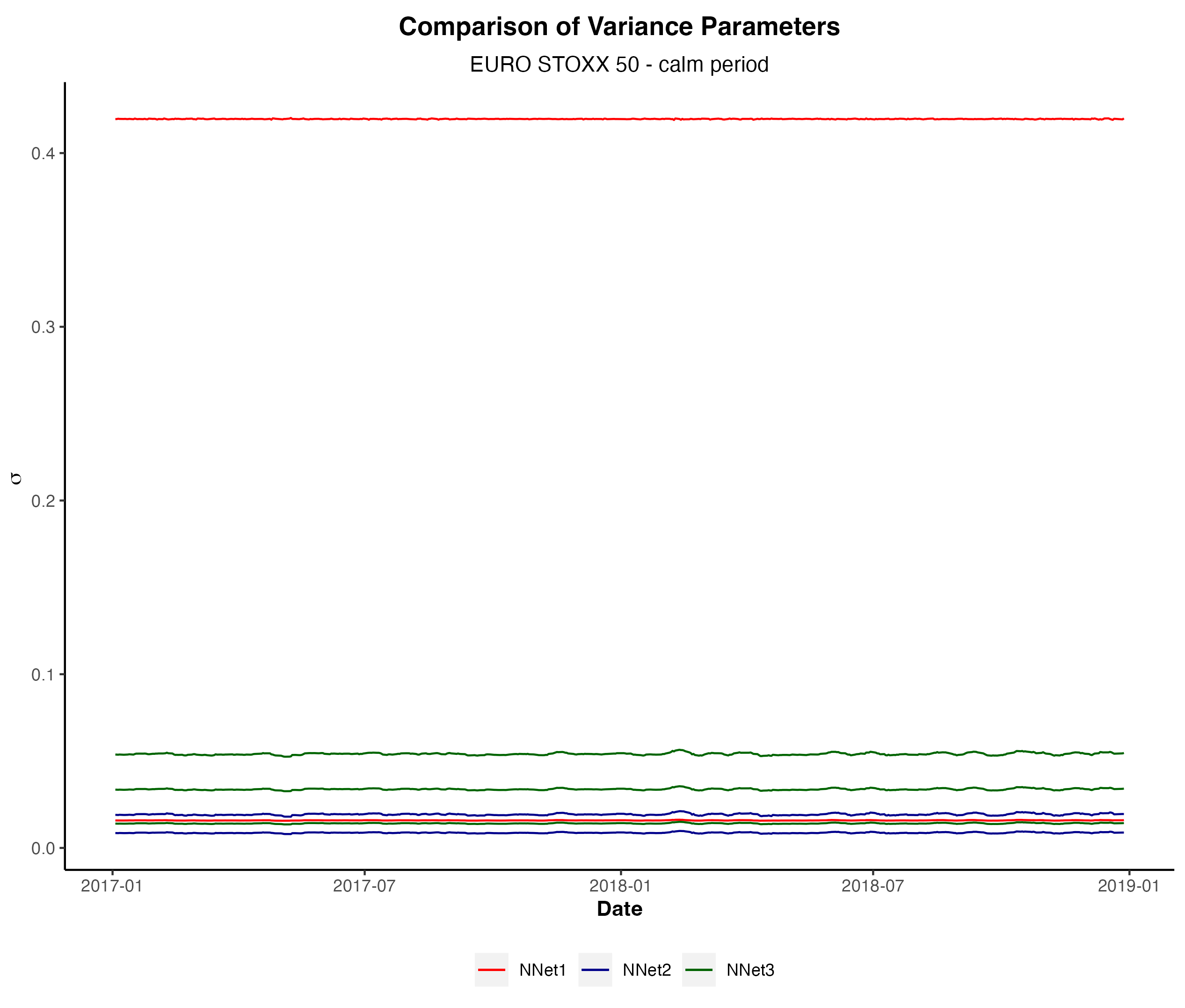}}
\hfill
\subfigure[Variance Parameters - EUSTOXX (turbulent period)]{\includegraphics[width=7.9cm]{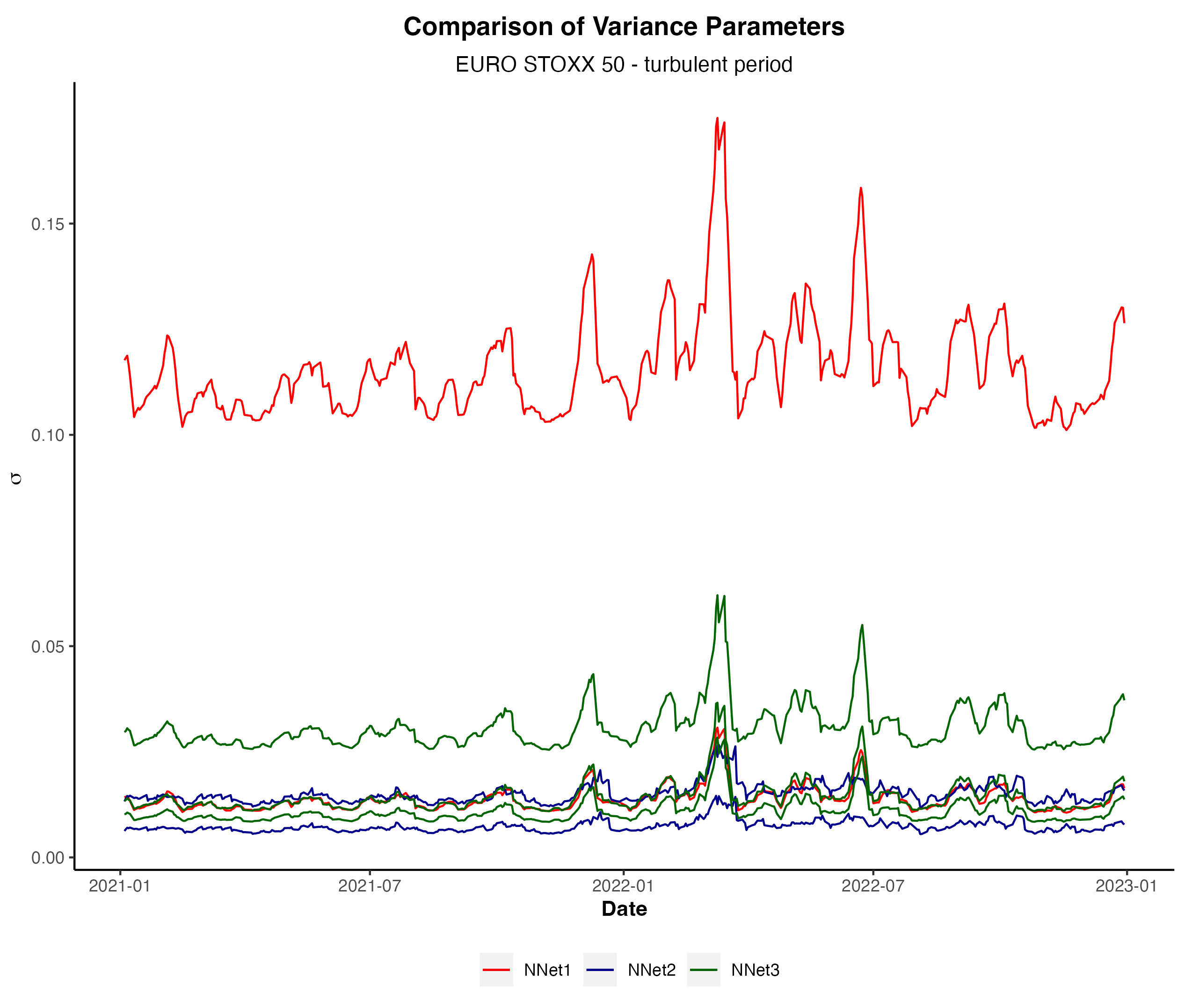}}
\hfill
\caption{Variance Parameters (2/2)}
\end{figure}
\chapter{Appendix 3 - Location Parameter Estimates} \label{Appendix3}
\begin{figure}[H]
\subfigure[Location Parameters - FTSE (calm period)]{\includegraphics[width=7.9cm]{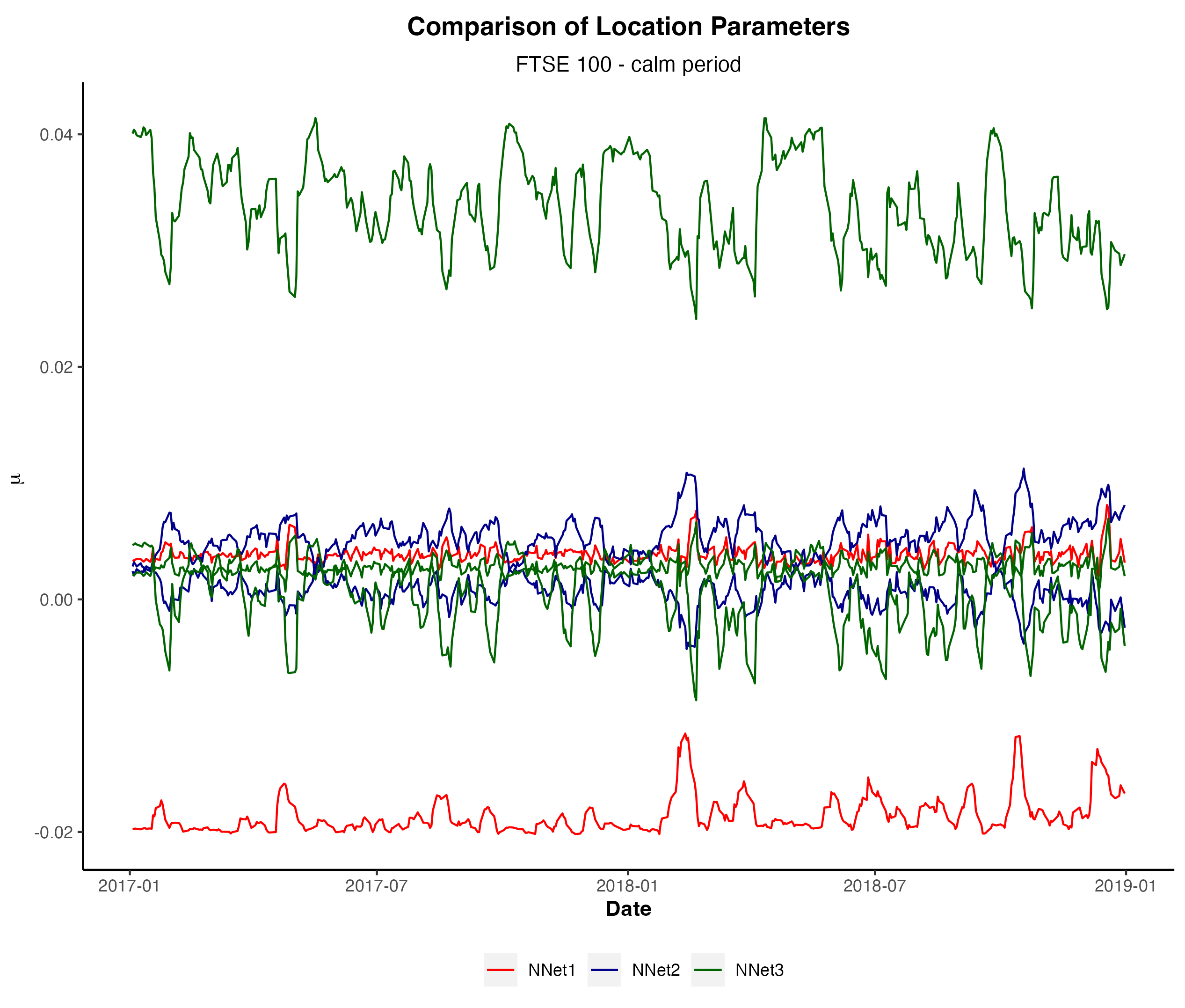}}
\hfill
\subfigure[Location Parameters - FTSE (turbulent period)]{\includegraphics[width=7.9cm]{Images/mu_params_FTSE_covid.png}}
\hfill
\subfigure[Location Parameters - S\&P (calm period)]{\includegraphics[width=7.9cm]{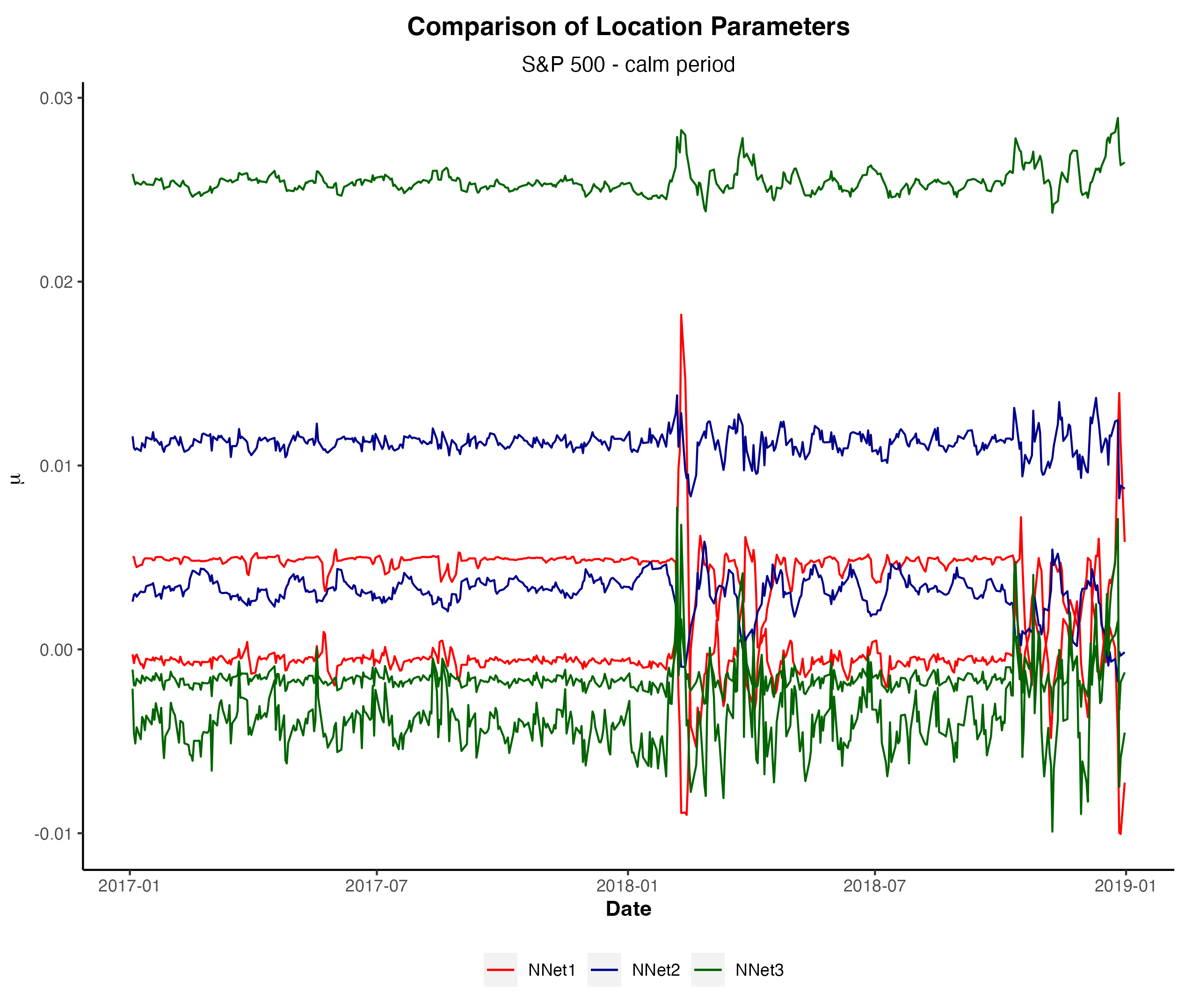}}
\hfill
\subfigure[Location Parameters - S\&P (turbulent period)]{\includegraphics[width=7.9cm]{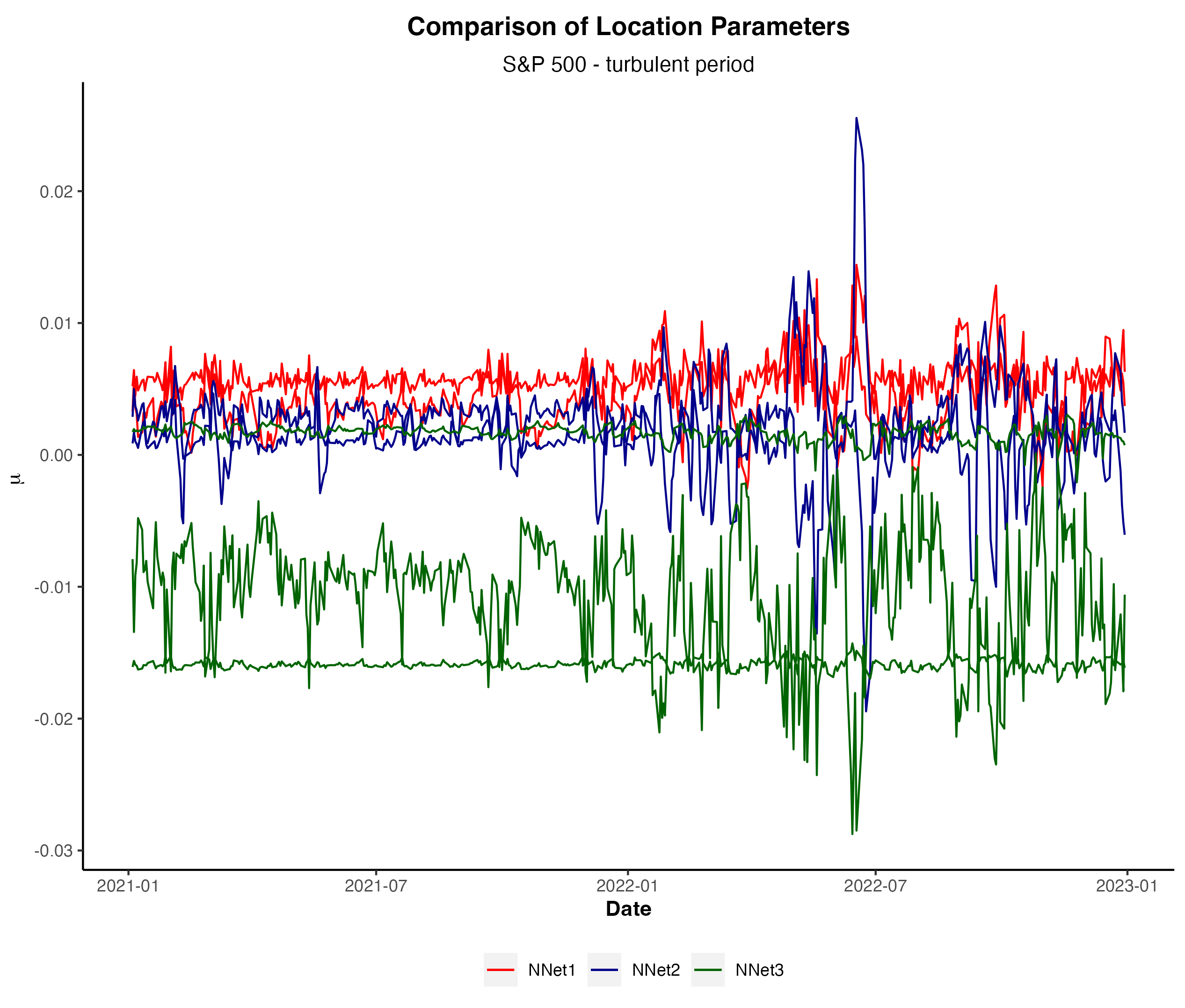}}
\hfill
\caption{Location Parameters (1/2)}
\end{figure}

\begin{figure}[H]
\subfigure[Location Parameters - EUSTOXX (calm period)]{\includegraphics[width=7.9cm]{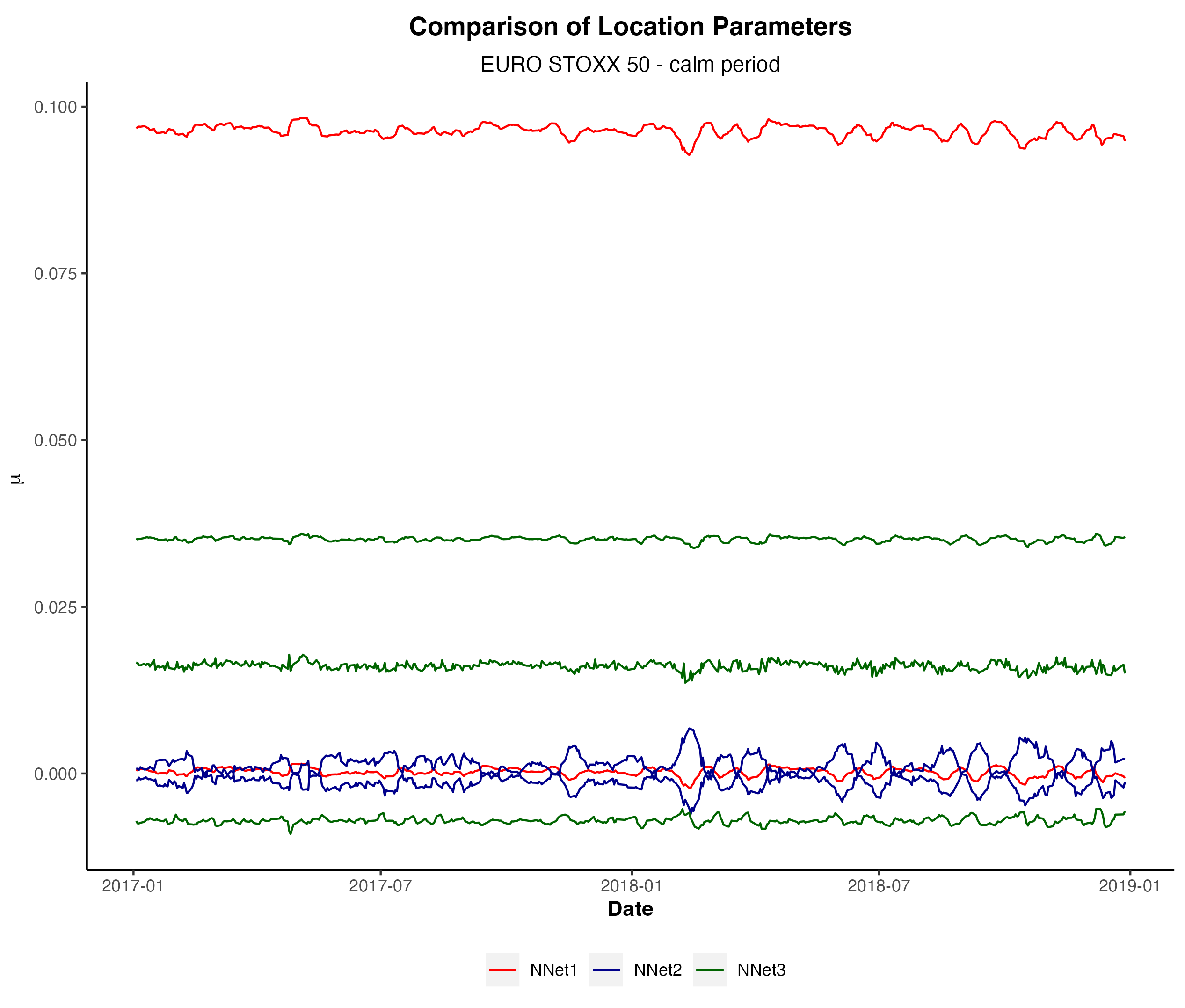}}
\hfill
\subfigure[Location Parameters - EUSTOXX (turbulent period)]{\includegraphics[width=7.9cm]{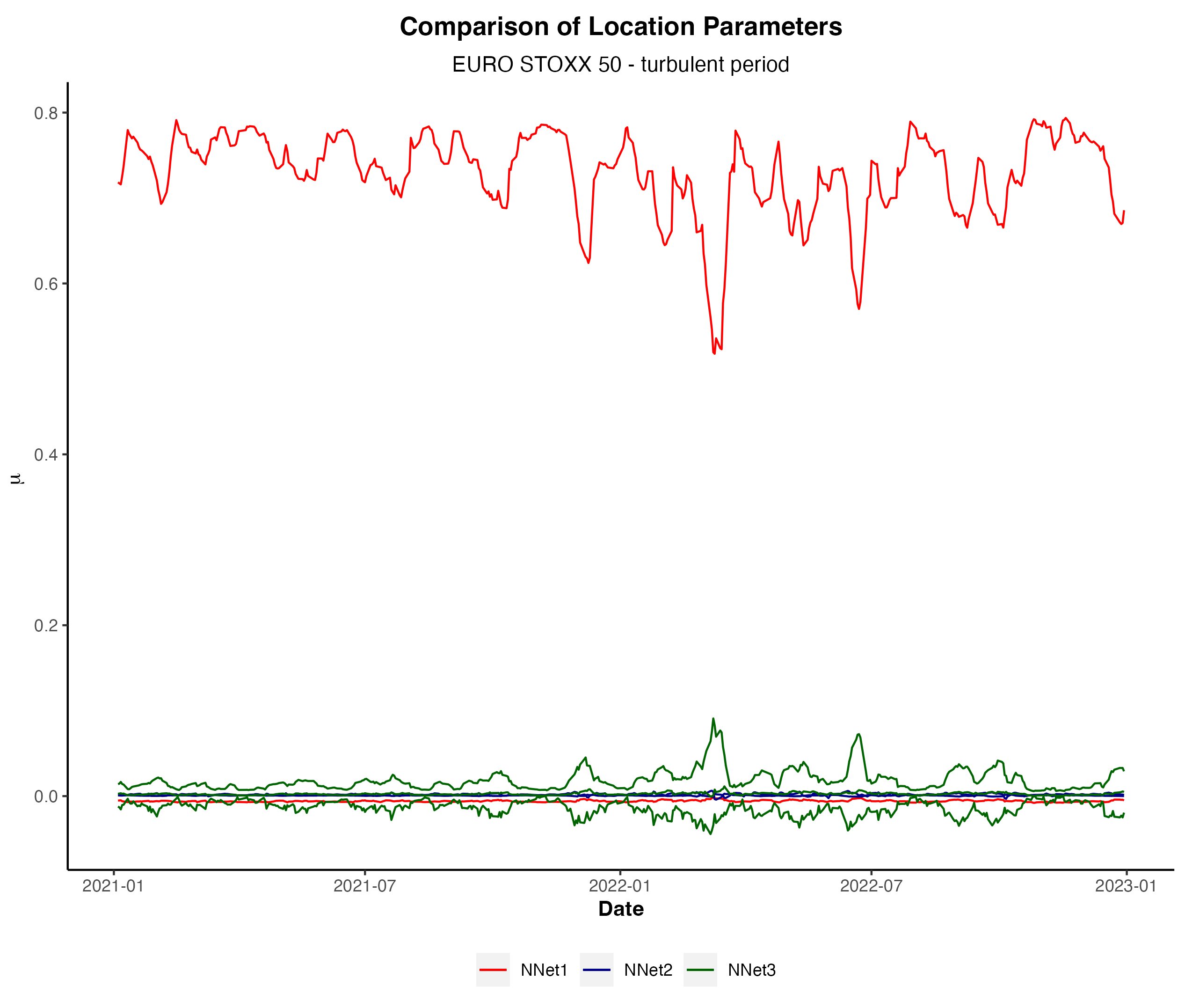}}
\hfill
\caption{Location Parameters (2/2)}
\end{figure}

% BIBLIOGRAPHY 
\bibliographystyle{apalike} % APA stle
\bibliography{main} % including Bibliography

\end{document}